\documentclass[english,final,3p,times,sort&compress]{elsarticle}

\usepackage{graphicx,bm,mdwlist}

\usepackage{float}
\usepackage{amsmath}
\usepackage{graphicx}
\usepackage{amssymb}
\usepackage[titletoc,title]{appendix}
\usepackage{bigstrut,array}

\numberwithin{equation}{section}

\def\be{\begin{equation}}
\def\ee{\end{equation}}
\def\e#1{\label{#1}\end{equation}}
\def\bea{\begin{eqnarray}}
\def\eea{\end{eqnarray}}
\def\ea#1{\label{#1}\end{eqnarray}}
\def\bes#1{\begin{subequations}\label{#1}}
\def\ese{\end{subequations}}

\journal{Physics Reports}

\begin{document}

\newcommand{\mss}[1]{\mbox{\scriptsize #1}}

\begin{frontmatter}



\title{Quantum feedback: theory, experiments, and applications}

\author[DA,CQI,RIKEN]{Jing Zhang}

\ead{jing-zhang@mail.tsinghua.edu.cn}

\author[IM,CQI,RIKEN]{Yu-xi Liu}

\author[DA,CQI,RIKEN]{Re-Bing Wu}

\author[UMD,RIKEN]{Kurt Jacobs}

\author[RIKEN,UM]{Franco Nori}

\address[DA]{Department of Automation, Tsinghua University,
Beijing 100084, P. R. China}

\address[CQI]{Center for Quantum
Information Science and Technology, TNList, Beijing 100084, P. R.
China}

\address[RIKEN]{CEMS, RIKEN, Saitama 351-0198, Japan}

\address[IM]{Institute of Microelectronics, Tsinghua University,
Beijing 100084, P. R. China}

\address[UMD]{Department of Physics, University of Massachusetts at
Boston, Boston, MA 02125, USA}

\address[UM]{Physics Department, The University of Michigan, Ann Arbor, MI 48109-1040, USA}
\begin{abstract}
The control of individual quantum systems is now a reality in a
variety of physical settings. Feedback control is an important
class of control methods because of its ability to reduce the
effects of noise. In this review we give an introductory overview
of the various ways in which feedback may be implemented in
quantum systems, the theoretical methods that are currently used
to treat it, the experiments in which it has been demonstrated
to-date, and its applications. In the last few years there has
been rapid experimental progress in the ability to realize quantum
measurement and control of mesoscopic systems. We expect that the
next few years will see further rapid advances in the precision
and sophistication of feedback control protocols realized in the
laboratory.
\end{abstract}


\end{frontmatter}

\tableofcontents

\section{Introduction}
\label{s1}

\subsection{History and background}\label{s11}

The subject of \textit{control} is concerned with methods to
manipulate the evolution of dynamical systems. As such it is
relevant to many fields both inside and outside
physics~\cite{Alessandro:2007,HMabuchiIJRNC:2005,PRouchon:2008,CBrif:2010,DYDong:2010,CAltafini:2012,JBechhoeferRMP:2005,GHuang:1983,Clark:1985,DDongNJP:2009,FMotzoiPRA:2011,JMGambettaPRA:2011,PRebentrostPRL:2009,JQYouPhysiscsToday:2005,JQYouNature:2011,PDNationRMP:2012,SNShevchenkoPR:2010,IBulutaScience:2009,IBulutaRPP:2011,IGeorgescuRMP:2014,JMaPR:2011,AGKofmanPR:2012,IGeorgescuPW:2012,CEmaryRPP:2014,SCong:2014,SSGeSJCO:2012,TLVuPRA:2012,ZGXueTAC:2013,FGaoPRA:2014,THZhangIJTP:2014,MJiangPRA:2013,CChenSWJ:2013,HZShenPRA:2013,XYLvPRA:2013,XYLvSR:2013,JQLiaoPRA:2013,JQLiaoPRA:2012,SBXuePRA:2012,XWXuPRA:2013}.
Control has a long history, but it emerged as a modern scientific
discipline only after the pioneering work of Norbert Wiener in the
1940's~\cite{Wiener:1965}. Up until the 1960's, control was
largely studied by analyzing dynamical systems in the frequency
domain (that is, the Fourier transform of the evolution). This was
a reasonable approach because people were mainly interested in
steady-state behavior. So long as the fluctuations about the
steady-state are sufficiently small, even nonlinear systems can be
well-approximated by linear dynamics, and are thus amenable to
frequency-domain methods. For networks of linear systems in which
the outputs of some systems are connected to the inputs of others,
frequency-space methods are also extremely useful, because complex
exponentials are the eigenfunctions of all linear systems.

Frequency-domain methods were less helpful in understanding how
control systems should make use of real-time information, and for
this reason control theorists turned to the time domain. The
techniques that were developed include the Kalman-Bucy
filter~\cite{Kalman:1963} and the Hamilton-Jacobi-Bellman
equation~\cite{Bellman:1960}. These are referred to as state-space
methods, and are often referred to as ``modern'' control theory.
Much of modern control theory is concerned with feedback control,
in which a control system, or ``controller'' obtains a stream of
information about the trajectory of the system, or ``plant'', and
uses this information in real time to control it. The term
``feedback'' comes from the notion that the controller is
``feeding'' the information it obtains ``back" to the system.
Feedback control is also referred to as ``closed-loop'' control,
because the flow of information to the controller, together with
the action taken by the controller to affect the system is thought
of as a loop that starts and ends at the
system~\cite{Athans2006,Isidori1995}.

One usually speaks of a controller as trying to achieve some
\textit{objective}. This objective may be to have the system reach
a given state at a given time, or to have it evolve in a precise
way, despite the presence of noise in its inputs or slight errors
in its construction. Given an objective, the central problem in
feedback control is to obtain a rule (or mapping) that the
controller can use to select the action it should take based on
the data it has received. Traditionally such a rule was referred
to as a ``feedback algorithm'', but in quantum control theory the
term ``feedback protocol" is usually used instead, so as to avoid
confusion with the algorithms of quantum
computing~\cite{MANielsen:2000}.

The idea of controlling systems using feedback has been around for
a long time; the first device on record employing feedback appears
to have been the water clock of Ktesibios in the first half of the
third century B.C.~\cite{Diels:1920,Mayr:1970,Lepschy:1992}
Another successful example of a feedback mechanism is the Watt
governor developed in the 1780's. This centrifugal governor was a
core component of the Watt steam engine which fueled the
industrial revolution. Incidentally it was Maxwell that first
performed a mathematical analysis of this control
mechanism~\cite{Maxwell:1868}. By introducing feedback control one
can speed up transient processes, tune the stationary output of a
system, and most importantly, reduce the effects of disturbances.
The importance of using feedback in controlling a system is that
it is the only way to reduce the effects of noise. Noise
introduces uncertainty into the system dynamics, and the only way
to reduce this uncertainty is to transfer it to another system. To
understand this better, consider what happens if we make a
measurement on a system. This reduces our uncertainty and allows
us to correct the motion. In doing so, we reduce the spread in the
state of the system, and thus the randomness in the system. But
note that the measuring device must record the result of the
measurement, and this result is necessarily as random as the
quantity being measured. Thus the measurement and feedback
transfers randomness from the system to the memory of the
measuring device. More fundamentally, randomness is entropy, and
because all physical laws are reversible, the only way to reduce
entropy in any system is to transfer it to another system, which
might be a thermal bath.  In the study of quantum feedback
control, it is natural to refer to any process that transfers
entropy from the system to the controller as a feedback process.

Feedback control was introduced into quantum dynamics in the early
1980's~\cite{HMWiseman:2009,MRJames:2005,HMabuchiScience:2002,SHabibLAS:2002,GFZhangCSB:2012,ASerafiniISRNOptics:2012}, but it was not until the 1990's that it began to be studied and
applied in earnest. Naturally the prerequisite for describing continuous-time measurement-based
feedback in quantum systems was a description of continuous quantum measurement. In the late 80's and early
90's a number of authors independently derived equations describing the continuous measurement of quantum systems. Srinivas and Davies~\cite{Srinivas:1981} obtained a stochastic equation for the quantum state an optical cavity under photon detections, Gisin~\cite{Gisin:1984} introduced a stochastic equation for a state vector that reproduced certain properties of measurement, and Diosi~\cite{Diosi:1986} introduced a stochastic equation for the quantum state of an open system. We note also that, at a similar time, Barchielli and Lupieri obtained an equation describing continuous measurement in the Heisenberg picture~\cite{Barchielli85}. Subsequently and independently, Belavkin~\cite{Belavkin:1987} (building on the work of Stratonovich), Diosi~\cite{Diosi88a, Diosi88b}, and Wiseman and Milburn~\cite{HMWisemanPRA_MF:1993,HMWisemanPRL_MF:1993} (building on the ``quantum trajectory'' work of Carmichael~\cite{Carmichael:1989, Carmichael:1993}) obtained an equation that described the evolution of the state of a quantum system under a continuous measurement with Gaussian noise. This is the \textit{stochastic master equation} that describes the continuous update in the observer's state of knowledge, and is the quantum equivalent of the Stratonovich equation that describes a Gaussian continuous measurement on a classical system~\cite{Maybeck:1982}. (The Kalman-Bucy filter is the special case of the Stratonovich equation for linear systems, in which the measurement is restricted to linear functions of the dynamical variables.) While Diosi considered only measurements on pure states, the other authors considered also mixed states, which makes possible the description of inefficient measurements, and shows how the quantum measurement obtains classical information about the state, thus purifying it in a way that is closely analogous to the action of classical measurements.

It was Belavkin who first presented a mathematical theory of feedback control in quantum systems~\cite{VPBelavkinJMA:1992,VPBelavkinCMP:1992}. Since Belavkin had derived quantum continuous measurements as an extension of the theory of classical continuous measurements, where the latter are used heavily in control theory, it was natural for him to consider the application of continuous measurements to feedback control of quantum systems, and to consider adapting the techniques from classical control~\cite{Belavkin:1983, Belavkin:1987}. The highly mathematical nature of Belavkins work, however, prevented it from being absorbed by the physics community, where applications for these ideas for were to arise in the following decade.

Taking a very different approach to quantum feedback, in 1994 Wiseman and Milburn showed that a Markovian master equation could be derived to describe continuous feedback in quantum systems, if the feedback was given by a particularly simple function of the stream of measurement results~\cite{HMWisemanPRA_MF:1994} (This kind of feedback is now referred to as
Markovian feedback). In 1998, Yanigasawa and
Kimura~\cite{Yanagisawa:1998} and Doherty and
Jacobs~\cite{Doherty:1999} introduced the notion of performing
feedback using estimates obtained from the SME, in the control
literature and physics literature, respectively. Both sets of
authors showed that for linear systems this class of feedback
protocols was equivalent to modern classical feedback control, so
that standard results for optimal control could be transferred to
quantum systems. This method was in fact that proposed by Belavkin
in 1983 in analogy to that used in classical control
theory~\cite{Belavkin:1983, Belavkin:1987}. In quantum control,
using estimates obtained from the SME is often referred to as
Bayesian feedback to distinguish it from Markovian feedback. In
the former the measurement results are processed (``filtered") to
obtain an estimate of properties of the current state, whereas in
the latter the measurement stream is fed back directly.

In Wiseman's seminal work on feedback control, he showed that
feedback mediated by continuous measurements can in fact be
implemented without measurements~\cite{HMWisemanPRA:1994}. To see
how this works, let us consider two parallel mirrors between which
a single mode of the electromagnetic field is trapped (the two
mirrors are referred to as an ``optical cavity''). The light that
leaks out through one of the mirrors can be detected, and the
information is used to manipulate the optical mode. Alternatively,
the output light can be directed to a mirror of another optical
cavity, and thus forms an input for this cavity. If we then
connect an output from the second cavity back to the first we have
a loop, and light can be made to travel only one way around the
loop by the use of optical circulators~\cite{Kenyon}. For
describing this situation the quantum input-output theory
developed by Collet and Gardiner is
invaluable~\cite{CWGardinerPRA:1985,CWGardinerBook:2004}. The
process of connecting quantum systems together via free-space
one-way traveling-wave fields was first considered by
Gardiner~\cite{CWGardinerPRL:1993} and
Carmichael~\cite{HJCarmichaelPRL:1993}, where the former called it
a ``cascade connection''. Wiseman showed that cascade connections
can implement the same feedback control processes as Markovian
measurement-based feedback and can perform tasks that the latter
cannot~\cite{HMWisemanPRA:1994}.

A second notion of feedback control without explicit measurements
was introduced by Lloyd in 2000. He suggested that a unitary
interaction between two quantum systems could be used to implement
feedback control~\cite{SLloydPRA:2000,RJNelsonPRL:2000}. This can
be achieved, for example, by choosing the interaction so as to
correlate the two systems, i.e., the controlled system and the
controller, whereby the state of the controller is dependent on
the state of the system. One then chooses a second interaction in
which the evolution of the system depends on the state of the
controller. This particular process is equivalent to a measurement
followed by a unitary feedback operation that depends upon the
measurement result, although coherent feedback processes are not
restricted to this form~\cite{Jacobs14c,BQiSCL:2010}. Both kinds
of ``measurement-free'' feedback, that mediated by cascade
connections and that which uses unitary interactions are now
referred to as \textit{coherent feedback control} (CFC), and the
latter is often called ``direct'' coherent-feedback. All control
involving explicit measurements is usually called
\textit{measurement-based} feedback control, or just
\textit{measurement feedback control} (MFC).

In the 2000's James and his collaborators studied ``feedback
networks'' of linear quantum systems connected by one-way
fields~\cite{MRJamesTAC:2008}, and Gough and James built on
input-output theory to construct a compact and convenient
formalism to handle arbitrarily complex
networks~\cite{GoughTAC:2009}. More recently a number of authors
have considered the use of nonlinear coherent-feedback networks
for various control
tasks~\cite{HMabuchiAPL:2011,JKerckhoffPRL:2012,JZhangTAC:2012,ZPLiuPRA:2013}.
In 2009, Nurdin, James, and Peterson showed that linear coherent
feedback networks could out-perform linear measurement-based
feedback~\cite{HINurdinAutomatica:2009}, suggesting that
measurement-based feedback was limited by the need to reduce the
information about a system to classical numbers. It is also shown
quite recently that coherent feedback can achieve more for
generating quantum
nonlinearity~\cite{JZhangTAC:2012,ZPLiuPRA:2013} and
cooling~\cite{RHamerlyPRL:2012} compared with the
measurement-based feedback. The relationship between
measurement-based and coherent feedback is a topic of current
research~\cite{Jacobs14c}.

There are not only fundamental differences between
measurement-based and coherent feedback, but also important
practical differences. Making measurements on quantum systems,
often possessing only a few quanta, usually requires a tremendous
amplification of the signal. This is because the measurement
results, by definition, are well-defined classical
numbers~\cite{CBrifPRA:2010}. To robustly store and manipulate
such numbers requires states with energies much greater than a
single quantum. Amplifying signals at the single-quantum scale
without swamping them with noise is a great challenge, and is one
major practical disadvantage of measurement-based feedback. A
second disadvantage is the timescale required to obtain and then
process the measurement results (usually on a digital device). On
the other hand, measurement-based feedback has the advantage that
the processing of the information is essentially noise-free. By
contrast, if a quantum system is used as a controller it will
likely be subject to noise processes from its environment. It may
also be less clear how to use the quantum system to process the
information to achieve a control objective.

It is important to note that the method of ``adaptive feedback'',
in which the term ``feedback'' is used, is not the feedback
control that we are concerned with in this review. Adaptive feedback~\cite{CBrif:2010}
is a method for obtaining control protocols, not a class of
protocols for controlling a system. In this method, one chooses an
arbitrary control protocol, tries it out on the system, and based
on the result make a modification to the protocol and tries it
again. In this way one can use one of many search algorithms to
look for a good protocol. People who refer to adaptive feedback as
a feedback method distinguish the feedback control we consider
here by calling it ``real-time feedback control''.

It is also important to note that we do not discuss in detail here all the
ways in which feedback can be realized. One could, for example,
perform a series of ``single-shot'' measurements with a discrete
set of outcomes, and perform a unitary action on the system for
each outcome. While there are certainly a range of interesting and
non-trivial questions regarding such feedback, such as controlling
thermal
dynamics~\cite{KMaruyamaRMP:2009,TSagawaPRL:2008,TSagawaPRE:2012,DAbreuPRL:2012,KJacobsPRA:2009}
and quantum error
correction~\cite{PWShorPRA:1995,AMSteanePRSLSA:1996,EKnillPRA:1997,JICiracScience:1996,WHZurekPRL:1996,LMDuanPRL:1997,SLloydPRL:1998},
the mathematical machinery required to analyze it does not require
stochastic differential equations. This is also true of coherent
feedback implemented via unitary interactions. This latter topic
has only recently begun to be explored in earnest, and there are
certainly many open questions~\cite{KJacobsarxiv:2013}. However in
this review we focus on continuous-time measurement-based feedback control, coherent feedback mediated by continuous fields that cary the information between system and controller. Both of these require the use of
stochastic (Ito) calculus, something that is less familiar to many
researchers in quantum theory. While measurement-based feedback
requires only the usual Ito stochastic calculus, field-mediated coherent
feedback requires a quantum version of Ito calculus developed by
Gardiner and Collett as part of their input-output
theory~\cite{Collett84,CWGardinerPRA:1985}. This quantum
stochastic calculus was also developed independently by Hudson and
Parthasarathy in a more rigorous measure-theoretic
way~\cite{RIHudsonCMP:1984}. A readily accessible introduction to
Ito calculus can be found in~\cite{JacobsSP}, and the quantum
version is described in~\cite{CWGardinerPRA:1985,
CWGardinerBook:2004,Jacobs14}.

To distinguish between experiments that realize quantum feedback
control rather than classical control, we apply the following criteria. Measurement-based feedback is
\textit{quantum} feedback if the dynamics of the system under the feedback loop cannot be
explained merely by using Bayes' theorem. Another way to say this is that the \textit{quantum back-action} from the measurement, which is the \textit{dynamical} effect of the measurement on the system, plays a significant role in the system evolution. Coherent feedback if the joint dynamics of the system and controller cannot be described by a classical model. For linear systems, the
only distinction between quantum and classical motion is that the
joint-uncertainty of position and momentum is limited by
Heisenberg's uncertainty principle. A measurement introduces noise
because a reduction in the uncertainty of one canonical variable
tends to increase the uncertainty of the conjugate variable.
Feedback control of a quantum harmonic oscillator can thus be
considered quantum mechanical if either (i) the ``back-action''
noise from the measurement must be taken into account in
understanding the behavior, or (ii) one of the
canonical variables has its uncertainty reduced below that of the
vacuum state, so as to produce a so-called ``squeezed state''.

Experiments implementing measurement-based feedback in the quantum
regime were realized initially in quantum optics, where it first
became possible to measure individual microscopic degrees of
freedom with sufficient fidelity. These were followed by
experiments involving trapped atoms and ions, and very recently it
has become possible to realize measurement-based feedback control
in mesoscopic superconducting circuits. We review experiments in
these various physical settings in Section~\ref{s5}. Experiments
involving continuous coherent feedback were performed prior to
those realizing continuous measurement-based feedback, although at
the time these experiments were not thought of as involving
feedback. An example is the cooling of trapped ions using the
``resolved sideband'' cooling method~\cite{Wineland75,
Diedrich89}. In Section~\ref{s53} we summarize recent experiments
implementing coherent feedback cooling of mechanical resonators
(using the resolved-sideband technique) and in Sections~\ref{s51}
and \ref{s54} we summarize recent experiments whose primary
purpose is the demonstration of coherent feedback.

In the remainder of this section we give a brief introduction to
classical feedback control. In Sec.~\ref{s2} we discuss quantum
continuous (weak) measurements and filtering, and their
application to quantum measurement-based feedback. In particular
we discuss the two ends of the spectrum of measurement-based
feedback: the simplest in which the measurement signal is not
processed at all before it is fed back to the system
(``Markovian'' feedback), and that in which the measurement signal
is fully processed to obtain the observer's complete
state-of-knowledge of the system as it evolves (``Bayesian''
feedback). We complete Sec.~\ref{s2} by giving an overview of most
of the applications of measurement-based feedback that have been
considered in the literature to-date. In Sec.~\ref{s3} we turn to
coherent feedback. We discuss the two primary ways in which it can
be implemented, and the formalism used to describe them. As with
measurement-based feedback, we then review the majority of
applications of coherent feedback that have been considered
to-date. In Sec.~\ref{s4} we review two further topics that
involve feedback, but not in the way envisioned in the traditional
notion of feedback control. In Sec.~\ref{s5} we turn to
experiments, and give an overview of all experiments to-date that
have realized continuous feedback control in the quantum regime.
These experiments cover a range of physical settings from quantum
optics to superconducting circuits. We also review all experiments
realized to-date whose purpose is to demonstrate coherent
feedback, as well as recent experiments that use coherent feedback
to cool mechanical resonators. In Section~\ref{s6} we give a
perspective on the current state of quantum feedback control and
discuss some open questions.

\subsection{A glance at classical feedback control}\label{s12}

In the engineering discipline called \textit{control theory}, a
control system is always broken into three
parts~\cite{AGolnaraghi2009}:
\begin{itemize}
  \item The system (or ``plant''): the device we want to control, having inputs and
  outputs;
  \item The input(s) to the system (or ``control''): the entity that we have freedom to choose to affect the
  system;
  \item The output(s) of the system (or ``yield''): this includes the quantity we want to control, and any quantities we can measure to obtain information about the plant.
\end{itemize}
As explained above, the explicitly causal structure, in which the
control system first obtains information from a measurement and
uses this to determine the input to the system, is a way of
thinking about the interaction between two systems that is
conceptually useful for feedback control. One can think about the
interaction in this way even if this structure is not explicit in
the mathematical description of the interaction. An example in
which it is not explicit is in the Hamiltonian description of an
interaction.
\begin{figure}
  \centering
  \includegraphics[width=10 cm]{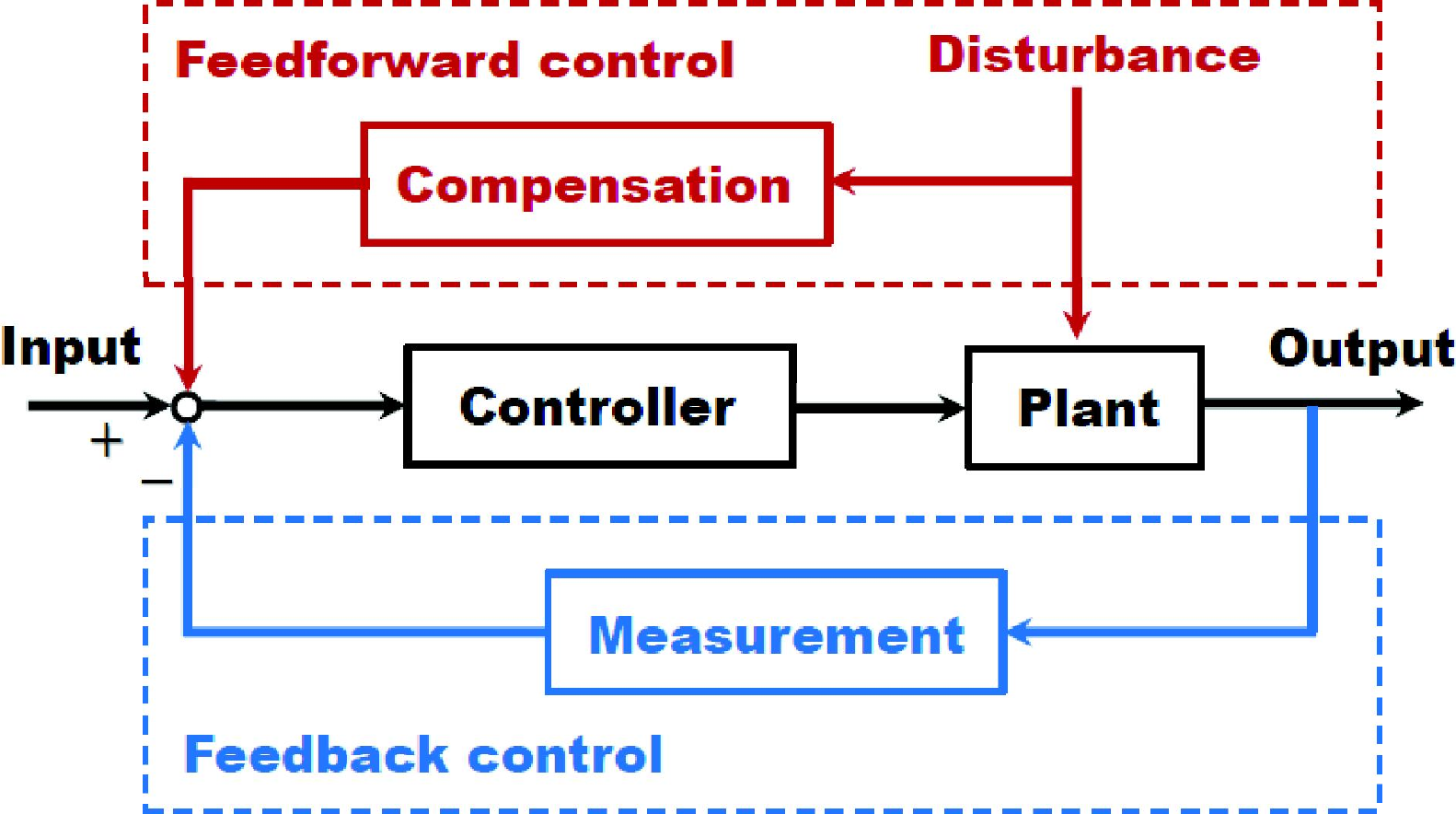}\\
  \caption{(Color online) Basic components in a control system with feedback (bottom loop, in blue) and feedforward (top branch, in red) controllers.}\label{Flowchart}
\end{figure}

The usual objective of control is to steer one or more outputs of
the plant toward a prescribed behavior against unknown
disturbances or noises. As shown in Fig.~\ref{Flowchart}, when the
disturbance is known (or can be precisely measured), a feedforward
controller can be used to cancel the effects of disturbances at
the input side. Otherwise, if the disturbance is unknown (or
cannot be precisely measured), a feedback controller must be
introduced to adjust the input according to the outputs of the
prescribed measurements imposed on the controlled system.

\subsubsection{Classical model of a control system}
We are concerned with dynamical systems that change continuously
in time, and are therefore modeled by differential equations. The
standard model of a control system can be written as
\begin{eqnarray}
    \dot{x}(t)&=&F\left[x(t),u(t)\right],\quad x(0)=x_0, \label{Eq.Classical Control system}\\
    y(t)&=&G\left[x(t),u(t)\right], \label{Eq.Classical Output}
\end{eqnarray}
where $F$ and $G$ are arbitrary vector-valued functions, the
vector $x(t)$ is the state of the system, $u(t)$ is the control
(set of inputs) that drives the system, and $y(t)$ is the set of
outputs, which is allowed to be some algebraic function of both
the state and the input.

In practice, the following linear control system model is favored
by control engineers:
\begin{eqnarray}
    \dot{x}(t)&=&Ax(t)+Bu(t),\quad x(0)=x_0, \label{Eq.Classical Linear Control system}\\
    y(t)&=&Cx(t),\label{Eq.Classical Linear Output}
\end{eqnarray}
in which $A$, $B$, and $C$ are constant matrices. This model can
be solved analytically, which makes it convenient for design
purposes. Many nonlinear systems can be transformed to a linear
system under a proper nonlinear coordinate transformation. Such a
procedure is called linearization, which is broadly used in the
literature. For those systems that cannot be precisely linearized,
one can still often approximately linearize them in a small
neighborhood of the ``working point'', using a perturbative
technique.

The relation between the input and output of a linear control
system can be alternatively characterized in the frequency domain
by a transfer function obtained by taking the Laplace transform of
Eq.~(\ref{Eq.Classical Control system}):
\begin{equation}\label{Eq.CCS-LT}
    Y(s)=G(s)U(s),\quad G(s)=C(sI-A)^{-1}B,
\end{equation}
where $U(s)$ and $Y(s)$ are the Laplace transforms of the input
$u(t)$ and the output $y(t)$, respectively, and $G(s)$ is the
transfer function of the system. Transfer function models are
popular in control engineering because they can be constructed
directly from the input-output data without having to know the
internal structure of the system. The corresponding analysis and
design are conceptually simple and can be visualized using Nyquist
or Bode plots, which require few computational
resources~\cite{JBechhoeferRMP:2005}.

\subsubsection{System analysis and design}
Consider the linear control system given by
Eqs.~(\ref{Eq.Classical Linear Control system}) and
(\ref{Eq.Classical Linear Output}). To implement feedback control,
we make the control, $u(t)$, a function of the state $x(t)$,
assuming that $x(t)$ can be determined from the measured outputs.
To keep the system dynamics linear, we set $u(t)=Kx(t)+r(t)$,
where $K$ is a constant matrix and $r(t)$ is the external input
signal. The matrix $K$ is called the feedback \textit{gain}, and
the function $r(t)$ is called the \textit{reference signal} or
\textit{command signal}, which is sometimes chosen as the
reference signal to be tracked by the system output under control.
With this choice for $u(t)$, the dynamics of the resulting
\textit{closed-loop} system becomes
\begin{equation}\label{Eq.Classical Closed-loop Control system}
    \dot{x}(t)=(A-BK)x(t)+Br(t), \quad y(t)=Cx(t).
\end{equation}
Correspondingly, the closed-loop transfer function from the
reference signal, $r(t)$, to the output we wish to control,
$y(t)$, becomes
\begin{eqnarray}
G_{\rm CL}(s)=C(sI-A-BK)^{-1}B. \label{Eq.CCS-FBLT}
\end{eqnarray}
The gain matrix and reference signal are together called the
\textit{control law}.

Two common control tasks are:
\begin{itemize}
  \item Regulation: to find a control law that keeps $y(t)$ close to some predetermined function of time.
  \item Tracking: to find a control law that keeps $y(t)$ close to a time-varying signal that is not known beforehand.
\end{itemize}


A prerequisite for accomplishing these tasks is that the
controlled system is \textit{stable}, and the systems stability
considered here is usually quantified by using the concept of
\textit{Lyapunov stability}. For linear systems, stability is
ensured by choosing the gain matrix $K$ so that the poles of the
transfer function lie in the left half of the complex plane, and
sufficiently far from the imaginary axis.  A central result of
control theory for linear systems, referred to as the \textit{pole
assignment theorem}, states that one can choose $K$ to place the
poles at arbitrary locations in the complex plane if and only if
the system is fully controllable, meaning that it is possible to
choose $K$ and $r(t)$ to steer the system from any state to the
origin.

\subsubsection{Optimal control}

In applications, one is often interested in obtaining a given
output while using minimal resources~\cite{Athans2006}. We can
formulate this goal as a typical constrained optimization problem.
If we define a function that measures how far the output is to the
desired output (the error incurred by the control law), and
another function that quantifies the resource cost of the inputs,
we can attempt to minimize the latter under a constraint on the
former. The dynamics of the system is essentially another
constraint in this optimization problem. We can alternatively
define a single ``cost'' function that combines the error and
resource cost, and attempt to minimize it.  A well-motivated form
for the cost function $J$ is
\begin{equation}\label{Eq: Cost}
  J[u(t)] =\Phi[x(T)]+\!\int_0^{\mss{T}}\!\! L[x(t),u(t)]\,{\rm d} t,
\end{equation}
where the differential equation given by Eq.~(\ref{Eq.Classical
Control system}) is the dynamical constraint.

The theory of optimal control is a beautiful part of modern
control theory that can be analyzed with variational methods. In
fact, with the above  form for $J$, this theory has the same
structure as that of Lagrangian and Hamiltonian mechanics. The
reason for this is that the Langrange equations give the
conditions for the minimization of an action, which has the same
form as $J$.

Subject to the restriction given by Eq.(\ref{Eq.Classical Control
system}), one can introduce a Lagrangian multiplier $\lambda(t)$,
and this turns out to be the ``momentum coordinate'' conjugate to
$x(t)$ in the sense of Hamiltonian mechanics. The (pseudo)
Hamiltonian to which this conjugate coordinate corresponds is
\begin{equation}\label{Eq: Pseudo Hamiltonian}
  \mathbf{H}[x(t),u(t)]=L[x(t),u(t)]+\lambda(t)^{\mss{T}}F[x(t),u(t)].
\end{equation}
One can prove that the necessary condition for a control $u(t)$ to
be optimal is
\begin{equation}\label{Eq: OCT condition}
  \frac{\partial\mathbf{H}[x(t),u(t)]}{\partial u(t)}\Big|_{u(t)=u_{\rm opt}(t)}=0,
\end{equation}
and $x(t)$ and $\lambda(t)$ can be obtained by solving the
following conjugate equations
\begin{eqnarray}\label{Eq: Hamiltonian Eq}
  \dot x(t)&=&\frac{\partial\mathbf{H}[x(t),u(t)]}{\partial \lambda(t)},\\
  \dot \lambda(t)&=&-\frac{\partial\mathbf{H}[x(t),u(t)]}{\partial x(t)}.
\end{eqnarray}
From the viewpoint of Lagrangian mechanics, the evolution of the
system under the control $u(t)$ minimizes the ``action'' $J$.

If the set of admissible controls $u(t)$ is not an open set, the
condition (\ref{Eq: OCT condition}) must be replaced by a more
general condition, due to the fact that $u(t)$ can no longer be
taken at any point in this set. In this case, the necessary
condition under which $u(t)$ is optimal is that $u(t)$ minimizes
$\mathbf{H}[x(t),u(t)]$. We can write this condition as
\begin{equation}\label{Max Principle}
  u_{\rm opt}(t) = \arg \left[ \min_{u(t)}\;\mathbf{H}[x(t),u(t)] \right] ,
\end{equation}
which is referred to as the ``Maximum Principle''. An alternative
technique called dynamical programming can be used to locate the
global optimal solutions for $u(t)$ by solving the so-called
Hamiltonian-Jacobi-Bellman equation~\cite{Jacobs14}, but requires
much higher computation resources than solving Eq. (\ref{Eq: OCT
condition}) or (\ref{Max Principle}). All these approaches merely
provide necessary conditions for optimality.

\subsubsection{Fighting disturbances and uncertainties}

So far we have not included the effects of uncertainty or noise in
the system control model, but we have to do so if the controller
is to combat them. If the system is driven by a noise of which the
spectrum is known, we can include this noise in the model of the
system, and explicitly calculate the observer's estimate of the
system state derived from a measurement process in real-time.
Feedback control can then be implemented based on this estimate.
Sometimes it is possible to minimize the noise in a subset of
system variables at the expense of others, which is analogous to
the squeezing of an optical beam or a quantum oscillator.

Alternatively, there may be uncertainties in the parameters that
determine the dynamics of the system. If we consider a linear
system whose equation of motion has the form of $\dot x = A x +
f(t)$, there are two distinct ways to show the uncertainty in the
system parameters. First, the driving term $f(t)$ may contain
unknown or partially known parameters, such as the phase or
amplitude of a sinusoidal drive. This is very similar to a
time-variant noise driving the system, except that in this case a
continuous measurement imposed on the system will provide
sufficient information to estimate the values of the unknown
parameters, and thus reduce the parametric uncertainty. Second,
there may exist uncertainty in the system dynamical matrix $A$,
due to which the oscillation frequencies of the system are
partially unknown. Such parametric uncertainty is often referred
to as \textit{model uncertainty}. Once again a measurement can be
used to extract information about $A$, and then one can implement
a feedback control with a gain matrix $K$ and reference signal
$r(t)$ to minimize the effects induced by the model uncertainty. A
control law that maintains a specific performance under (bounded)
variations in the system parameters is referred to as being
\textit{robust}. Such robust control problems have stimulated a
rich body of studies in the literature~\cite{Zhou1995}.

\section{Quantum Measurement-Based Feedback}
\label{s2}

As a first example we consider feedback based on the results of a
von Neumann measurement. Each outcome of  a von Neumann
measurement projects a system into one of a set of basis states.
For each of these states we are then able to perform a different
unitary operation on the system. This form of measurement-based
feedback does have important applications, an example of which is
quantum error correction~\cite{PWShorPRA:1995,AMSteanePRSLSA:1996,EKnillPRA:1997,JICiracScience:1996,WHZurekPRL:1996,LMDuanPRL:1997,SLloydPRL:1998}.
The simplest  example of error-correction is the three-qubit
``bit-flip'' code, in which the state of a single logical qubit is
encoded in three physical qubits by using the mapping
$|0\rangle\rightarrow|\bar{0}\rangle =|000\rangle$ and
$|1\rangle\rightarrow|\bar{1}\rangle=|111\rangle$. If any one of
the physical qubits suffers from an error that flips the states
$|0\rangle$ and $|1\rangle$, then this error can be corrected
without otherwise disturbing the joint state of the three qubits.
This correction is achieved by making a measurement that tells us
about the total parity of each pair of adjacent qubits. To do this
we make two measurements, one of which projects the state onto one
of the eigenstates of $M_0=\sigma_z \otimes \sigma_z \otimes I$
and the other onto the eigenstates of $M_1=I \otimes\sigma_z
\otimes\sigma_z$. From the two measurement outcomes we can
determine which bit has flipped, and thus apply a $\sigma_x$
operator to that qubit to correct the error. For example, if the
two measurements return even and odd parity, respectively, then it
is the third qubit that has flipped, and the feedback operation
that corrects the error is the unitary operator $ I \otimes I
\otimes \sigma_x$~\cite{PWShorPRA:1995,
CAhnPRA:2002,CAhnPRA:2003}. Since this error-correction code can
correct a flip error on any single qubit, it is only if an error
occurs on two or more of the physical qubits that the logical
qubit will be corrupted. If the errors on each of the physical
qubits are independent, and occur with a probability $p \ll 1$,
then the probability of an error on two or more qubits is
proportional to $p^2$, which is much less than $p$.

Maxwell's famous demon is another simple example of quantum
feedback, and one that can be usefully analyzed in terms of von
Neumann measurements. The ``demon'' is a device that makes
measurements on a system and uses the information obtained to
extract
work~\cite{KMaruyamaRMP:2009,TSagawaPRL:2008,TSagawaPRE:2012,DAbreuPRL:2012,KJacobsPRA:2009}.


\subsection{Continuous quantum measurements}\label{s21}

\subsubsection{Quantum filtering}\label{s211}

When we make a continuous measurement on a quantum system, if we
know the dynamics of the system, then we can derive an equation of
motion for our full state-of-knowledge of the system determined by
the continuous stream of measurement results. For a classical
system, an observer's state-of-knowledge is given by a probability
distribution over the state-space. For a quantum system, it is the
density matrix that captures all the information that an observer
has about a system. Control theorists refer to the process by
which an observer calculates his or her state of knowledge of a
system from a series of measurement results \textit{filtering},
and the quantum version of this \textit{quantum
filtering}~\cite{VPBelavkinJMA:1992,VPBelavkinCMP:1992,Bouten:2007}.
In classical control theory, when we can only obtain partial
information of the system state from the measurement output (e.g.,
we can measure the position but not the momentum of a mechanical
system), we can introduce a dynamical system called a filter,
using the measurement results as inputs to recover the whole
system dynamics. However, quantum filtering is not just a trivial
extension of classical filtering. In quantum filtering, we feed
the measurement output (which is a classical signal) into a
classical system to generate an estimated state of the measured
quantum system. Thus, we use a classical system to mimic a quantum
system, and some particular quantum effects, such as quantum
coherence, may be lost during this process by the action of the
measurement on the system. Quantum filtering is a bridge between a
quantum system and a classical controller, since the classical
controller can use the resulting state-of-knowledge to decide how
to apply control forces to the system. For certain systems,
results for optimal control from classical control theory can be
directly applied to obtain optimal protocols for quantum
system~\cite{Doherty:1999,Doherty:2000}.

We now present the theory of quantum measurement and filtering by
first extending classical probability theory to quantum mechanics.
A different approach to deriving the quantum filtering equation,
which uses the Schr\"{o}dinger picture and avoids the use of
measure-theoretic probability theory, can be found
in~\cite{Jacobs1:2006}. The main difference between classical and
quantum mechanics is that quantum mechanics is
\textit{noncommutative}, by which we mean that the operators that
represent different physical variables do not always commute with
each other. Heisenberg's uncertainty principle, for example, is a
direct result of the commutation relation $[x,p]=i\hbar$, between
the position $x$ and the momentum $p$ of a quantum degree of
freedom. Because of this, quantum probability theory is a
noncommutative version of classical probability theory. Recall
that classical probability theory consists of the triple,
$\left(\Omega,\mathcal{F},\mathbb{P}\right)$, referred to as a
probability space Here $\Omega$ is the sample space which is the
set of all elementary, mutually exclusive outcomes. For example,
for probability experiment of throwing a coin, the sample space is
the set $\Omega=\left\{{\rm head}, {\rm tail}\right\}$. The second
item in the triple, $\mathcal{F}$, is the set of ``events", where
each event is some subset of the possible outcomes (a subset of
the set $\Omega$). This makes $\mathcal{F}$ a so-called
$\sigma$-algebra, which satisfies the following conditions: (i)
the empty set $\emptyset$ belongs to $\mathcal{F}$; (ii)
$\mathcal{F}$ is closed under complement: $\Omega\backslash
A\in\mathcal{F}$ if $A\in\mathcal{F}$; and (iii) $\mathcal{F}$ is
closed under countable unions:
$\bigcup_{n=1}^{\infty}A_n\in\mathcal{F}$ if all
$A_n\in\mathcal{F}$. The elements of $\mathcal{F}$ can also be
equivalently expressed as a function defined on $\Omega$ (that is,
something that associates a value with every outcome), which is
called a random variable. In fact, for any $A \in \mathcal{F}$, we
can define a random variable
$\chi_A:\,\Omega\rightarrow\mathcal{R}$ such that
$\chi_A\left(\omega\right)=1$ if $\omega\in A$, and
$\chi_A\left(\omega\right)=0$ if it is not. The final item in the
triple is the probability measure $\mathbb{P}$, which is a
function that associates a probability with every subset of
$\mathcal{F}$: ($\mathbb{P}:\mathcal{F}\rightarrow[0,1]$) such
that: (i) $\mathbb{P}$ is countably additive, i.e.,
$\mathbb{P}\left(\bigcup_{n=1}^{\infty}A_n\right)=\sum_{n=1}^{\infty}\mathbb{P}\left(A_n\right)$
for any sequence $A_1,\,A_2,\,\cdots,\,A_n$ of disjoint sets in
$\mathcal{F}$, and (ii) the measure of the whole sample space
$\Omega$ is normalized so that $\mathbb{P}\left(\Omega\right)=1$.

Quantum probability theory, developed in the
1980s~\cite{RIHudsonCMP:1984,VPBelavkinJMA:1992,VPBelavkinCMP:1992,Doherty:1999,Doherty:2000,Bouten:2007}
is a non-commutative analog of classical probability theory. In
quantum probability theory, there is no longer an underlying
sample space, and so the quantum probability model can be
described by a pair $\left(\mathcal{N},\mathbb{P}\right)$. The
first item, $\mathcal{N}$, is an algebra, and is defined as the
set of all Hermitian operators in the Hilbert space of the system.
An element $A \in \mathcal{N}$ is an observable of the quantum
system which can be looked as the quantum version of the random
variable. The ``events" of quantum probability are defied as all
the projection operators $P \in \mathcal{N}$. These project onto
the subspaces of the Hilbert space. Thus each possible set of
outcomes is given by a subspace of the Hilbert space. This is
simply the projection postulate of quantum measurement theory. The
function $\mathbb{P}:\mathcal{N}\rightarrow\mathcal{C}$, where
$\mathcal{C}$ is the complex space, is called a state on
$\mathcal{N}$. In fact, we can always find a system density
operator $\rho$ such that $\mathbb{P}\left(A\right)={\rm
tr}\left(A\rho\right)$ for any $A\in\mathcal{N}$.

By comparing the classical probability model $\left( \Omega,
\mathcal{F}, \mathbb{P} \right)$ and the quantum probability model
$\left( \mathcal{N}, \mathbb{P} \right)$, the main difference is
that the algebra $\mathcal{N}$, called the von Neumann algebra, is
noncommutative (the Hermitian operators may not commute with each
other) while the $\sigma$-algebra $\mathcal{F}$ in classical
probability is a commutative algebra. As an example, let us
consider a quantum measurement of the observable $A$. Before this
measurement, the quantum system can be described by the quantum
probability model $\left( \mathcal{N}, \mathbb{P} \right)$. After
this measurement, the quantum state collapse occurs. The
measurement output corresponds to a classical probability model
$\left(\mathcal{A},\mathbb{P}\right)$, where
\begin{equation}
\mathcal{A}=\{X:X=f\left(A\right),f:\mathcal{R}\rightarrow\mathcal{C}\}
\end{equation}
forms a commutative algebra. Thus, the quantum measurement of the
observable $A$ converts a quantum probability model into a
classical probability model. More generally, in the following
discussions, we will show that the quantum filtering process,
which is based on quantum measurement, is merely a bridge between
a quantum probability model and a classical probability model.

To better  understand quantum filtering, let us consider an
indirect quantum measurement, which is achieved by interacting the
measured system with a bath via a system operator $L$, and then
making a measurement on the bath. The bath is a continuum of
harmonic oscillators of different frequencies. The bath also
describes a field, such as the electromagnetic field, in which the
oscillators are the modes of the field. The Hamiltonian of the
total system composed of the measured system and the bath is given
by
\begin{eqnarray}\label{General Hamiltonian}
H&=&H_s+H_b+H_{\rm int}, \nonumber\\
H_b&=&\int_{-\infty}^{+\infty}\!\!d
\omega\;\omega\;b^{\dagger}\!\left(
\omega \right)\;b\left( \omega \right),\nonumber\\
H_{\rm int}&=&i\int_{-\infty}^{+\infty}d \omega \left[ \kappa
\left( \omega \right) b^{\dagger}\left( \omega \right) L - {\rm
h.c.} \right],
\end{eqnarray}
where $H_s$ is the free Hamiltonian of the measured system, $b^{\dagger}\left(\omega\right)$ and $b \left( \omega
\right)$ are the creation and annihilation operators of the bath
mode with frequency $\omega$, and satisfy
\begin{equation}\label{Commutation relation of continuous variable}
\left[b\left(\omega\right),b^{\dagger}\left(\tilde{\omega}\right)\right]=\delta\left(\omega-\tilde{\omega}\right).
\end{equation}
The bath mode with frequency $\omega$ interacts with the system
via the system operator $L$, where $\kappa \left( \omega \right)$
is the corresponding coupling strength. Hereafter we set
$\hbar=1$. The total Hamiltonian $H$ can be re-expressed in the
interaction picture as
\begin{equation}\label{Effective Hamiltonian of the total system}
H_{\rm eff}\;=\;\exp{\left(i H_b t\right)} \left( H_s + H_{\rm
int} \right) \exp{\left(-i H_b
t\right)}\;=\;H_s+i\int_{-\infty}^{+\infty} d \omega \left[ \kappa
\left( \omega \right) e^{i \omega t} b^{\dagger} \left( \omega
\right) L -{\rm h.c.}\right].
\end{equation}
We now introduce the Markovian assumption
\begin{equation}\label{Markovian assumption}
\kappa \left( \omega \right) = \sqrt{ \frac{\gamma}{2\pi} },
\end{equation}
which allows the Hamiltonian $H_{\rm eff}$ to be expressed as
\begin{equation}\label{Effective Hamiltonian under the Markovian assumption}
H_{\rm eff}=H_s+i \sqrt{\gamma} \left[ b_{\rm in}^{\dagger} \left(
t \right) L - L^{\dagger} b_{\rm in} \left( t \right) \right],
\end{equation}
where
\begin{equation}\label{Input field}
b_{\rm in}\left( t
\right)=\frac{1}{\sqrt{2\pi}}\int_{-\infty}^{+\infty}\!\!d\omega\;e^{-i
\omega t}\;b \left( \omega \right)
\end{equation}
is the Fourier transform of the bath modes. The operator $b_{\rm in}\left( t
\right)$ is, in fact, the
time-varying field that is incident on, and thus the input to, the system, and satisfies~\cite{Jacobs14, CWGardinerPRA:1985, CWGardinerBook:2004}
\begin{equation}\label{Commutation relation of continuous
variable in the time domain} \left[ b_{\rm in} \left( t \right),
b_{\rm in}^{\dagger} \left( \tilde{t} \right) \right] = \delta
\left( t - \tilde{t} \right).
\end{equation}
We now define a new bath operator
\begin{equation}\label{Quantum Wiener process}
B_{\rm in} \left( t \right) = \int_0^{\mss{t}} b_{\rm in} \left(
\tau \right) d \tau
\end{equation}
which is called a {\it quantum Wiener process}. If we assume that the bath is
initially in a vacuum state, the increment of the quantum Wiener
process $dB_{\rm in}$ and its conjugate $dB_{\rm in}^{\dagger}$
satisfy the following algebraic conditions:
\begin{equation}\label{Quantum Ito rule}
d B_{\rm in}\;d B_{\rm in}^{\dagger} = dt,\quad\quad d B_{\rm
in}^{\dagger}\;d B_{\rm in} = d B_{\rm in}^{\dagger}\;d B_{\rm
in}^{\dagger} = d B_{\rm in}\;d B_{\rm in} = 0.
\end{equation}
These are the quantum version of {\it It\^{o}
rule}~\cite{RIHudsonCMP:1984}. With the above notation, in the
Heisenberg picture, an arbitrary system operator $X \left( t
\right) $ satisfies the following quantum stochastic differential
equation~\cite{CWGardinerPRA:1985,CWGardinerBook:2004}
\begin{equation}\label{Quantum stochastic
differential equation} d X = - i \left[ X, H_s \right] dt +
\frac{\gamma}{2} \left\{ L^{\dagger} \left[ X, L \right] + \left[
L^{\dagger}, X \right] L \right\} dt +\sqrt{\gamma} \left\{ d
B_{\rm in} \left[ L^{\dagger}, X \right] + \left[ X, L \right] d
B_{\rm in}^{\dagger} \right\}.
\end{equation}
It is then possible to define an output filed $b_{\rm out} \left( t \right)$ which describes the field leaving the system after it has interacted with it, and we can similarly define its Ito increment
\begin{eqnarray*}
B_{\rm out} \left( t \right) = \int_0^{\mss{t}} \!\!d \tau\;b_{\rm
out} \left( \tau \right).
\end{eqnarray*}
The celebrated input-output relation for the system can then be written as~\cite{CWGardinerPRA:1985,CWGardinerBook:2004}
\begin{equation}\label{Input-output relation in the differential form}
d B_{\rm out} = d B_{\rm in} + \sqrt{\gamma}\;L\;dt.
\end{equation}
If homodyne detection is performed on the output field $B_{\rm out} \left( t \right) $, then the operator corresponding to the measured output is
\begin{eqnarray*}
d Y_{\rm out} = \frac{1}{\sqrt{\gamma}}\left( d B_{\rm out} + d
B_{\rm out}^{\dagger} \right) ,
\end{eqnarray*}
and satisfies the following equation
\begin{equation}\label{Output equation of the balanced homodyne detection}
d Y_{\rm out} = \left( L + L^{\dagger} \right) dt +
\frac{1}{\sqrt{\gamma} } \left( d B_{\rm in } + d B_{\rm
in}^{\dagger} \right).
\end{equation}

With the above preparation, we can now present the main results of
quantum filtering theory~\cite{Bouten:2007}. The purpose of
quantum filtering is to provide an estimate $ \pi \left( X \right)
$ of the value of the system observable $X$, at time $t$, given
the stream of measurement results up until that time. We will
define this estimate as the expectation value of $X$ given the
measurement results. To obtain $ \pi \left( X \right)$ we first
define
\begin{equation}\label{Commutative algebra of measurement output}
\mathcal{Y}_{\rm out}=\left\{ X:X=f\left(Y_{\rm out}
\right),f:\mathcal{R}\rightarrow\mathcal{C} \right\},
\end{equation}
which is the smallest commutative algebra generated by the
observation process $Y_{\rm out}$, and denote $ \mathbb{P} $ as
the probability measure on $\mathcal{Y}_{\rm out}$. The estimate $\pi
\left( X \right) $ is then the conditional expectation of
$X$ on $\mathcal{Y}_{\rm out}$
\begin{equation}\label{Quantum filtering estimation of X}
\pi \left( X \right) = \mathbb{P} \left( X | \mathcal{Y}_{\rm out}
\right).
\end{equation}
From the definition of $ \pi \left( X \right) $ given in
Eq.~(\ref{Quantum filtering estimation of X}), it can be proved
(see, e.g., the derivations in Ref.~\cite{Bouten:2007}) that we can obtain the following
dynamical equation for $\pi \left( X \right) $ and the
corresponding output equation from Eqs.~(\ref{Quantum stochastic
differential equation}) and (\ref{Output equation of the balanced
homodyne detection})
\begin{eqnarray}
d \pi \left( X \right)&=&\pi \left[ \mathcal{L} \left( X
\right) \right] dt + \sqrt{\gamma} \left[ \pi \left( L^{\dagger} X
+ X L \right) - \pi \left( L + L^{\dagger} \right) \pi \left( X
\right) \right]\;d W, \label{Quantum filtering equation in the Heisenberg picture} \\
d Y_{\rm out} & = & \pi \left( L + L^{\dagger} \right) dt +
\frac{1}{ \sqrt{\gamma} }\;dW,\label{Output equation for quantum
filtering}
\end{eqnarray}
where $\mathcal{L} \left( X \right) $ is the Liouville
superoperator of the system defined as
\begin{eqnarray*}
\mathcal{L} \left( X \right) = -i \left[ X, H_s \right] + \gamma
\left( L^{\dagger} X L - \frac{1}{2} L^{\dagger} L X - \frac{1}{2}
X L^{\dagger} L \right).
\end{eqnarray*}
The process $ W \left( t \right) $ in Eqs.~(\ref{Quantum filtering
equation in the Heisenberg picture}) and (\ref{Output equation for
quantum filtering}) is called the innovation process of quantum
filtering, and has been shown to be a classical Wiener
process~\cite{Bouten:2007}. The increment of $ W \left( t \right)
$ satisfies the following classical It\^{o} relations
\begin{equation}\label{Classical Ito rule}
E \left( dW \right) = 0, \quad\quad \left( dW \right)^2 = dt,
\end{equation}
where $E \left( \cdot \right)$ is the ensemble of the
stochastic process induced by $dW$. The dynamical equation
(\ref{Quantum filtering equation in the Heisenberg picture}) of
$\pi \left( X \right) $ is called the {\it quantum filtering equation}.
The filtering equation (\ref{Quantum
filtering equation in the Heisenberg picture}) and the output
equation (\ref{Output equation for quantum filtering}) are the
main results of quantum filtering theory.

Additionally, we can convert the filtering equation (\ref{Quantum
filtering equation in the Heisenberg picture}) from the Heisenberg
picture to the Schr\"{o}dinger picture, and thus obtain a stochastic equation for the evolution of the density matrix. To show this, we use the fact that the density operator $\rho_{\mss{c}}$ satisfies
${\rm tr} \left[ \pi \left( X \right) \rho_0 \right] = {\rm tr}
\left( X_0 \rho_{\mss{c}} \right)$, where $\rho_0$ is the initial density
operator of the system and $X_0$ is the corresponding system
observable in the Schr\"{o}dinger picture. Substituting Eq.~(\ref{Quantum
filtering equation in the Heisenberg picture}) into the above relation, the system
density operator $\rho_{\mss{c}}$ evolves according to the following
stochastic master equation:
\begin{equation}\label{Stochastic master equation}
d \rho_{\mss{c}} = -i \left[ H_s, \rho_{\mss{c}} \right] dt +
\frac{\gamma}{2} \left( 2 L \rho_{\mss{c}} L^{\dagger} -
L^{\dagger} L \rho_{\mss{c}} - \rho_{\mss{c}} L^{\dagger} L
\right) dt + \sqrt{\gamma}\left\{ L \rho_{\mss{c}} +
\rho_{\mss{c}} L^{\dagger} - {\rm tr} \left[ \left( L +
L^{\dagger} \right) \rho_{\mss{c}} \right] \rho_{\mss{c}} \right\}
dW.
\end{equation}
From Eq.~(\ref{Quantum filtering estimation of X}), we have
\begin{equation}\label{Quantum filtering estimation of the system density operator}
\rho_{\mss{c}} = \mathbb{P} \left( \rho | \mathcal{Y}_{\rm out} \right).
\end{equation}
That is, $\rho_{\mss{c}}$ is the conditional expectation of the density
operator $\rho$ which is defined by ${\rm tr} \left[ X \rho_0
\right] = {\rm tr} \left( X_0 \rho \right)$. The stochastic master
equation~(\ref{Stochastic master equation}) is also often referred to as {\it
quantum filtering equation}.

To summarize, the quantum stochastic differential
equation~(\ref{Quantum stochastic differential equation}) and the
output equation~(\ref{Output equation of the balanced homodyne
detection}) give the dynamics of the operators that describe the measured quantum system.
These equations are driven by the quantum Wiener noise $d B_{\rm
in}$, and are thus defined on a quantum probability space. As
a comparison, the quantum filtering equation (\ref{Quantum
filtering equation in the Heisenberg picture}) (or the stochastic
master equation~(\ref{Stochastic master equation})) and the output
equation (\ref{Output equation for quantum filtering}) give the
observers state-of-knowledge of the measured quantum system based on the
information extracted by the quantum measurement.
These equations are driven by the classical Wiener noise $dW$ and
thus defined on a classical probability space. Thus in
quantum filtering theory we use a classical stochastic system
to mimic the dynamics of a quantum stochastic model, which is why
we refer to quantum filtering as a bridge between a quantum
probability model and a classical probability model.

\subsubsection{Another point of view: quantum trajectories}

An alternative, albeit less rigorous way to obtain the quantum filtering equation~(\ref{Stochastic master equation}) and
output equation~(\ref{Output equation for quantum filtering}), is to work in the Schr\"{o}dinger
picture~\cite{Carmichael:1993, Plenio:1998, Jacobs1:2006}. In this approach the evolution of the density matrix
conditioned on the stream of measurement results (the ``measurement record''), is often referred to as a \textit{quantum trajectory}, a term coined by \textit{Camichael}~\cite{Carmichael:1993}. Before the development
of the quantum trajectory approach, most of the initial studies involving quantum systems interacting with a bath considered only the ensemble description, in which one discards the measurement record, and thus calculates only the evolution of the system averaged over all possible records. This was all that was required before experimental techniques made it possible  to observe single quantum systems in real-time.  However, with the experimental progress, especially in optical systems and ion traps in the 1990s, it became necessary to describe the evolution of a system for an individual measurement record.

One of the early approaches to obtaining a quantum trajectory for
a given measurement record was to express the quantum master
equation as an average over a stochastic equation for the
evolution of a pure quantum state. This  is equivalent to the use
of a ``Monte Carlo'' method to simulate the master
equation~\cite{Molmer93}. Carmichael referred to the process of
expressing a master equation as the average of a stochastic
equation as \textit{unravelling} it. For a single master equation
there is more than one stochastic equation that will unravel it,
and it can be unravelled by stochastic equations driven either by
Gaussian white noise (Wiener
noise)~\cite{Gisin:1992,Percival:1998,Jacobs1:2006} or by a
``point
process''~\cite{Plenio:1998,Srinivas:1981,Dalibard:1992,GPZ:1992}.
A point process consists of intervals of smooth (deterministic)
motion, punctuated by instantaneous events in which the state of
the system changes discontinuously. The Poisson process is an
example of a point process. The different stochastic equations
correspond to different ways in which the system can be
continuously monitored.

\noindent 1. \textit{Quantum jumps:}

Consider the master equation
\begin{equation}\label{General master equation for quantum unravelling}
\dot{\rho}=-i[H,\rho]+\frac{1}{2}\sum_{\mu} \left(2c_{\mu}\rho
c_{\mu}^{\dagger}-c_{\mu}^{\dagger}
c_{\mu}\rho- \rho c_{\mu}^{\dagger} c_{\mu}\right).
\end{equation}
A stochastic equation that unravels this master equation, and that is driven by a point process, is
\begin{equation}\label{Quantum-jump unravelling-stochastic Schrodinger equation}
d|\psi_{\mss{c}}\rangle=\left[-iH+\frac{1}{2}\sum_{\mu}\left(\langle
c_{\mu}^{\dagger}c_{\mu}\rangle\left(t\right)-c_{\mu}^{\dagger}c_{\mu}\right)\right]|\psi_{\mss{c}}\rangle\;
dt+\sum_{\mu}\left(\frac{c_{\mu}}{\sqrt{\langle
c_{\mu}^{\dagger}c_{\mu}\rangle\left(t\right)}}-1\right)|\psi_{\mss{c}}\rangle\;
dN_{\mu} .
\end{equation}
Here, for each $\mu$, the increment $dN_\mu$ is an increment of a point process, and takes only two values, either 0 or 1. The value 1 corresponds to an instantaneous event, and thus $dN_\mu$ is equal to 1 only at a set of discrete points. The rest of the time $dN_\mu =0$. The events occur randomly and independently, and the probability per unit time that an event occurs for the process labelled by $\mu$ is $\langle c_{\mu}^{\dagger}c_{\mu}\rangle\left(t\right)$. This means that the probability for an event in the time interval $[t,t+dt]$ is $\langle c_{\mu}^{\dagger} c_{\mu} \rangle dt$. The point-process increments satisfy the relations
\begin{equation}\label{Quantum jump increment term}
E\left[d N_{\mu}\left(t\right)\right]=\langle c_{\mu}^{\dagger}
c_{\mu} \rangle dt,\quad\quad dN_{\mu} dN_{\nu}=dN_{\mu}
\delta_{\mu\nu}.
\end{equation}
Since Eq.~(\ref{Quantum-jump unravelling-stochastic Schrodinger
equation}) is a stochastic equation for the state vector, it is
usually called a \textit{stochastic Schr\"{o}dinger equation}. We
can alternatively write down a \textit{stochastic master equation}
for the density matrix $\rho_{\mss{c}} = |\psi_{\mss{c}}\rangle
\langle \psi_{\mss{c}} |$, which is
\begin{equation}\label{Quantum-jump unravelling-stochastic master equation}
d\rho_{\mss{c}}=\sum_{\mu}\mathcal{G}\left[c_{\mu}\right]\rho_{\mss{c}}
dN_{\mu}\left(t\right)+\mathcal{H}\left[-iH-\frac{1}{2}\sum_{\mu}c_{\mu}^{\dagger}c_{\mu}\right]\rho_{\mss{c}}
dt .
\end{equation}
The superoperators $\mathcal{G}\left[c\right]\rho_{\mss{c}}$ and
$\mathcal{H}\left[c\right]\rho_{\mss{c}}$ are defined as
\begin{eqnarray}\label{Superoperators G and H}
\mathcal{G}\left[c\right]\rho_{\mss{c}}&=&\frac{c\rho_{\mss{c}} c^{\dagger}}{{\rm
tr}\left[c\rho_{\mss{c}} c^{\dagger}\right]}-\rho_{\mss{c}},\nonumber\\
\mathcal{H}\left[c\right]\rho_{\mss{c}}&=&c\rho_{\mss{c}}+\rho_{\mss{c}}
c^{\dagger} - \langle c+c^{\dagger}\rangle \rho_{\mss{c}}.
\end{eqnarray}
The point process (quantum jump) stochastic Schr\"{o}dinger equation (SSE) describes, for example, an optical cavity in which the light that leaks out of the cavity is measured with a photon-counter~\cite{HMWiseman_quantumjump:1993}. In this case there is a single Lindblad operator $c=\sqrt{\gamma}a$, where $\gamma$ and $a$ are the damping rate and annihilation operator for the cavity, respectively. The events at which $dN=1$ correspond to the detection of a photon by the photo-detector.

If the photon-detector is ideal, meaning that it never misses a photon, and never clicks when there
is no photon, then an initially pure state remains pure, and the SSE is sufficient to describe the observer's state-of-knowledge as the measurement proceeds. But if the photon-detector is not perfect, the observer no longer has full information about the quantum state as the measurement proceeds. The observer's state-of-knowledge is then necessarily given by a density matrix, and we must use a stochastic master equation (SME), rather than a Schr\"{o}dinger equation. We must also modify the master equation to include imperfect detection. If the photon-detector is inefficient, so  that it records only a fraction $\eta$ of the photons emitted by the cavity, and does not record any non-existent photons, then the SME in Eq.(\ref{Quantum-jump unravelling-stochastic master
equation}) becomes
\begin{equation}\label{Stochastic master equation for inideal photon detection: quantum jump}
d\rho_{\mss{c}}=\left\{dN\mathcal{G}\left[\sqrt{\gamma\eta}a\right]+dt
\mathcal{H}\left[-iH-\frac{\eta\gamma}{2}a^{\dagger}a\right]+dt\left(1-\eta\right)\mathcal{D}\left[\sqrt{\gamma}a\right]\right\}\rho_{\mss{c}},
\end{equation}
where
\begin{equation}\label{Superoperator D}
\mathcal{D}\left[c\right]\rho=c\rho
c^{\dagger}-\frac{1}{2}c^{\dagger}c\rho-\frac{1}{2}\rho
c^{\dagger}c.
\end{equation}

\noindent 2. \textit{Quantum diffusion:}

The master equation given by Eq.~(\ref{General master equation for
quantum unravelling}) is also unravelled by the
SSE~\cite{HMWisemanPRA_MF:1993,
Percival:1998,Jacobs1:2006,Gisin:1992,HMWisemanCP:2001}
\begin{equation}\label{Quantum state diffusion-unravelling stochastic Schrodinger equation}
d|\psi\rangle=-i H|\psi\rangle dt+\sum_{\mu}\left(\langle
c_{\mu}^{\dagger} \rangle
c_{\mu}-\frac{1}{2}c_{\mu}^{\dagger}c_{\mu}-\frac{1}{2}\langle
c_{\mu}^{\dagger} \rangle \langle c_{\mu} \rangle \right)
|\psi\rangle dt+\sum_{\mu}\left(c_{\mu}-\langle c_{\mu} \rangle
\right)|\psi\rangle dW_{\mu},
\end{equation}
where the $dW_\mu$ are a set of mutually independent Wiener noises satisfying
\begin{equation}\label{Quantum Wiener noise increment}
E\left(dW_{\mu}\right)=0,\quad\quad dW_{\mu}
dW_{\nu}=\delta_{\mu\nu}dt.
\end{equation}
The equivalent stochastic master equation is
\begin{eqnarray}\label{Quantum state diffusion-unravelling stochastic master equation}
d\rho_{\mss{c}}=-i\left[H,\rho_{\mss{c}}\right]dt+\sum_{\mu}\left(\mathcal{D}\left[c_{\mu}\right]\rho_{\mss{c}}
dt+\mathcal{H}\left[c_{\mu}\right]\rho_{\mss{c}} dW_{\mu}\right).
\end{eqnarray}
Stochastic SSE's and SME's driven by Wiener noise correspond to
measurement techniques that are quite different from
photon-counting. If, instead of detecting the light from a cavity
with a photo-detector directly, one first interferes the light
with a laser whose intensity is much greater than the cavity
output, the result is a measurement containing Wiener noise. This
measurement technique is sensitive to the phase of the cavity
output, whereas direct photo-detection is not, and is called
\textit{homodyne detection}~\cite{HMWisemanPRA_MF:1993}.

In the limit that the power of the laser is infinite, the dynamics of a single mode of an optical cavity measured by homodyne detection is given by Eq.(\ref{Quantum state diffusion-unravelling stochastic master equation}) with a single
Lindblad operator $c=\sqrt{\gamma}a$. The measured output is
\begin{equation}\label{Measurement output of a homodyne detection}
dy=\langle x \rangle dt+\frac{1}{\sqrt{2\gamma}}dW,
\end{equation}
where $x=\left(a+a^{\dagger}\right)/\sqrt{2}$ is the normalized
position operator of the cavity mode, and $dW$ is the same Wiener
noise increment that appears in the SME. If the photo-detector is
inefficient, then the SME becomes
\begin{equation}\label{Stochastic master equation for inideal photon detection: quantum state diffusion}
d
\rho_{\mss{c}}=-i\left[H,\rho_{\mss{c}}\right]dt+\gamma\mathcal{D}\left[a\right]\rho_{\mss{c}}+\sqrt{\eta\gamma}\mathcal{H}\left[a\right]\rho_{\mss{c}}
dW,
\end{equation}
and the measurement output is
\begin{equation}\label{Measurementoutput of a homodyne detection: unideal case}
dy=\langle x \rangle dt+\frac{1}{\sqrt{2\eta\gamma}}dW ,
\end{equation}
where $\eta$ is the detection efficiency. Continuous measurements
containing Wiener noise are also sometimes referred to as weak
measurements. We prefer to call them continuous measurements
because (i) weak measurements are not necessarily continuous, and
(ii) it can lead to confusion with the ``weak values'' of
Aharonov, Albert and Vaidman~\cite{Aharonov:1988,Aharonov:1990}.

More generally, a continuous measurement of the quantum variables
$A_l$ $\left(l=1,\cdots,m\right)$ can be expressed as the
stochastic master equation~\cite{Jacobs1:2006}
\begin{equation}\label{Quantum stochastic master equation of multiple observables}
d\rho_{\mss{c}}=-i\left[H,\rho\right]dt+\sum_{l=1}^m
\left(\Gamma_l\,\mathcal{D}\left[A_l\right]\,\rho_{\mss{c}}\,dt +
\sqrt{\eta_l\Gamma_l}\,\mathcal{H}\left[A_l\right]\,dW_l\right),
\end{equation}
and output equation
\begin{equation}\label{Output equation of multiple observables}
dy_l=\langle A_l \rangle\,dt+\frac{1}{\sqrt{2\eta_l
\Gamma_l}}\,dW_l,
\end{equation}
where $\Gamma_l$ and $\eta_l$ represent the measurement strengths
and measurement efficiencies.

The stochastic master equation~(\ref{Quantum stochastic master
equation of multiple observables}) and the equation for the stream
of measurement results, Eq.~(\ref{Output equation of multiple
observables}), can be derived from the quantum filtering equations
(Eqs.~(\ref{Stochastic master equation}) and (\ref{Output equation
for quantum filtering})). The quantum filtering equations give the
evolution of the system and the output field before any
measurement is made on the output field. Making a measurement on
the output field turns the quantum filtering equations into a
stochastic master equation.

As mentioned above, we can simultaneously make more than one
continuous measurement on a system, and we can simultaneously
measure observables that do not commute. Since the respective
dynamics induced by the continuous measurements of two different
observables commute to first order in $dt$, we can think of the
measurements of the two observables as being  interleaved --- the
process alternates between infinitesimal measurements of each
observable. Note that a von Neumann measurement cannot
simultaneously project a system onto the eigenstates of two
non-commuting observables, but continuous measurements do not
perform instantaneous projections. The effect of simultaneously
measuring the position and momentum of a single particle, for
example~\cite{Scott:2001, JGough:2004}, is to feed noise into both
observables. Measuring noncommuting observables therefore in
general introduces more noise into a system than is necessary to
obtain a given amount of information. The optical measurement
techniques of heterodyne detection~\cite{HMWiseman:2009} and
eight-port homodyne detection~\cite{Leonhardt:93} are very similar
to simultaneous measurements of momentum and position.


\subsection{Markovian quantum feedback}\label{s22}

The continuous collapse of the quantum state in continuous quantum
measurement means that we can execute real-time quantum feedback
control before the quantum state collapses to a completely
classical state. That is the starting point of continuous
measurement-based feedback control. The key questions in feedback
control are usually (i) what observable should we measure? and
(ii) how should we chose the feedback forces as a function(al) of
the stream of measurement results? Optimal feedback strategies can
always be obtained by using the SME to determine the observer's
full state of knowledge (the density matrix) given the stream of
measurement results up to the present time, and using this to
determine the choice of Hamiltonian at each time. But solving the
SME can take significant numerical resources, and it may not be
possible to do so in real-time. In that, one can attempt to
approximate the SME with a simpler differential equation, which
may be possible depending on the dynamics of the
system~\cite{DASteckPRL:2004, DASteckPRA:2006, Jacobs11}.
Alternatively we can take the opposite approach, and see what can
be achieved with quantum feedback when we perform no processing of
the measurement results, and merely engineer a term in the
Hamiltonian of the system that, at each time, is proportional to
the  measurement result at that time. This is the kind of feedback
protocols that were introduced by Wiseman and
Milburn~\cite{HMWisemanPRA_MF:1993,HMWisemanPRL_MF:1993,HMWisemanPRA_MF:1994},
and are now referred to as Markovian feedback. The reason for this
name is that for this kind of feedback, if we average the
evolution over all trajectories, the result is a Markovian master
equation. This is not usually true for feedback protocols.

Let us consider a quantum continuous measurement of the operator $A$ with
efficiency $\eta$. From Eqs.~(\ref{Quantum stochastic master
equation of multiple observables}) and (\ref{Output equation of
multiple observables}), the measurement and output equations of
this measurement can be expressed as
\begin{equation}\label{Quantum stochastic master equation of a single observable}
d\rho_{\mss{c}}=-i\left[H,\rho_{\mss{c}}\right]dt+\Gamma_A\,\mathcal{D}\left[A\right]\,\rho_{\mss{c}}\,dt
+
\sqrt{\eta\Gamma_A}\,\mathcal{H}\left[A\right]\,\rho_{\mss{c}}\,dW,
\end{equation}
and
\begin{equation}\label{Output equation of a single observable}
dy=\langle A \rangle\,dt+\frac{1}{\sqrt{2 \eta \Gamma_A}}\,dW.
\end{equation}
These two equations can also be expressed equivalently by
\begin{equation}\label{Quantum stochastic master equation of a single observable with white noise}
\dot{\rho}_{\mss{c}}=-i\left[H,\rho_{\mss{c}}\right]+\Gamma_A\,\mathcal{D}\left[A\right]\,\rho_{\mss{c}}
+
\sqrt{\eta\Gamma_A}\,\mathcal{H}\left[A\right]\,\rho_{\mss{c}}\,\xi\left(t\right),
\end{equation}
and
\begin{equation}\label{Output equation of a single observable with white noise}
I_A\left(t\right)=\langle A \rangle+\frac{1}{\sqrt{2 \eta
\Gamma_A}}\,\xi\left(t\right),
\end{equation}
where $\xi\left(t\right)$ is the white noise satisfying
\begin{equation}\label{Classical white noise}
E\left(\xi\left(t\right)\right)=0,\quad\quad
E\left(\xi\left(t\right)\xi\left(t'\right)\right)=\delta\left(t-t'\right).
\end{equation}
Formally, we can convert Eqs.~(\ref{Quantum stochastic master
equation of a single observable}) and (\ref{Output equation of a
single observable}) into Eqs.~(\ref{Quantum stochastic master
equation of a single observable with white noise}) and
(\ref{Output equation of a single observable with white noise}) by
setting $\xi\left(t\right)=dW/dt$.

The main object of measurement-based quantum feedback is to use
the output signal $I_A\left(t\right)$ to engineer the system
dynamics given by Eq.~(\ref{Quantum stochastic master equation of
a single observable}). The most general form of the system
dynamics, modified based on the output signal $I_A\left(t\right)$,
can be expressed as~\cite{HMWiseman:2009}
\begin{equation}\label{General form of quantum output feedback}
\dot{\rho}_f=\mathcal{F}\left[t,\{I_A\left(\tau\right)|\tau\in\left[0,t\right]\}\right]\rho_f,
\end{equation}
where $\mathcal{F}\left[t,\{I_A\left(\tau\right)|\tau\in
\left[0,t\right]\}\right]$ is the superoperator depending on the
output signal $I_A\left(t\right)$ for all past times. In this
general form of the response of the feedback control loop, the
control induces both unitary dynamics and dissipation effects on
the controlled system. However, for most of the existing studies,
quantum feedback control is introduced coherently by varying the
parameters in the system Hamiltonian, which leads to the following
modified closed-loop stochastic master equation
\begin{equation}\label{General form of quantum output feedback by varying the system Hamiltonian}
\dot{\rho}_f=-i\left[H+H_{f}\left(t,\{I_A\left(\tau\right)|\tau\in\left[0,t\right]\}\right),\rho_f\right]+\Gamma_A
\mathcal{D}\left[A\right]\rho_f+\sqrt{\eta
\Gamma_A}\mathcal{H}\left[A\right]\rho_f\xi\left(t\right).
\end{equation}
As discussed above, in Markovian quantum feedback a term in the
Hamiltonian is made proportional to the output signal. Denoting
this term by $H_f$, we set $H_f=I_A\left(t\right)F$ for some
Hermitian operator $F$. Then, by averaging over the noise term and
using the Ito rule of the white noise $\xi\left(t\right)$, we can
derive the following Wiseman-Milburn master
equation~\cite{HMWiseman:2009} from Eq.~(\ref{General form of
quantum output feedback by varying the system Hamiltonian}):
\begin{equation}\label{Wiseman-Milburn master equation}
\dot{\rho}=-i\left[H,\rho\right]+\Gamma_A
\mathcal{D}\left[A\right]\rho-i\sqrt{\Gamma_A}\left[F,A\rho+\rho
A\right]+\frac{1}{\eta}\mathcal{D}\left[F\right]\rho.
\end{equation}
The effects induced by the feedback loop are clearer in this form:
(i) the first feedback term $-i\left[F,A\rho+\rho A\right]$ plays
a positive role to steer the system dynamics to achieve the
desired effects; and (ii) the second feedback term
$\mathcal{D}\left[F\right]\rho/\eta$ represents the decoherence
effects induced by feedback, which tends to play a negative role
for purposes of control. The master equation (\ref{Wiseman-Milburn
master equation}) can be reexpressed as the traditional Lindblad
form~\cite{HMWisemanPRA_MF:1994,HMWiseman:2009}
\begin{equation}\label{Lindblad form of the Wiseman-Milburn master equation}
\dot{\rho}=-i\left[H+\frac{\sqrt{\Gamma_A}\left(AF+FA\right)}{2},\rho\right]+\mathcal{D}\left[\sqrt{\Gamma_A}A-iF\right]\rho+\frac{1-\eta}{\eta}\mathcal{D}\left[F\right]\rho.
\end{equation}
Although the Markovian quantum feedback given by
Eq.~(\ref{Wiseman-Milburn master equation}) is the simplest
measurement-based quantum feedback approach, it can be used to
solve various problems by choosing $A$ and $F$ appropriately.
Markovian quantum feedback has been used to stabilize arbitrary
one-qubit quantum
states~\cite{JWangPRA:2001,HFHofmannOptExp:1998,HFHofmannPRA:1998},
manipulate quantum
entanglement~\cite{JWangPRA:2005,JGLiPRA:2008,NYamamotoPRA:2005,SManciniEPJD:2005,YliPRA:2011,ARRCarvalhoPRL:2007,CViviescsPRL:2010,ARRCarvalhoPRA:2007,ARRCarvalhoPRA:2008,SCHouPRA:2010,RNStevensonEPJD:2011,LCWangEPJD:2010,DXueJPB:2010,JSongJOSAB:2012},
generate and protect Schr\"{o}dinger cat states~\cite{PTombesiPRA:1995,PGoetschPRA:1996,DVitaliPRL:1997,DVitaliPRA:1998,MFortunatoPRA:1999},
and induce optical, mechanical, and spin
squeezing~\cite{HMWisemanPRA_Squeezing:1994,HMWisemanPRA_Squeezing:1992,GJMilburnPRA_Squeezing:1993,PTombesiPRA_Squeezing:1994,HMWisemanPRA_Squeezing:1995,HMWisemanPRL_Squeezing:1998,RRuskovPRB_Squeezing:2005,AVinantePRL_Squeezing:2013,LKThomsenPRA_Squeezing:1995}.

\subsection{Feedback via time-averaging}

Markovian quantum feedback is simple to describe analytically, but
is also rather limited. Further, feeding back the measurement
signal at each instant of time does not make optimal use of the
information extracted by the measurement. To do that we must
process the measurement results using the SME. It is worth pausing
at this point to understand a little more how the measurement
results, given by Eq.~(\ref{Output equation of multiple
observables}), provide information about the measured operator and
the state of the system. If we process the measurement results so
that we know $\rho_{\mss{c}}$ at each time, then we also know the
expectation value of the measured operator, $\langle A \rangle$,
at each time. The first term in Eq.~(\ref{Output equation of
multiple observables}) is therefore already known, and provides no
new information about the system. It is the noise term $dW$ that
carries the new information, and that modifies our
state-of-knowledge. In fact, by definition we always know the
expectation value $\langle A \rangle = \mbox{Tr}[
A\rho_{\mss{c}}]$ at the start of the continuous measurement ,
because $\rho_{\mss{c}}$ is our state-of-knowledge. But the system
might really be in some pure state $|\psi\rangle$, so that the
true mean value of $A$ is $\bar{A} = \langle \psi | A | \psi
\rangle$. As the measurement proceeds, the conditional expectation
value $\langle A \rangle$ tends to $\bar{A}$ and $\rho_{\mss{c}}$
tends to $|\psi\rangle$.

Now consider what happens if $A$ is a Hermitian observable, and
$|\psi\rangle$ is an eigenstate of both the system Hamiltonian and
$A$. In this case, assuming that the system is not driven by other
noise sources, it remains in the state $|\psi\rangle$ as the
measurement proceeds, and $\bar{A}$ is constant. In that case we
can obtain an estimate of $\bar{A}$ of ever increasing accuracy
without solving the SME. All we need to do is to average the
measurement results obtained so far, and divide by the total
time~\cite{JKStocktonPRA:2004}. If we define
\begin{equation}\label{Time-average measurement output}
Y_A\left(t\right)= \frac{1}{t}\int_0^{\mss{t}} dy =
\frac{1}{t}\int_0^{\mss{t}} \langle A \rangle dt +   \frac{1}{t
\sqrt{2 \eta \Gamma_A}} \int_0^{\mss{t}}   dW ,
\end{equation}
then as $t\rightarrow\infty$ the second term tends to zero and
$Y_A\left(t\right) \rightarrow \bar{A}$. The reason that the
second term, being the average of the noise, tends to zero is that
it has equally positive and negative fluctuations and these
average to zero over time.

The mean value of the measured observable,  $\bar{A}$, is usually
not constant for a system that we are trying to control.
Nevertheless we can still use an averaging procedure to obtain an
estimate of $\bar{A}$ and use this to chose our feedback forces.
This method is not as complex as processing the measurements using
the SME, but more complex than Markovian
feedback~\cite{ZhangJPRA:2010,VGiovannettiPRA:1999,KNishioPRA:2009,CEmaryPTRSA:2013}.
To do this we average the signal over a time $T$ that is long
enough to reduce the noise but not so long that $\bar{A}$ changes
too much during $T$. We can also include a weighting function,
$f(t)$, to smoothly reduce the dependence on our estimate of
$\bar{A}$ on measurement results that are too far in the past. For
example, if we use an exponential weighting function, our estimate
of $\bar{A}$ at time $t$ is
\begin{equation}\label{Exponentially convergent filter}
\tilde{Y}_A\left(t\right)=\frac{1}{T}\int_{t-T}^{\mss{t}}
e^{-\gamma_f t}\left(\langle A \rangle dt+\frac{1}{\sqrt{2 \eta
\Gamma_A}}dW\right).
\end{equation}
When $T\ll 1/\gamma_f$, the estimate converges as
\begin{equation}\label{Exponential convergence of the output filter}
\tilde{Y}_A\left( t \right)-\bar{A}
\left(t\right)=\exp{\left(-\gamma_f t\right)}
\left[\tilde{Y}_A\left(0\right)-\bar{A} \left( 0 \right)\right].
\end{equation}
Such an exponentially-convergent filter has been introduced in the
literature to stabilize two-qubit entanglement~\cite{MSarovarPRA:2005, LiXQPRA:2010} and a three-qubit
GHZ state~\cite{LiXQPRA:2011} both in optical systems and in
superconducting circuits. It has also been applied experimentally to the
adaptive estimation of the optical phase~\cite{WheatleyPRL:2010}.

\subsection{Bayesian quantum feedback}\label{s23}


To make full use of the information provided by the measurement,
we must process the measurement results using the SME
(Eq.~(\ref{Quantum stochastic master equation of multiple
observables})) to obtain the conditional density matrix. Since
this density matrix, along with the knowledge of the dynamics of
the system, determines the probabilities of the results of any
measurement on the system at any time in the future, any optimal
strategy for controlling the system can ultimately be specified as
a rule for choosing the Hamiltonian at time $t$ as a function of
the density matrix at that time and possibly the time itself:
$H(t) = f(\rho_{\mss{c}}(t),t)$. Feedback control in which the
feedback protocol is specified in this way is sometimes referred
to as ``Bayesian feedback'' because the SME is the quantum
equivalent of processing the measurement record using Bayes'
theorem~\cite{Wiseman02}.

As we have mentioned above, the SME, since it requires simulating
the full dynamics of the system, may be impractical to solve in
real-time. Sometimes it is possible to approximately, or even
exactly, reduce the computational overhead by choosing an ansatz
for $\rho_{\mss{c}}$ that contains only a small number of
parameters. The SME then reduces to a stochastic differential
equation for these parameters. Two examples in which an
approximate ansatz provides an effective control protocol can be
found in~\cite{DASteckPRL:2004, DASteckPRA:2006, Jacobs11}. There
is one class of systems in which an ansatz with a small number of
parameters provides an \textit{exact} solution to the SME, that of
linear systems. A quantum system is referred to as linear if its
Hamiltonian is no more than quadratic in the position and momentum
operators, any Lindblad operators that describe the noise driving
the system are linear in the position and momentum operators, and
any measurements are (i) driven by Wiener noise, and (ii) of
operators that are linear in the position and momentum.

The noise that drives linear systems reduces all initial states to
Gaussian states (states that are Gaussian in the position and
momentum bases, and thus have Gaussian Wigner functions), and
Gaussian states remain Gaussian under the evolution. No proof of
the first of these statements exists, but experience leads us to
believe it. The second statement is not difficult to show, and
implies immediately that if the state of a linear system is
Gaussian, the SME reduces to a stochastic differential equation
for the means and (co-)variances of the position and
momentum~\cite{Doherty:1999}. What is more, the dynamics of these
variables are exactly reproduced by those of a classical linear
system driven by Gaussian noise, and subjected to continuous
measurements of the same observables. To correctly reproduce the
quantum dynamics, for each continuous measurement made on the
system a noise source must be added to the classical system to
mimic Heisenberg's uncertainty principle.

\begin{figure}
\centerline{\includegraphics[width=16.8
cm]{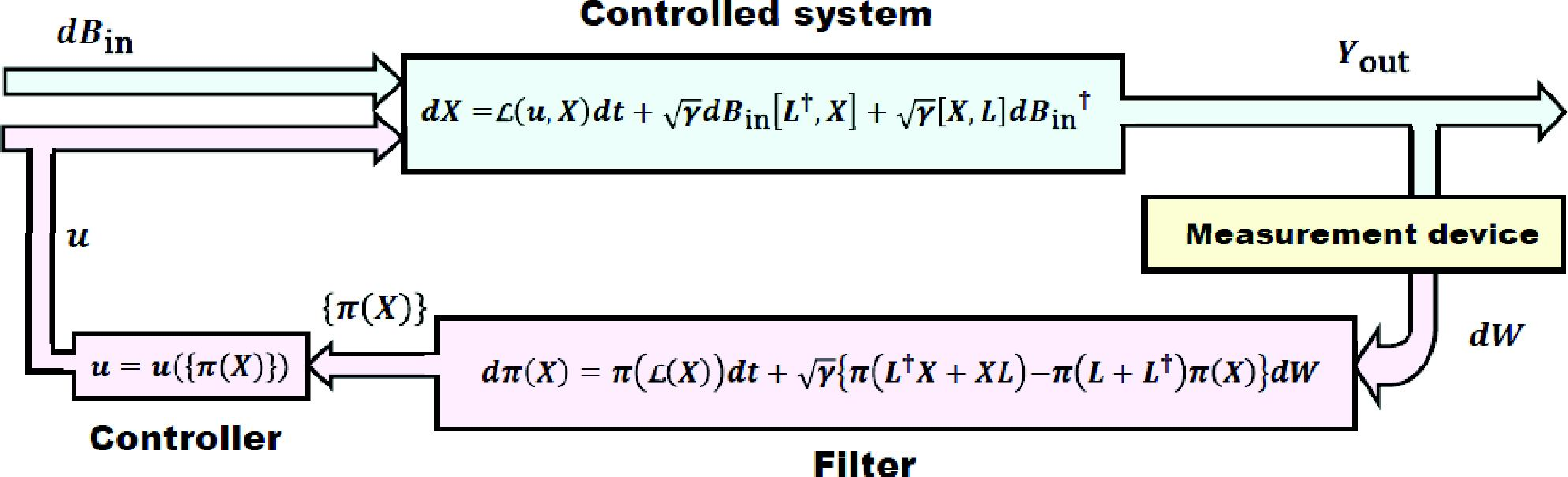}} \caption{(Color online) Diagram for
state-based quantum feedback. The controlled system (top branch,
in blue) is described by a quantum stochastic differential
equation driven by the quantum Wiener noise $dB_{\rm in}$. Part of
the quantum output field $Y_{\rm out}$ from the controlled system
is converted into a classical signal $dW$ by a measurement device
(shown in yellow) and then fed into the filter. The dynamics of
the filter is determined by the quantum filtering equation driven
by the classical Wiener noise, i.e., the innovation process $dW$.
The estimated quantum state $\left\{\pi\left(X\right)\right\}$ is
fed into a classical controller to obtain a control signal $u$,
which is then fed back to steer the dynamics of the controlled
system. The filter and controller which form the classical control
loop (in pink) can be realized by a classical Digital Signal
Processor (DSP).}\label{Fig of the State-based feedback}
\end{figure}


Consider a linear quantum system with $N$ degrees of freedom~\cite{WisemanPRL:2005, NYamamotoPRA:2006, NYamamotoPRA:2007, AChiaPRA1:2011, AChiaPRA2:2011}, and write the $N$ position and momentum operators, denoted respectively by $q_n$ and $p_n$, in the vector
\begin{eqnarray}
\mathbf{x} = \left( q_1,p_1,\cdots,q_N,p_N \right)^{\mss{T}} .
\end{eqnarray}
We scale these operators so that $[q_n,p_n]=i$. If $x_m$ is the $m^{\mbox{\textit{\scriptsize th}}}$
element of the vector $\mathbf{x}$, then we have $ \left[ x_n, x_m
\right] = i \Sigma_{nm} $, where
\begin{eqnarray*}
\Sigma=\bigotimes_{n=1}^N \left(%
\begin{array}{cc}
  0 & 1 \\
  -1 & 0 \\
\end{array}%
\right).
\end{eqnarray*}
For linear quantum systems, the system Hamiltonian $H_s$ and the
dissipation operator $L$ in Eq.~(\ref{Quantum stochastic differential equation}) can be written
as~\cite{WisemanPRL:2005, NYamamotoPRA:2006, NYamamotoPRA:2007}
\begin{equation}\label{Hamiltonian and dissipation operator of linear quantum systems}
H_s = \frac{1}{2} \mathbf{x}^{\mss{T}} G\,\mathbf{x} -
\mathbf{x}^{\mss{T}} \Sigma\,\mathbf{b}\,u, \quad\quad L=
\mathbf{l}^{\mss{T}} \mathbf{x} ,
\end{equation}
where $G$ is a real and symmetric matrix, and
$\mathbf{b}$, $\mathbf{l}$ are real and complex vectors, respectively. The second
term in $H_s$, including the time-dependent function $ u \left( t
\right) $, describes the force applied by the feedback controller (see Fig.~\ref{Fig of the State-based feedback}). This feedback Hamiltonian must be linear in the conditional mean values of the position and momentum operators, in order to ensure that the system remains linear. This also means that there is a linear map from the measurement output $ Y_{ \rm out } $ to $ u \left( t \right) $, and thus a linear input-output relation for the controlled system. From
Eq.~(\ref{Quantum stochastic differential equation}), the dynamics of the controlled system can be expressed as the following linear quantum stochastic differential equation:
\begin{equation}\label{Quantum stochastic differential equation of linear quantum systems}
d \mathbf{x}\,= A\,\mathbf{x}\,dt + \mathbf{b}\,u\,dt + i
\sqrt{\gamma}\,\Sigma \left[ \mathbf{l}\,d B_{\rm in}^{\dagger} -
\mathbf{l}^*\,d B_{\rm in} \right],
\end{equation}
where the matrix $ A = \Sigma \left[ G + {\rm Im}
\left( \mathbf{l}^* \mathbf{l}^{\mss{T}} \right) \right] $. The output equation
(\ref{Output equation of the balanced homodyne detection}) can be
written as
\begin{equation}\label{Output equation of linear quantum systems}
d Y_{\rm out} = F\,\mathbf{x}\,dt + \frac{1}{\sqrt{\gamma}} \left(
d B_{ \rm in } + d B_{ \rm in }^{ \dagger } \right) , \quad \quad
F = \mathbf{l}^{\mss{T}} + \mathbf{l}^{ \dagger }.
\end{equation}
After quantum measurement, the dynamics of this linear quantum
system can be fully described by the conditional means $ \pi
\left( \mathbf{x} \right) $ and variances $ V_t = \mathbb{P}\left(
P_t | \mathcal{Y} \right)$, where $ P_t $ is the covariance matrix
of the position and momentum variables with the $ \left( i, j
\right) $-element being $P_{ij} = \left( \Delta x_i \Delta x_j +
\Delta x_j \Delta x_i \right) / 2 $, and $ \Delta x_i = x_i - \pi
\left( x_i \right) $. The conditional mean values $ \pi \left(
\mathbf{x} \right) $ obey the filtering equation
\begin{equation}\label{Quantum filtering equation for linear system}
d \pi \left( \mathbf{x} \right)=A\,\pi \left( \mathbf{x} \right)
dt + B\,u\,dt + \left[ V_t\,F^{\mss{T}} + \Sigma^{\mss{T}}\,{\rm
Im} \left( l \right) \right] \times \left[ d Y - F \pi \left(
\mathbf{x} \right) dt \right] ,
\end{equation}
and the conditional covariance matrix satisfies the deterministic
Riccati differential equation
\begin{equation}\label{Riccati differential equation for linear quantum systems}
\dot{V}_t=A\,V_t + V_t\,A^{\mss{T}} + D - \left[ V_t\,F^{\mss{T}}
+ \Sigma^{\mss{T}}\,{\rm Im} \left( l \right) \right]\times \left[
F\,V_t + {\rm Im} \left( l^{\mss{T}} \right) \Sigma \right],
\end{equation}
where $ D = \Sigma\,{\rm Re} \left( l^* l^{\mss{T}} \right)
\Sigma^{\mss{T}} $. Thus, the filtering equation (\ref{Quantum
filtering equation in the Heisenberg picture}) or (\ref{Stochastic
master equation}) is equivalent to the closed set of filtering
equations (\ref{Quantum filtering equation for linear system}) for
the first-order quadrature and the Riccati differential equation
(\ref{Riccati differential equation for linear quantum systems}),
which is finite-dimensional and thus simulated with relative ease.
The quantum filter given by Eqs.~(\ref{Quantum filtering equation
for linear system}) and (\ref{Riccati differential equation for
linear quantum systems}) is called a quantum Kalman
filter~\cite{Doherty:1999,Doherty:2000,WisemanPRL:2005,NYamamotoPRA:2006,NYamamotoPRA:2007}.

For linear quantum feedback control systems, many objectives, such
as cooling and squeezing, can be reduced to the optimization of
the following quadratic cost function of the system state
$\mathbf{x}$
\begin{equation}\label{Quantum cost function}
J_q = \frac{1}{2} \mathbf{x}_T^{\mss{T}}\,S\,\mathbf{x}_T +
\frac{1}{2} \int_0^{\mss{T}} \left[
\mathbf{x}_{\tau}^{\mss{T}}\,Q\,\mathbf{x}_{\tau} +
u_{\tau}^{\mss{T}}\,R\,u_{\tau} \right] d \tau.
\end{equation}
To obtain a closed-form control problem, we should first take the
expectation value over the conditioned state and then average over
all the stochastic trajectories to define a new quadratic cost
function $ J = \left\langle \mathbb{P} \left( J_q |
\mathcal{Y}_{\rm out} \right) \right\rangle_{\mss{c}} $, where
$\langle \cdot \rangle_{\mss{c}}$ is the average taken over the
classical Wiener noise $dW$. From Eq.(\ref{Quantum cost function})
we have
\begin{equation}\label{Quadratic cost function}
J=\left\langle \frac{1}{2} \int_0^{\mss{T}} \left[ \pi \left(
\mathbf{x}_{\tau} \right)^{\mss{T}} Q \pi \left( \mathbf{x}_{\tau} \right) +
{\rm tr} \left( Q V_{\tau} \right) + u_{\tau}^{\mss{T}} R u_{\tau} \right]
d \tau \right\rangle_{\mss{c}} + \left\langle \frac{1}{2} \pi \left( \mathbf{x}_T \right)^{\mss{T}} S
\pi \left( \mathbf{x}_T \right) + \frac{1}{2} {\rm tr} \left( S V_T
\right) \right\rangle_{\mss{c}}.
\end{equation}
Here the control $u_t = u \left( \pi\left( \mathbf{x}_t \right), V_t
\right)$ is a function of the conditional means and variances $\pi \left( \mathbf{x}_t \right) $ and $V_t$. The
optimization of the quadratic cost function~(\ref{Quadratic cost
function}) subject to the quantum filtering equations
(\ref{Quantum filtering equation for linear system}) and
(\ref{Riccati differential equation for linear quantum systems})
is a standard classical Linear-Quadratic-Gaussian (LQG) control
problem which can be solved by the Kalman filtering theory well
developed in the field of classical control.

\subsection{Applications}\label{s24}

\subsubsection{Noise reduction and quantum error correction}\label{s241}

\begin{figure}
\centerline{\includegraphics[width = 13
cm]{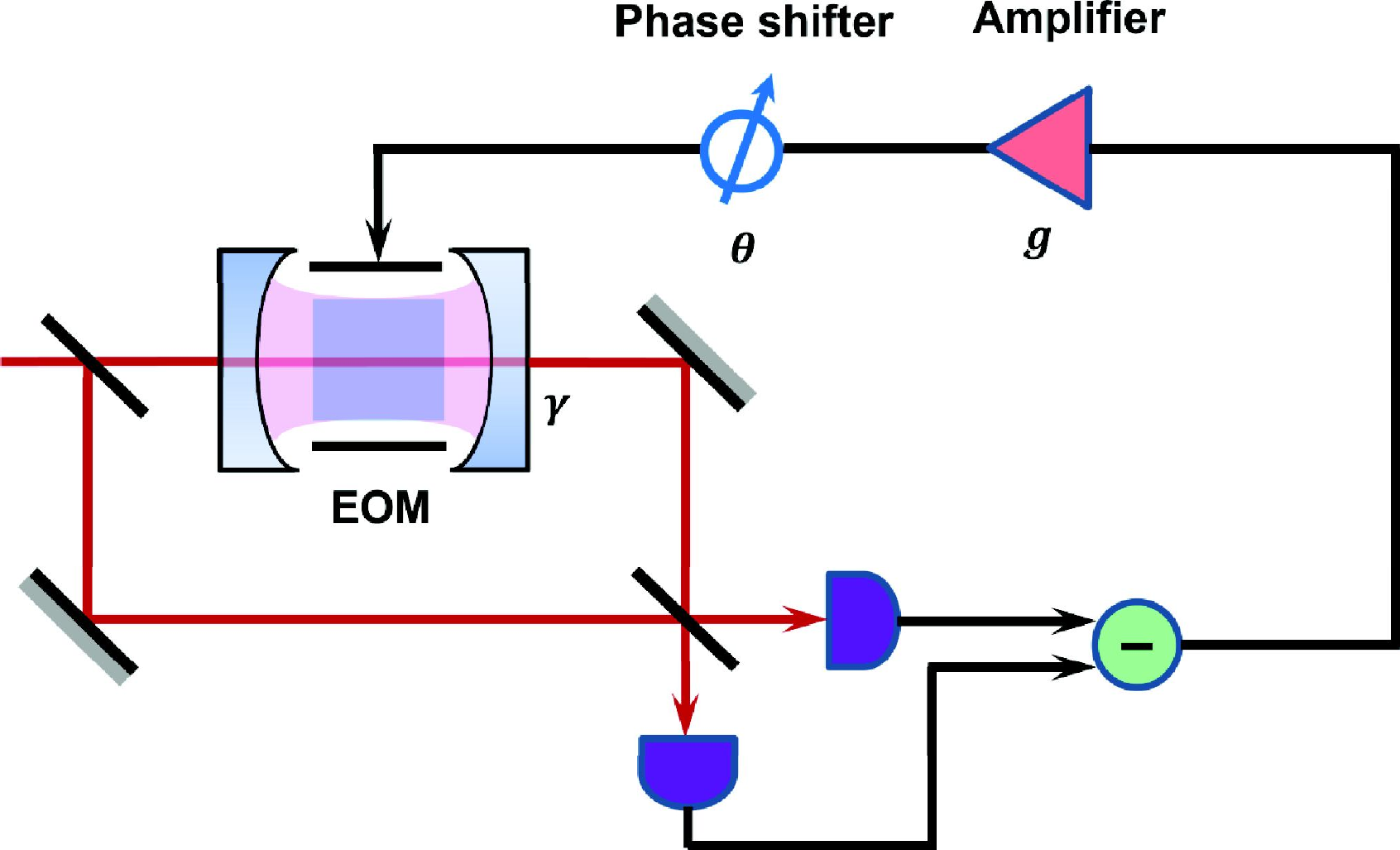}} \caption{(Color online)
Diagram of a proposal by Goetsch \textit{et
al.}~\cite{PGoetschPRA:1996} to extend the lifetime of
superpositions of macroscopically distinguishable coherent states.
Such a ``Schr\"{o}dinger-cat'' state is initially prepared inside
an optical or superconducting cavity. The output field from the
cavity is measured by homodyne detection and then the homodyne
photocurent is fed back to control the transitivity of an
electro-optic modulator. The feedback gain and phase are denoted
by $g$ and $\theta$, respectively. The negative sign in the
triangle indicates that the two signals are subtracted.}
\label{Fig of Schrodinger cat state protection}
\end{figure}

Similar to classical feedback, one of the most important
applications of quantum feedback is to suppress the effects of
noise, which in quantum systems causes decoherence. Markovian
quantum feedback can be used to suppress the decoherence of
macroscopic-superposition states (so-called ``Schr\"{o}dinger
cat''
states)~\cite{PTombesiPRA:1995,PGoetschPRA:1996,DVitaliPRL:1997,DVitaliPRA:1998,MFortunatoPRA:1999}
if we measure the output channel that is causing the decoherence.
As an example, if we prepare the following superposition of two
coherent states,
 \begin{equation}\label{Superposition of macroscopically
distinguishable coherent states}
|\psi\rangle=\frac{|\alpha_0\rangle+|-\alpha_0\rangle}{\sqrt{2}} ,
\end{equation}
in an optical cavity, then by making a Homodyne measurement of the
light that leaks out of the cavity we can use Markovian feedback
to extend the time over which the coherence survives.
Without quantum feedback, the timescale over which the coherence
between the two coherent states survives is
$\tau=1/\left(2\gamma|\alpha_0|^2\right)$, where $\gamma$ is the
decay rate of the cavity~\cite{PGoetschPRA:1995}. If the signal
from the Homodyne measurement is used to control the
transmissivity of an electro-optic modulator (EOM), as depicted in
Fig.~\ref{Fig of Schrodinger cat state protection}, then the
timescale over which the coherence survives
is~\cite{PGoetschPRA:1996}
\begin{equation}\label{Prolonged decoherence time of Schrodinger-cat state}
\tau_{\rm
fb}=\frac{\tau}{\left(1-g\sin\theta\right)^2} ,
\end{equation}
where $\theta$ and $g$ are the phase shift and gain of the
feedback, respectively.

Another example of the use of Markovian quantum feedback is to reduce the phase
noise in an atom laser~\cite{HMWisemanPRL:2001, LKThomsenPRA:2002}. The primary source of this phase noise
is collisions between atoms. A single-mode atom laser can be described by the master equation
\begin{equation}\label{Master equation for atom laser without feedback}
\dot{\rho}=-i[C(a^{\dagger})^2a^2,\rho]+\kappa\mu\int_0^{\infty}\!\!dq\,\mathcal{D}[a^{\dagger}e^{-qaa^{\dagger}/2}]\,\rho+\kappa\,\mathcal{D}[a]\,\rho,
\end{equation}
where as usual $a$ is the annihilation operator for the mode,
$\kappa,\,\mu\gg 1$ are respectively the damping rate and the
stationary mean number of atoms in the laser mode. The nonlinear
Hamiltonian
\begin{eqnarray*}
H_{\rm coll}= \hbar C (a^{\dagger})^2  a^2
\end{eqnarray*}
describes the collisions between atoms at a rate $C$. From
Eq.~(\ref{Master equation for atom laser without feedback}) the
linewidth of the atom laser without feedback can be shown to be
\begin{equation}\label{Atom laser linewidth withouut feedback}
l=\left\{%
\begin{array}{lll}
    {\displaystyle \frac{\kappa}{2\mu}\left(1+\chi^2\right)}, & & \chi\ll\sqrt{\mu}, \\
    \\
    {\displaystyle \frac{\kappa\chi}{\sqrt{\pi\mu/2}} }, & & \chi\gg\sqrt{\mu}, \\
\end{array}%
\right.
\end{equation}
where $\chi=4\mu C/\kappa$ is a dimensionless atomic interaction
strength. We now make a quantum nondemolition measurement of the
atomic number, and use the photocurrent (the stream of measurement
results) from this measurement to modulate the external field
applied to the condensate of atoms to generate the output of the
laser. With this feedback the linewidth of the atom laser becomes
\begin{equation}\label{Atom laser linewidth with feedback}
l=\frac{\kappa}{2\mu}\left(1+\frac{\chi}{\sqrt{\eta}}\right) .
\end{equation}
The feedback is thus capable of eliminating the effect of atomic collisions on the linewidth of the laser~\cite{HMWisemanPRL:2001,LKThomsenPRA:2002}.

Continuous-time feedback has also been applied to quantum error
correction, a technique that is able to slow the decoherence of
unknown quantum
states~\cite{CAhnPRA:2002,CAhnPRA:2003,MSarovarPRA:2004,BAChasePRA:2008,HMabuchiNJP:2009,KKeanePRA:2012,SSSzigetiPRL:2014}.
By ``unknown'', we mean that the controller is able to preserve
the initial state without knowing what the state is. This requires
that the state is initially encoded in a larger Hilbert space
before the error-correction can be applied. As an example, we
return to the three-qubit bit-flip code given in the beginning of
this section. The idea is to replace the projective measurements
that extract the error syndrome with continuous
measurements~\cite{CAhnPRA:2002}. Recall that the state of a
single logical-qubit is encoded (stored) in three physical qubits.
To extract the information about the error, we make a continuous
measurement of the three operators $ZZI$, $IZZ$, and $ZIZ$, all
with the same measurement strength, $\kappa$. We also apply three
control Hamiltonians, $H_1 = \hbar \lambda_1 XII$, $H_2 = \hbar
\lambda_2 IXI$, and $H_3 = \hbar \lambda_3 IIX$. These three
Hamiltonians apply the corrections for the three possible errors,
and thus the rates $\lambda_1$, $\lambda_2$, and $\lambda_3$ are
to be determined by the measurement results. Recall that Bayesian
feedback involves integrating the stochastic master equation,
which in this case is
\begin{eqnarray}\label{Stochastic master equation for quantum error correction}
d\rho_{\mss{c}}&=&-i[\lambda_1XII+\lambda_2IXI+\lambda_3IIX,\rho_{\mss{c}}]dt+\gamma\left(\mathcal{D}[XII]+\mathcal{D}[IXI]+\mathcal{D}[IIX]\right)\rho_{\mss{c}}
dt\nonumber\\
&&+\kappa\left(\mathcal{D}[ZZI]+\mathcal{D}[IZZ]+\mathcal{D}[ZIZ]\right)\rho_{\mss{c}}
dt+\sqrt{\kappa}\left(\mathcal{H}[ZZI]dW_1+\mathcal{H}[IZZ]dW_2+\mathcal{H}[ZIZ]dW_3\right)\rho_{\mss{c}} .
\end{eqnarray}
The feedback control parameters $\{\lambda_j\}$ are then
determined from the conditional density matrix $d\rho_{\mss{c}}$.
By minimizing a cost function, which is defined as the distance
between the conditional density matrix and the space in which the
logical qubit should reside (the codespace), the optimal feedback
is determined to be
\begin{eqnarray}\label{Optimal feedback coefficients for error correction}
\lambda_1&=&\lambda\,{\rm sgn}\langle YZI+YIZ \rangle_{\mss{c}},\nonumber\\
\lambda_2&=&\lambda\,{\rm sgn}\langle ZYI+IYZ \rangle_{\mss{c}},\nonumber\\
\lambda_3&=&\lambda\,{\rm sgn}\langle ZIY+IZY \rangle_{\mss{c}},
\end{eqnarray}
where $\lambda$ is the maximum available feedback strength. Here,
${\rm sgn}(x)=+1$ if $x>0$, ${\rm sgn}(0)=0$, and ${\rm
sgn}(x)=-1$ if $x<0$. It is shown in Ref.~\cite{CAhnPRA:2002} that
this quantum error correction protocol can efficiently increase
the fidelity of the encoded quantum states beyond that achieved by
traditional projective-measurement-based error-correction, so long
as the time delay induced by the feedback loop is small enough.

It was shown in~\cite{MSarovarPRA:2004} that averaging the stream of measurement results is sufficient to perform quantum error-correction, thus replacing the highly complex Bayesian feedback with time-averaged feedback. Further efficient methods for continuous-time quantum error-correction are presented in~\cite{BAChasePRA:2008}.

A simpler situation occurs if the environment that is causing the errors can itself be measured. An example of this is when the light that leaks out of an optical cavity is detected. In this case the measurement provides direct information about what error has occurred, reducing the resources required for quantum error correction. It is shown in~\cite{CAhnPRA:2003} that when the bath that causes the errors is detected, Markovian feedback is all that is required to perform error-correction, and only $n+1$ physical qubits are required to encode $n$ logical qubits.

Quantum feedback has been combined with open-loop control
protocols to reduce errors in quantum systems. The open-loop
technique of dynamical decoupling allows errors to be reduced if
they are due to noise that has a sufficiently long correlation
time~\cite{MJBiercukNature:2009,LViolaPRL:1999,
PZanardiPLA:1999,TGreenPRL:2012}. In~\cite{FTicozziPRA:2006},
feedback and dynamical decoupling are combined by feeding the
output of a dynamical decoupling protocol to a feedback
controller. It is shown that for a single qubit the combination of
quantum feedback and dynamical decoupling outperforms either
method when used alone.

\subsubsection{State reduction and stabilization}\label{s242}

In the previous section we discussed the use of feedback to protect quantum state temporarily. We refer to the indefinite protection of a quantum state as stabilization. Unknown states cannot be stabilized, but known states certainly can be. Sometimes open-loop control can be used to stabilize states, but only in certain circumstances, for example when an effectively zero-temperature environment is available~\cite{JWangPRA:2001}. Markovian quantum feedback can be used to stabilize the states of a single two-level atom when the source of decoherence is detected~\cite{JWangPRA:2001,HFHofmannOptExp:1998,HFHofmannPRA:1998}. In Ref.~\cite{JWangPRA:2001} it is shown that for a two-level atom,
states in an ellipse in the lower half part of the Bloch ball can be stabilized by open-loop control under the damping process. However, when we introduce Markovian quantum feedback and carefully choose the feedback gain and the strength of the driving field, it is only states on the equator of the Bloch sphere that cannot be stabilized.

More generally, Bayesian feedback has been applied to the
stabilization of quantum states in a variety of mesoscopic
systems~\cite{JZhangTAC:2010,NGanesanPRA:2007,AHopkinsPRB:2003,JZhangcoolingPRA:2009,MJWoolleyPRB:2010,RRuskovPRB:2002,ANKorotkovMicroJ:2005,ANKorotkovPRB:2005,QZhangPRB:2005,ANJordanPRB:2006,ANKorotkovPRB:2001,ANKorotkovPRB:1999,WCuiPRA:2012,HSGoanPRB:2001,NPOxtobyPRB:2005,NPOxtoby:2006,JSJinPRB:2006,WCuiPRA:2013,DRistePRL:2012,RVijayNature:2012,PCampagneIbarcqPRX:2013,SGustavssonPRL:2007,SHOuyangPRB:2007,ZZLiSR:2013}.
These include nano-mechanical
resonators~\cite{AHopkinsPRB:2003,JZhangcoolingPRA:2009,MJWoolleyPRB:2010},
quantum-dots~\cite{RRuskovPRB:2002,ANKorotkovMicroJ:2005,ANKorotkovPRB:2005,QZhangPRB:2005,ANJordanPRB:2006,ANKorotkovPRB:2001,ANKorotkovPRB:1999,WCuiPRA:2012,HSGoanPRB:2001,NPOxtobyPRB:2005,NPOxtoby:2006,JSJinPRB:2006,SGustavssonPRL:2007,SHOuyangPRB:2007,ZZLiSR:2013},
and superconducting
qubits~\cite{WCuiPRA:2013,DRistePRL:2012,RVijayNature:2012,PCampagneIbarcqPRX:2013}.

Immediately before a stabilization feedback process starts, the system to be controlled will likely be in some steady-state determined by the noise processes that drive it, and this state is often significantly mixed. When we first apply measurement-based feedback, the measurement will reduce the entropy of the system, a process often called \textit{purification}. If the state towards which the measurement projects the system is not one of the eigenstates of the initial density matrix, then the measurement necessarily also induces a ``collapse'' of the wave-function, a process often called ``state-reduction''. Thus purification may or may not involve state-reduction, although often these terms are used interchangeably.

If we make a continuous measurement of an observable, and perform no feedback, then the measurement will continually try to project the system onto one of the eigenstates of the observable, where the choice of eigenstate is random. In the absence of noise that interferes with the reduction process, the final state of the system will be one of these eigenstates. As shown in~\cite{RvanHandelIEEETAC:2005,CAltafiniIEEETAC:2007,RvanHandelJOB:2005,BQiTAC:2013,MMirrahimiSIAMJCO:2007}, if we perform feedback control during the state-reduction process, we can control which eigenstate that the measurement projects onto.

To examine this process further, we consider the single-qubit state-reduction protocol presented in
Ref.~\cite{RvanHandelIEEETAC:2005}. As shown in Fig.~\ref{Fig of
the qubit state stabilization}, a single two-level system, such as atom, is inserted into an
optical cavity, and the output of the cavity is measured by a homodyne detection. When the cavity damping rate is much faster than the dynamics of the qubit, this measurement procedure realizes a continuous measurement of the $\sigma_z$ operator of the two-level system.

The stream of measurement results is fed
into a digital controller and then back to control a classical field that allows us to rotate the two-level system around the $y$-axis. The control Hamiltonian is thus $H=B\left(t\right)\sigma_y$, where $B(t)$ is the control parameter that we can vary. Writing the state of the two-level system using the Bloch vector $\mathbf{v}=\left(v_x,v_y,v_z\right)^{\mss{T}}= \left(\langle \sigma_x \rangle,\langle \sigma_y\rangle ,\langle \sigma_z\rangle\right)^{\mss{T}}$, the dynamics under the feedback process is
\begin{eqnarray}\label{Bloch equation for qubit state stabilization}
d
v_x&=&\left[B\left(t\right)v_z-\frac{\Gamma_z}{2}v_x\right]dt-\sqrt{\Gamma_z}\,v_x
v_z\,dW, \nonumber\\
d v_y&=&-\frac{\Gamma_z}{2}
v_y\,dt-\sqrt{\Gamma_z}\,v_y v_z\,dW, \nonumber\\
d v_z&=&-B\left(t\right) v_x\,dt +
\sqrt{\Gamma_z}\left(1-v_z^2\right) dW .
\end{eqnarray}
If we assume that $v_y=0$, the control field $B\left(t\right)=G v_x$ with $G>0$ can
steer the system to the spin-up state with unit probability. This method is extended in~\cite{MMirrahimiSIAMJCO:2007} to systems with any fixed total angular momentum $J$, such as a dilute gas of two-level atoms. In this work a piecewise-continuous
control protocol is designed to stabilize any selected eigenstate of the $J_z$ operator, and can also be used to stabilize the symmetric and antisymmetric states in a system with two qubits.

\begin{figure} \centerline{\includegraphics[width =
14 cm]{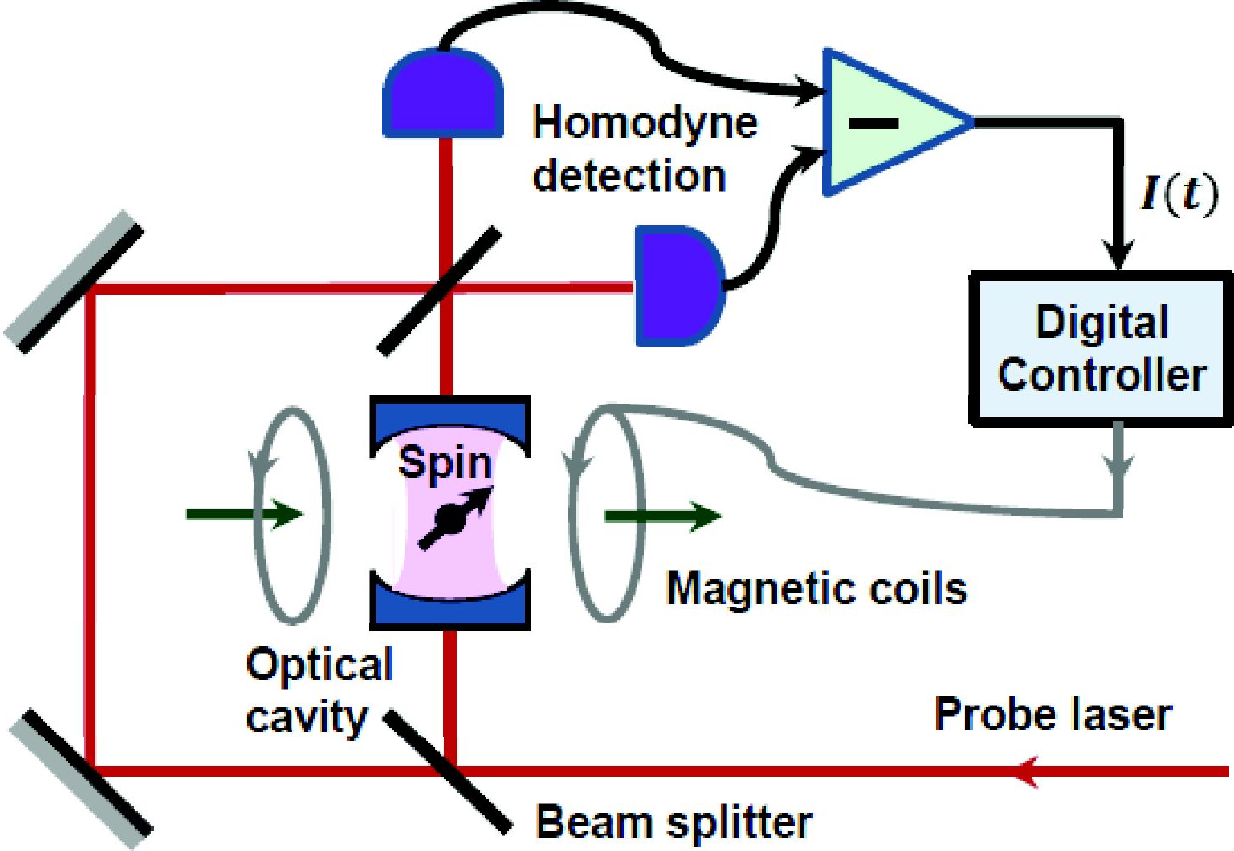}} \caption{(Color online)
Diagram of a proposal by van Handel \textit{et
al.}~\cite{RvanHandelIEEETAC:2005} for the stabilization of qubit
states. The information of the spin inside the cavity is
continuously extracted by an optical probe field and then detected
by homodyne measurement. The electric output signal
$I\left(t\right)$ of the homodyne measurement is fed into a
digital controller and then fed back to control the magnetic field
(shown in green) imposed on the spin.}\label{Fig of the qubit
state stabilization}
\end{figure}

As our final example of stabilization, we now present the
approaches introduced
in~\cite{WPSmithPRL:2002,JEReinerPRA:2003,JEReinerPRA:2004} in
which feedback is used to stabilize a particular dressed state in
a strongly-coupled cavity-QED system. For a weakly-driven
single-cavity mode strongly coupled to $N$ atoms with coupling
strength $g$, the steady state is
\begin{equation}\label{Steady state of strongly coupled cavity QED system}
|\psi_{\rm ss}\rangle=|0,g\rangle+\lambda\left(|1,g\rangle-\frac{2g\sqrt{N}}{\gamma}|0,e\rangle\right)+O\left(\lambda^2\right) .
\end{equation}
Here $\gamma$ is the damping rate of the atom, and $\lambda \ll 1$
is proportional to the ratio $\varepsilon_d/\kappa$, where
$\varepsilon_d$ and $\kappa$ are the driving strength and damping
rate of the cavity, respectively. The state $|j,g\rangle$ denotes
$j$ photons in the cavity mode, and all the atoms in the their
ground states, while $|j,e\rangle$ denotes $j$ photons in the
cavity mode and all but one of the atoms are in their ground
states.

Since the steady-state has almost no photons, photon detections are relatively rare. When a photon is detected, the conditional state changes abruptly, and evolves as
\begin{equation}\label{Steady state of strongly coupled cavity QED system with truncation}
|\psi_{\mss{c}}\left(\tau\right)\rangle=|0,g\rangle+\lambda\left[\xi\left(\tau\right)|1,g\rangle+\theta\left(\tau\right)|0,e\rangle\right],
\end{equation}
where $\xi\left(\tau\right)$ and $\theta\left(\tau\right)$ are oscillatory functions of time. It turns out that by adjusting the driving field at a specific time after the detection, the state of the system can be frozen indefinitely (see Section~\ref{secSmith02} below). Once the driving field is returned to its original value, the evolution of the conditioned state continues as if it had never been interrupted, and returns to the steady-state. This feedback scheme was realized experimentally in Ref.~\cite{WPSmithPRL:2002}.

It has also been shown that the same atom-cavity system can be stabilized in the opposite regime of strong-driving~\cite{JEReinerPRA:2003}. In this case one of the dressed states is stabilized by flipping the atomic state using a $\pi$-pulse when a measurement-induced quantum jump is observed. The signature of this stabilization process appears in the atomic fluorescence spectrum as the enhancement of one sideband.

Not all feedback involves changing the Hamiltonian of a system conditional upon the measurement stream. If the purpose is to engineer a specific kind of measurement, for example, one may change the \textit{measurement} conditional on the stream of measurement results. The result of this feedback-modified measurement is referred to as an \textit{adaptive} measurement. We will give examples of adaptive measurements below. In Refs.~\cite{RIKarasikPRL:2011,RIKarasikPRA:2011}, the authors consider how to design an adaptive measurement to minimize the classical memory required to track the state of the measured system. They show that for a $d$-dimensional system it is possible to modify the measurement with time so that the system is restricted to jumping between a discrete set of just
$k\geq\left(d-1\right)^2+1$. This means that a $d$-dimensional system can be tracked by a classical $k$-state machine, and a special case shows that a qubit can be tracked by a classical bit.

\subsubsection{Squeezing via feedback}\label{s244}

A quantum harmonic oscillator obeys Heisenberg's uncertainty
principle, meaning that the momentum variance can only be reduced
below $\hbar m\omega/2$ at the expense of increasing the position
variance above $\hbar/(2m\omega)$. Here $m$ and $\omega$ are
respectively the mass and frequency of the oscillator. For an
optical mode, the equivalent conjugate variables are referred to
as the amplitude quadrature $X$ and phase quadrature $Y$. A state
in which the variance of one conjugate variable is decreased at
the necessary expense of the other is called a \textit{squeezed}
state. For a single mode of an optical or electrical cavity it is
not only the state of the mode that is of interest, but the state
of the travelling-wave light or electrical signal that is emitted
from the cavity. We may want to squeeze either the oscillator or
this output. Methods for squeezing both have been investigated
quite extensively.

Assuming that the thermal noise on an oscillator is negligible,
its state can be squeezed merely by modulating its frequency
$\omega$ at $2\omega$. This is called \textit{parametric
amplification} because it amplifies one quadrature, and because
$\omega$ is a ``parameter'' in the Hamiltonian. If we want to
squeeze the oscillator in the presence of thermal noise, then we
need to reduce this noise, for which measurement-based feedback is
one option. Neither measuring a single quadrature, nor making a
homodyne measurement are sufficient for this
purpose~\cite{HMWisemanPRA_Squeezing:1994}. There are two methods
presently known to do this. One is to make a measurement of a
single quadrature, but do so in the rotating frame of the
oscillator. Braginsky \textit{et al.}~\cite{Braginsky80}  were the
first to devise a method to do this, which effectively involves
making a measurement of an oscillator's position, and turning the
measurement off and on at the frequency $2\omega$ (see
also~\cite{RRuskovPRB_Squeezing:2005,AClerkNJP:2008,
JZhangcoolingPRA:2009}). This ``strobing'' of the measurement can
be achieved merely by modulating the interaction with the
measuring device sinusoidally at this frequency. In the
interaction picture the quadrature variables are unchanging, so
that Braginsky's measurement is a \textit{quantum non-demolition}
(QND) measurement. Other methods for making QND measurements of
the quadratures have be devised~\cite{HMWisemanPRA_Squeezing:1994,
PTombesiPRA_Squeezing:1994,
HMWisemanPRA_Squeezing:1992,GJMilburnPRA_Squeezing:1993,
HMWisemanPRA_Squeezing:1995}, but Braginsky's is probably still
the most practical. A QND measurement produces a squeezed
conditional state --- a squeezed state from the point of view of
someone who is able to have processed the measurement record to
determine the means of the quadratures. Since the means fluctuate,
without this information the state is not squeezed. But since a
linear feedback force can be used to stabilize the means, we can
use feedback from the QND measurement to produce an unconditional
squeezed steady-state~\cite{AClerkNJP:2008,
JZhangcoolingPRA:2009}.

The second method to produce squeezed states in the presence of
significant thermal or other noise is to combine a parametric
drive with a standard (unmodulated) measurement of position.  Here
the parametric drive creates the squeezing and the measurement is
much weaker, and extracts the entropy injected by the thermal
noise~\cite{Szorkovszky11, Szorkovszky14}. This method of
producing squeezing has been realized, although not in the quantum
regime,
in~\cite{Szorkovszky13,AVinantePRL_Squeezing:2013,APontinPRL:2014}.

We note that QND measurements have also been used to generate
squeezing in the collective spin state of a gas of two-level
atoms, both theoretically~\cite{LKThomsenPRA_Squeezing:1995} and
experimentally~\cite{Vuletic10, Vuletic10b, Vuletic12}.

\subsubsection{Controlling mechanical resonators}\label{s245}

In the previous section we discussed the use of feedback to
squeeze oscillators, mechanical or otherwise. We now focus on
mechanical oscillators, and consider the creation of other states.
The first task in bringing a mechanical resonator into the quantum
regime is to suck out all the thermal noise so as to put the
resonator in its ground state. From there we can use purely
open-loop control protocols to place it in a more interesting
nonclassical state. To make it possible to cool a mechanical
resonator to the ground state (see,
e.g.,~\cite{KXiaPRB:2010,KXiaPRL:2009}), it must have a very high
frequency so as to reduce the thermal noise, even when placed in a
dilution refrigerator. To attain the required frequencies the
resonators must have a very small mass, and there are a number of
ways to realize such microscopic and mesoscopic resonators. One
can merely make them small~\cite{AKubanekNature:2009,
TFischerPRL:2002,OArcizetNature:2006,SGiganNature:2006,TCorbittPRL:2007,PFCohadonPRL:1999,OArcizetPRL:2006,DKlecknerNature:2006,Thompson08,Schliesser08,Groeblacher09,Schliesser09,Groeblacher09b},
in which case their frequencies are still a little low to enter
the quantum regime. One can go even smaller by fabricating
resonators on layered structures using lithography, and these are
usually referred to as nano-resonators~\cite{Anetsberger09,
Eichenfield09, Eichenfield09b, Teufel11, Teufel11b,
Chan11,SGHoferPRL:2013}. One can also use microscopic systems such
as trapped ions~\cite{PBushevPRL:2006},
electrons~\cite{BDUrsoPRL:2003}, laser-trapped
nano-particles~\cite{JGieselerPRL:2012}, or a gas of neutral atoms
trapped in an optical
lattice~\cite{MVMorrowPRL:2002,AJBerglundAPB:2004,
AJBerglundOL:2007, AJBerglundOE:2007, Murch08, Purdy10}. Quite
recently, quantum feedback has also been applied to Bose-Einstein
condensates (BECs) for feedback-cooling an ultra-cold atomic
ensemble undergoing continuous weak
measurement~\cite{SSSzigetiPRA:2009,SSSzigetiPRA:2010}. It is
shown that, in certain regimes, full quantum-field simulations and
more exotic feedback controls are required in order to
successfully cool the BEC close to the ground
state~\cite{MRHushNJP:2013}, and the robustness of a control
scheme to corruption of the measurement signal by classical noise,
detector inefficiencies, parameter mismatches and a time delay is
considered~\cite{SSSzigetiPRA:2013}.

State-of-the-art cooling schemes for nano- and micro-mechanical
resonators currently use coherent feedback, to be discussed in
Section~\ref{s3}. Nevertheless measurement-based cooling methods
for mechanical resonators have been investigated quite
extensively. When the resonator is far from the quantum regime,
feedback cooling of resonators can be considered
classical~\cite{MYVilenskyPRA:2006,YKishimotoPRE:1997,GCZhangPRA:1999}.
The primary obstacle to cooling resonators to the ground state
using measurement-based feedback is the requirement that the
measurement has an efficiency near unity. At the time of writing,
measurements on nano-resonators do not have the required
efficiency, but we expect this to change in the near future.
Cooling via measurement has been investigated both for Markovian
feedback~\cite{JADunninghamPRA:1997,SManciniPRA_cooling:1997,SManciniPRA_cooling:2000}
and Bayesian
feedback~\cite{SDWilsonPRA:2007,AHopkinsPRB:2003,JZhangcoolingPRA:2009,
DASteckPRL:2004,DASteckPRA:2006}. Feedback can also be used to
create and control highly non-classical states of
resonators~\cite{Jacobs11}.

\subsubsection{Controlling transport in nano-structures}\label{s243a}

Measurement-based quantum feedback can be used to control the
quantum transport process in
nanostructures~\cite{GKieblichPRL:2011,GKieblichNJP:2012,TbrandesPRL:2010,CPoltlPRB:2011}.
This can be thought of as stabilizing the quantum state of the
transport device. The proposal in Ref.~\cite{TbrandesPRL:2010}
shows that a classical feedback control can freeze the
fluctuations of quantum transport by changing parameters in the
Hamiltonian conditional to the number of tunneled particles. This
feedback method can be further used to reconstruct the full
counting statistics of the transport device from the frozen
distribution. In~\cite{CPoltlPRB:2011}, this proposal was applied
to nonequilibrium-electron-transport through a double-quantum-dot,
which for this purpose can be treated as a two-level system. The
feedback is able to purify the transport state, represented by the
full counting statistics of the electron flow through the device,
and it is shown that half of the quantum states on the Bloch
sphere of the double-quantum-dot can be stabilized. The feedback
is also able to stabilize the coherent delocalized states of the
electrons.

\subsubsection{Entanglement creation and control}\label{s243}

It is shown in~\cite{SSchyneiderPRA:2002} that when a cavity
containing two two-level atoms is resonantly driven, the
steady-state of the atoms can be entangled. More specifically, if
(i) the cavity damping rate $\kappa$ is much faster than all other
timescales, so that the cavity can be adiabatically eliminated,
and (ii) the collective damping rate of the atoms induced by the
cavity is much larger than the atoms' spontaneous emission rates,
then one can recover a Dicke model for the atoms. The steady state
of this Dicke model can be written in the angular momentum basis
and analyzed in terms of the symmetric and antisymmetric
subspaces. When the initial state of the atoms is symmetric, the
stationary state is entangled, although this entanglement,
measured by the Wootters' concurrence~\cite{WKWoottersPRL:1998},
is only about $0.11$~\cite{SSchyneiderPRA:2002}.

It turns out that Markovian quantum feedback can be used to increase the steady-state entanglement of the atoms~\cite{JWangPRA:2005,JGLiPRA:2008,NYamamotoPRA:2005,SManciniEPJD:2005,YliPRA:2011,ARRCarvalhoPRL:2007,CViviescsPRL:2010,ARRCarvalhoPRA:2007,ARRCarvalhoPRA:2008,SCHouPRA:2010,RNStevensonEPJD:2011,LCWangEPJD:2010,DXueJPB:2010,JSongJOSAB:2012}. It has been shown that this is possible using both feedback based on photon detections (quantum jumps)~\cite{ARRCarvalhoPRA:2007,ARRCarvalhoPRA:2008,SCHouPRA:2010,RNStevensonEPJD:2011,LCWangEPJD:2010,DXueJPB:2010,JSongJOSAB:2012}, and feedback using homodyne detection (trajectories driven by Gaussian noise)~\cite{JWangPRA:2005,JGLiPRA:2008,NYamamotoPRA:2005,SManciniEPJD:2005,YliPRA:2011}.

\begin{figure}
\centerline{\includegraphics[width = 15
cm]{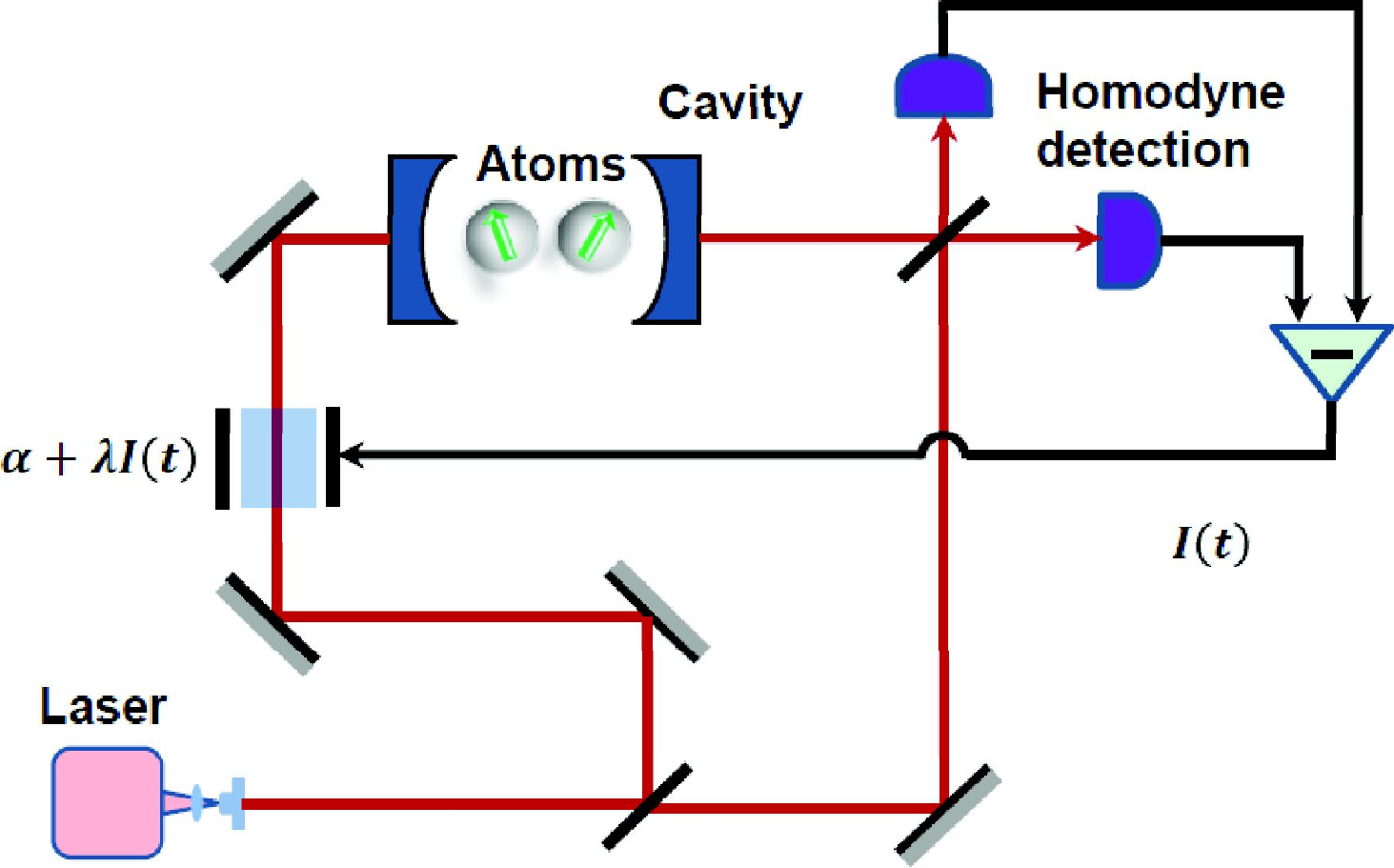}} \caption{(Color online) Diagram
of a theoretical proposal by Wang \textit{et
al.}~\cite{JWangPRA:2005} for two-atom entanglement creation by
homodyne-mediated feedback. The homodyne current
$I\!\left(t\right)$ from the damped cavity is directly fed back to
generate a control signal $\alpha+\lambda I\!\left(t\right)$,
which is converted into an optical signal by an electro-optic
modulator and then used to resonantly drive the cavity coupled to
the two atoms. These two atoms are entangled in the steady state.}
\label{Fig of the entanglement creation by MF}
\end{figure}

In Ref.~\cite{JWangPRA:2005} the authors show that for homodyne detection the stationary entanglement can be increased from $0.31$ under feedback that is symmetric for the two atoms (see Fig.~\ref{Fig of the entanglement creation by MF}). This stationary entanglement can be increased further to $0.82$ if local asymmetric feedback is introduced~\cite{JGLiPRA:2008}. Feedback based on photon detections is even better at maintaining entangled states, and for symmetric feedback is able to achieve a concurrence of  $0.49$~\cite{ARRCarvalhoPRA:2008}. This stationary entanglement is reduced by spontaneous emission,  but can be increased further by the use of local asymmetric feedback. Feedback based on photon-detections is also robust against fluctuations in various parameters, such as the detection efficiency, especially in the adiabatic regime~\cite{ARRCarvalhoPRA:2008}. Bayesian feedback has also been used to generate and protect entanglement in the above system~\cite{NYamamotoAutomatica:2007, JKStocktonPRA:2004}. Although the computational complexity for this kind of control is higher, it can also potentially produce higher entanglement.

As an extension of the above feedback schemes, the stabilization
of multipartite entanglement via feedback has also been
considered~\cite{RNStevensonPRA:2011, JSongPRA:2012}. Due to the
lack of a measure for multipartite entanglement, these studies
have focussed on stabilizing particular multipartite entangled
states, such as Dicke states~\cite{JKStocktonPRA:2004} or GHZ
states~\cite{LiXQPRA:2011}. More recently, the control of
entanglement via feedback has been discussed for solid-state
systems, especially superconducting
circuits~\cite{MSarovarPRA:2005,LiXQPRA:2010,LiXQPRA:2011}, and
has been demonstrated experimentally both in superconducting
circuits~\cite{DRisteNature:2013} and cavity QED
systems~\cite{SBrakhanePRL:2012}.

Continuous-variable entanglement of optical beams, which is closely related to multi-mode optical squeezing, can, and usually is, produced by using nondegenerate parametric oscillator in an optical cavity. This method is limited by the strength of the optical nonlinearity employed, which is usually very weak, and is further reduced by the cavity damping. It has been shown that feedback can be used to increase the  continuous-variable entanglement~\cite{SManciniPRA:2006,GXLiPRA:2006,SSKeJPBAMOP:2007,SManciniHMWisemanPRA:2007,MGGenoniPRA:2013,ASerafiniPRL:2010,MYanagisawaPRL:2006,NYamamotoPRA:2008,HINurdinPRA:2012} generated in this way.

Consider two optical modes with quadrature operators $X_1$, $Y_1$ and $X_2$, $Y_2$, respectively. In Ref.~\cite{SManciniPRA:2006} it is shown that a single-loop Markovian feedback scheme can be used to reduce the variance of the operator $X_1-X_2$ while preserving the variance of the operator $Y_1+Y_2$, thus improving the steady-state entanglement of the two modes. However, the entangled state in this case is a mixed state. A modified proposal in Refs.~\cite{GXLiPRA:2006,SSKeJPBAMOP:2007} is able to produce a pure steady-state two-mode entangled state by using two independent feedback loops to control the variances of $X_1-X_2$ and $Y_1+Y_2$ simultaneously.

It has been shown that the problem of finding the optimal homodyne
measurement and Markovian feedback to produce two-mode intracavity
Einstein-Podolsky-Rosen correlations for both a vacuum
environment~\cite{SManciniHMWisemanPRA:2007} and a thermal
bath~\cite{MGGenoniPRA:2013} is a semidefinite programming
problem~\cite{LVandenberghe}. This means that a global optimum can
be found numerically in a systematic way. A general upper bound
for the generation of steady-state entanglement for multi-mode
bosonic fields via feedback has also been
obtained~\cite{ASerafiniPRL:2010}. Quantum feedback has  been
applied to the problem of generating deterministic entanglement at
the single-photon level~\cite{MYanagisawaPRL:2006}, to avoid
entanglement sudden death~\cite{NYamamotoPRA:2008}, and to enhance
entanglement distributed between cavities including propagation
delays and photon loss~\cite{HINurdinPRA:2012}.

\subsubsection{Quantum state discrimination}\label{s246}

The efficient discrimination of nonorthogonal quantum states has
various applications in quantum and classical communications and
quantum-enhanced metrology. For discrete-variable quantum systems,
such as qubits, it has been shown
theoretically~\cite{AMBranczykPRA:2007} and
experimentally~\cite{GGGillettPRL:2010} that continuous
measurement and feedback can efficiently discriminate two
nonorthogonal states of a single qubit, as well as correct these
states against dephasing noise. It has also been demonstrated
experimentally that adaptive local measurement and feedback
performs much better than non-adaptive measurements for
discriminating  non-orthogonal states of qubits, given multiple
copies, and can efficiently suppress
noise~\cite{BLHigginsPRL:2009}.

Quantum feedback has also been used for the discrimination of
coherent states of oscillators or traveling-wave fields. Such
discrimination is useful for communication, because coherent
states are easy to prepare and manipulate. Adaptive measurement
schemes can maximize the information rate so as to achieve
Holevo's bound~\cite{VGiovannettiPRL:2004}, and allow for
long-distance communication. It has been shown both
theoretically~\cite{SGuhaJMO:2011} and
experimentally~\cite{KTsujinoPRL:2011,JChenNatPhoton:2012,FEBecerraNatPhoton:2013}
that joint-detection and adaptive feedback with
pulse-position-modulation codewords reduces the error probability
for both conditioned and unconditioned coherent-state
discrimination, compared with traditional direct detection methods
(e.g., homodyne detection). This method can also beat the standard
quantum limit to approach the Helstrom
limit~\cite{CWHelstrom:1976}, being the minimum achievable average
probability of error for discriminating quantum states.

\subsubsection{Quantum parameter estimation}\label{s247}

High-precision phase measurements of optical beams, especially
those in the quantum regime, have various important applications,
such as interferometric gravity-wave detection or quantum
communication. However, the phase of the electromagnetic field
mode is not a directly-measurable quantity, and phase measurement
protocols always measure some other quantity that introduces
excess uncertainty and noise into the estimation process.

The traditional method for measuring phase was to use heterodyne
detection, in which the field mode to be measured is combined with
a far-detuned strong local oscillator field. It is well-known that
phase measurements based on heterodyne detection can reach the
standard quantum limit, in which the phase sensitivity, being the
variance of the measured phase $\left(\delta\phi\right)^2$, scales
as $N^{-1}$ when a state with an average of $N$ photons is fed
into the input port to be measured. But this is not the
fundamental limit to phase estimation. The latter is the
Heisenberg limit, which gives a scaling of $N^{-2}$. The
Heisenberg limit could be achieved with a perfect measurement of
canonical
phase~\cite{BCSandersPRL_Adaptive:1995,BCSandersJMO_Adaptive:1997},
but this appears to be essentially impossible.

Wiseman and
collaborators~\cite{HMWisemanPRL_Adaptive:1995,HMWisemanPRA_Adaptive:1997,HMWisemanPRA_Adaptive:1998,TCRalphJOBQSO_Adaptive:2005,DWBerryPRA_Adaptive:2000,DTPopePRA_Adaptive:2004,DWBerryPRL_Adaptive:2000,DWBerryPRA_Adaptive:2001,DWBerryPRA_Adaptive:2002}
have shown that by using an adaptive measurement it is possible to
realize a measurement of phase that is very close to the
Heisenberg
limit~\cite{Pegg88,Pegg89,BCSandersPRL_Adaptive:1995,BCSandersJMO_Adaptive:1997}.
There are primarily two approaches. One can use an adaptive
homodyne
measurement~\cite{HMWisemanPRL_Adaptive:1995,HMWisemanPRA_Adaptive:1997,HMWisemanPRA_Adaptive:1998,TCRalphJOBQSO_Adaptive:2005,DWBerryPRA_Adaptive:2000,DTPopePRA_Adaptive:2004}
or an adaptive interferometric
measurement~\cite{DWBerryPRL_Adaptive:2000,DWBerryPRA_Adaptive:2001,DWBerryPRA_Adaptive:2002}
(see Fig.~\ref{Fig of the adaptive phase measurement}). As shown
in Fig.~\ref{Fig of the adaptive phase measurement}(a), the key
element of  an adaptive homodyne phase measurement is to feed back
the output of the homodyne detection to control the phase of a
local oscillator, and thus track the phase quadrature to be
estimated. It is shown in Ref.~\cite{HMWisemanPRA_Adaptive:1997}
for a semiclassical model and in
Ref.~\cite{HMWisemanPRA_Adaptive:1998} for a full quantum analysis
that an excess phase uncertainty scaling as $N^{-3/2}$ can be
reached. A modified approach in
Ref.~\cite{DWBerryPRA_Adaptive:2000} shows that a more
sophisticated feedback protocol can reach a better theoretical
limit, scaling as ${\rm ln}N/N^2$.

The adaptive interferometric phase measurement can perform even
better than the adaptive homodyne method. In an adaptive
interferometry phase measurement, a Mach-Zehnder interferometer is
introduced with the unknown phase to be estimated in one arm and
the controllable phase used to track the unknown phase in the
other arm. This achieves a phase sensitivity very close to the
Heisenberg
limit~\cite{DWBerryPRL_Adaptive:2000,DWBerryPRA_Adaptive:2001}. It
has also been shown that both the adaptive homodyne
method~\cite{DTPopePRA_Adaptive:2004} and the interferometric
method~\cite{DWBerryPRA_Adaptive:2002} can be used to estimate a
stochastically-varying phase. More complex feedback designs, such
as those based on time-symmetric
smoothing~\cite{MTsangPRL_parameter:2009,MTsangPRA_parameter:2009,MTsangPRA_parameter:2010},
can also be used in adaptive phase measurement. A number of
adaptive phase-measurement schemes have been demonstrated in
experiments~\cite{ArmenPRL:2002,HigginsNature:2007,NOON,GYXiangNatPhoton:2011,GYXiangNatPhoton:2010,WheatleyPRL:2010,HYonezawaScience:2012,ROkamotoPRL:2012}.

\begin{figure}
\centerline{\includegraphics[width = 16.8 cm]{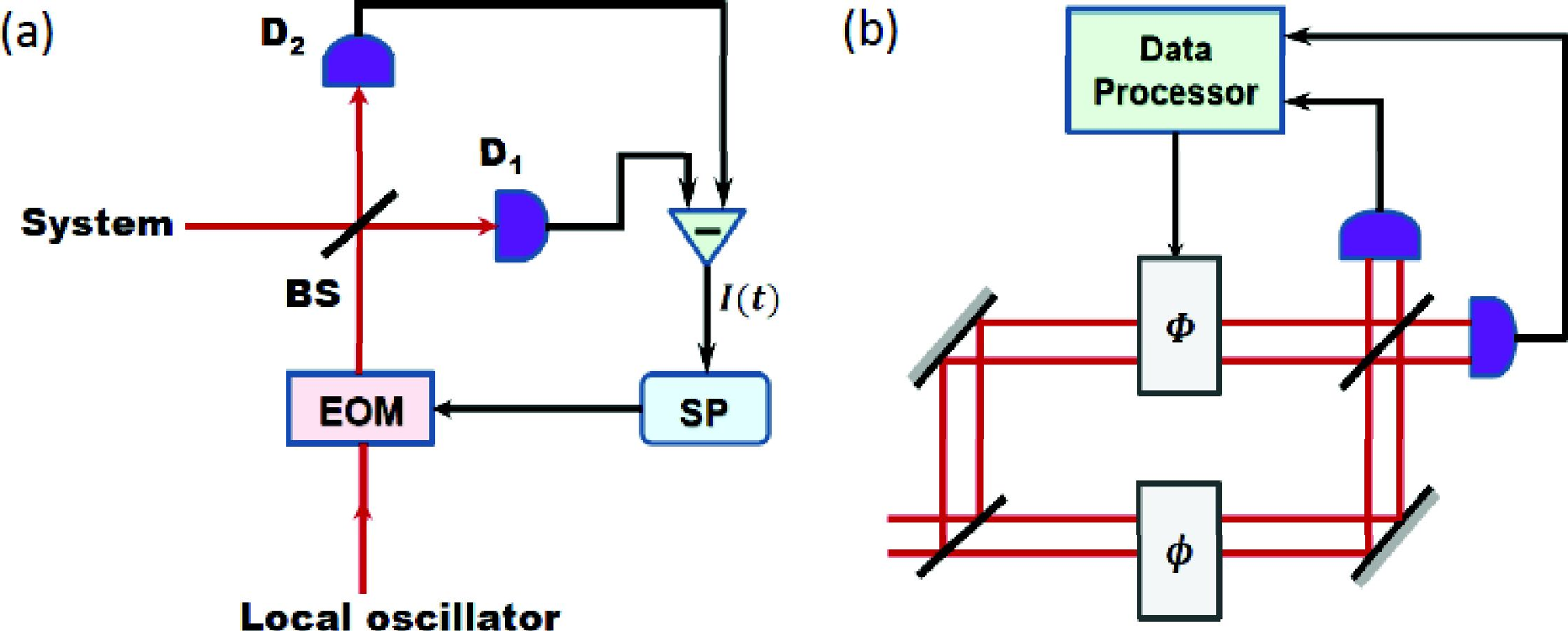}}
\caption{(Color online) (a) Theoretical proposal by
Wiseman~\cite{HMWisemanPRL_Adaptive:1995} for adaptive homodyne
phase measurement. The output $I\!\left(t\right)$ of the homodyne
detection is processed by a Signal Processor (SP) which then
controls the phase of the local oscillator by an electro-optic
modulator (EOM). (b) Theoretical
proposal~\cite{DWBerryPRA_Adaptive:2001} for adaptive
interferometry phase measurements. The figure shows a Mach-Zehnder
interferometer with the unknown phase $\phi$ to be estimated in
one arm and the controllable phase $\Phi$ (used to track the
unknown phase $\phi$) in another arm.}\label{Fig of the adaptive
phase measurement}
\end{figure}

More generally, we may wish to estimate one or more numbers that
parameterize the state, Hamiltonian, or overall evolution of a
system. Such parameters can be estimated by making a continuous
measurement on an evolving system and processing the measurement
results
~\cite{FVerstraetePRA_parameter:2001,JGambettaPRA_parameter:2001,PWarszawskiPRA_Adaptive:2004,BAChasePRA_parameter:2009,JFRalphPRA_parameter:2011,JKStocktonPRA_parameter:2004}.
Such a procedure has applications to metrology, such as the
detection of weak force by monitoring a harmonic
oscillator~\cite{JFRalphPRA_parameter:2011}, or estimating the
Rabi frequency of a two-level
atom~\cite{JGambettaPRA_parameter:2001}. Feedback control can be
used to make the estimation process more robust to the uncertainty
in the system parameters~\cite{JKStocktonPRA_parameter:2004}. The
basic method involved in parameter estimation and metrology via
continuous measurements to use a ``hybrid'' master equation that
evolves the observer's knowledge of the system as well as their
knowledge of the parameters~\cite{JFRalphPRA_parameter:2011}, or
an equivalent quantum particle filtering
equation~\cite{BAChasePRA_parameter:2009}.

\subsubsection{Rapid state-purification and measurement}\label{s248}

It is possible to use quantum feedback to speed up the rate at which a continuous measurement purifies, or provides information about a quantum system~\cite{KJacobsPRA:2003,JCombesPRL:2006,HMWisemanNJP:2006,ANJordanPRB:2006,CHillNJP:2007,EJGriffithPRB:2007,CHillPRA:2008,HMWisemanQIP:2008,JCombesPRL:2008,JCombesPRA:2010,JCombesPRA2:2010,JCombesPRX:2011,KMaruyamaPRA:2008,TTanamotoPRA:2008}. To understand this further, let us consider a
continuous measurement of a qubit which provides information about the basis
$\left\{|0\rangle,|1\rangle\right\}$. The dynamics of this measurement is given by the SME
\begin{equation}
\label{Stochastic master equation for measuring sigmaz}
      d\rho=\mathcal{D}[\sigma_z]\,\rho\,dt+\mathcal{H}[\sigma_z]\,\rho\,dW.
\end{equation}
In order to study how one can reduce the observer's uncertainty of
the measured quantum system, an algebraically simple measure of the observer's uncertainty, called
``linear entropy", $s=1-{\rm tr}\left[\rho^2\right]$, is useful. If we
assume without loss of generality that $y=0$, we can obtain from
Eq.~(\ref{Stochastic master equation for measuring sigmaz})
that~\cite{HMWisemanNJP:2006}
\begin{equation}\label{Evolution of the linear entropy for measuring sigmaz}
ds=-\left(8s^2+4x^2s\right)dt-4zs\,dW,
\end{equation}
where $\alpha={\rm tr}\left[\sigma_\alpha\rho\right],\,\alpha=x,y,z$. It can be seen
that $s$ will decrease more rapidly when $x$ is maximized. If we introduce an
ideal Hamiltonian feedback that rotates the qubit at each time step to maintain $z=0$, then
$ds$ is maximized and given by $ds=-4sdt$. As shown in Ref.~\cite{KJacobsPRA:2003}, if
we start from a maximally mixed state, then under this feedback, and in the long time limit, the time
required to achieve a purity of
$1-\epsilon$ is $\tau_q={\rm ln}\left(\epsilon^{-1}\right)/4$, which is half the time taken for the average purity to  reach this level without feedback. Here $\epsilon\ll
1$ denotes the error threshold value. Such an increase in the rate of purification is
a purely  quantum effect, and cannot be realized for an equivalent measurement on a classical
bit.

Although the above analysis shows that using feedback to
continually rotate the quantum state onto the plane orthogonal to
the measurement axis will speed up purification,
Ref.~\cite{HMWisemanNJP:2006} shows that keeping the quantum
states parallel to the measurement axis can reduce the average
time for the measured quantum system to reach a given purity.

As an extension of the results in Ref.~\cite{KJacobsPRA:2003}, a more general study in Ref.~\cite{JCombesPRL:2006} shows that Hamiltonian feedback can speed up the rate of purification, or state reduction, by
at least a factor of $2\left(d+1\right)/3$ for an observable with $d$ equispaced eigenvalues. However, the quantum feedback methods in these studies concentrate only on maximizing the purity of the measured quantum states, and do not care about how to obtain information about the initial state of the system. That is why they
are referred to as rapid purification protocols rather than rapid measurement protocols.

In contrast to quantum rapid purification, in Ref.~\cite{JCombesPRL:2008} the authors show
that quantum feedback can increase the rate of information gain about the initial preparation. It is found that the
information-extraction rate for a $d$-dimensional system can be
increased by a factor that scales as $d^2$. More exact bounds for rapid measurement protocols
are given in~\cite{JCombesPRA2:2010}, in which it is shown that feedback can increase the rate of information extraction by a factor $R$ by
\begin{equation}\label{Bounds for the purity rate for d systems}
\frac{2}{3}\left(d+1\right)\leq R \leq \frac{d^2}{2}
\end{equation}
for an observable with $d$ equispaced eigenvalues. Further results on quantum rapid purification and
measurement can be found in Ref.~\cite{JCombesPRX:2011}.

\subsubsection{Control-free control}
The term ``control-free control'' refers to measurement-based
feedback control in which the state of the system is controlled
without modifying the Hamiltonian of the system, but merely by
changing the measurement with time. This is an adaptive
measurement process designed to control the system. Control-free
control exploits the fact that, in general, quantum measurements
affect the dynamics of a system in ways that classical
measurements do not. Several approaches for control-free control
have been discussed to-date. In one of these, the quantum
anti-Zeno effect is used to drag the system in the direction of
one of the states onto which the measurement
projects~\cite{Pechen06, Shuang08}. However, this requires rather
strong measurements. In another approach, a small set of
measurements~\cite{Ashhab_You_NJP_2009,Ashhab_You_PS_2009,Ashhab_You_PRAS_2009}
is able to prepare a state by alternating between the
measurements~\cite{Ashhab_PRA_2010,Wiseman_Nature}. This process
is able to stochastically drive the system towards a final pure
state with unit probability. A third method involves making a
continuous measurement, and exploiting the fact that when the
measurement-basis is chosen in the right way, the measurement
generates diffusion of the state in Hilbert space. By changing the
measurement basis with time, a diffusion gradient can be created
in Hilbert space. This diffusion gradient will then stochastically
drive the system towards a single pure state~\cite{Jacobs10}. The
control-free control protocols~\cite{Ashhab_PRA_2010} have been
realized in recent experiments. For example, in
Ref.~\cite{MSBlok}, measurement-only state manipulations are
realized on a nuclear spin qubit in diamond by adaptive partial
measurements (see Fig.~\ref{Fig of the control-free control
experiment}). By combining a quantum non-demolition readout on the
electron ancilla qubit with real-time adaptation of the
measurement strength, the nuclear spin can be steered to a target
state by measurements alone. This interesting work~\cite{MSBlok}
shows that it is possible to implement measurement-based quantum
computing by quantum feedback.

\begin{figure}
\centerline{\includegraphics[width = 16.8
cm]{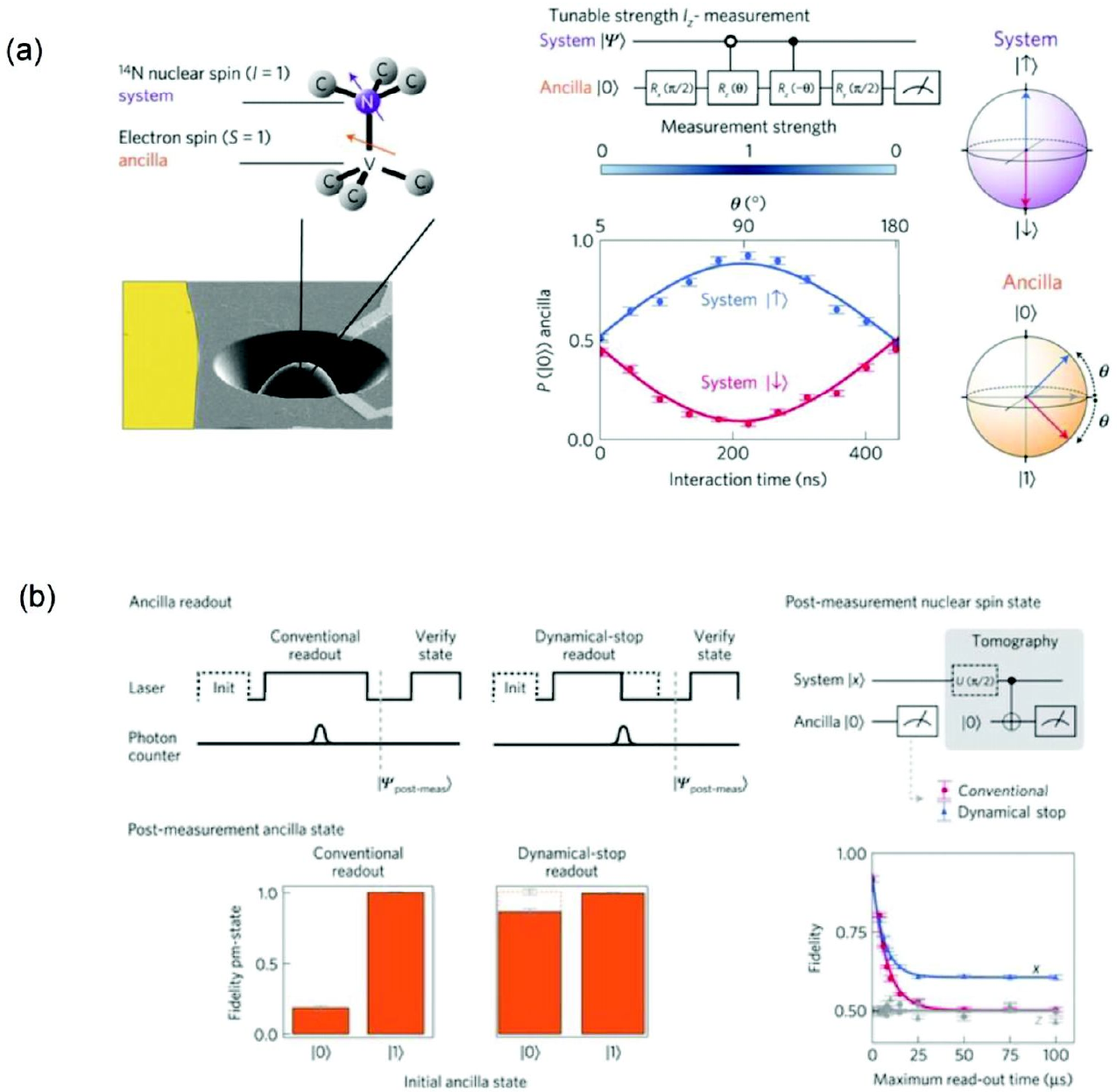}} \caption{(Color online)
Control-free control~\cite{Ashhab_PRA_2010}
experiment~\cite{MSBlok} on nuclear spin qubits in diamond. (a)
Partial measurement of a spin qubit in diamond. (b) Quantum
non-demolition measurement of the ancilla and system qubit
coherence during readout. The figure is from
Ref.~\cite{MSBlok}.}\label{Fig of the control-free control
experiment}
\end{figure}


\section{Coherent Quantum Feedback}\label{s3}

As explained above, measurement-based feedback involves using the
results of measurements on a quantum system to direct its motion.
When we make a measurement on a quantum system, we obtain
classical information. But we necessarily obtain only partial
information about the dynamical variables, and in general we
disturb the state at the same time. It is therefore interesting to
consider a feedback loop in which classical information is not
extracted. This concept, now referred to as \textit{coherent
feedback}, was first introduced by Lloyd in
2000~\cite{SLloydPRA:2000}, and it can be seen as the more general
case of the all-optical feedback proposed earlier, in 1994, in
quantum optical systems by Wiseman and
Milburn~\cite{HMWisemanPRA:1994}. The idea is that instead of
having a classical controller that makes a measurement on the
system, the controller is a quantum system, and the control is
achieved simply by having the two systems interact. To understand
this better, it is worth examining the Watt governor, which has a
very simple feedback mechanism. The purpose of the Watt governor
is to control the speed of an engine. To do this, the engine is
connected to a simple mechanical device so that it spins the
device. The device is designed so that the centrifugal force from
the spinning causes it to expand, so that the faster the engine
spins, the more it expands. This expansion is then used to reduce
the fuel supply to the engine, thus stabilizing the engine at some
chosen speed. The nice thing about this simple feedback system is
that we can think of it as a loop in which the control device
obtains information from the engine, and uses this to control it.
It is also clear that the engine and controller are merely two
coupled mechanical systems. In the Hamiltonian description of the
joint system, there is therefore no loop, but merely an
interaction between the two systems. A quantum controller can
therefore act in the same way, performing feedback control even
though the description of the system may not involve an explicit
loop.

In fact, there is a way to make the loop explicit for a quantum
controller in which there are no measurements. This is done by
coupling the system to a traveling-wave electrical (optical)
field that propagates in one direction from the system to the
controller. We then use a second traveling-wave field that
propagates from the controller to the system, thus closing the
loop. To do this, the two traveling fields must continue
propagating after they interact with the systems, and this
introduces an irreversible element to the dynamics. However, since
control systems are usually intended to introduce some kind of
damping to the system, this irreversibility need not be
detrimental. In what follows, we first discuss feedback control that
employs a unitary (Hamiltonian) interaction between the system and
controller, often referred to as \textit{direct coherent
feedback}, and then turn to feedback in which the interaction is mediated by traveling-wave
fields, often referred to as \textit{field-mediated} feedback.

\begin{figure} \centerline{\includegraphics[width =
16.8cm]{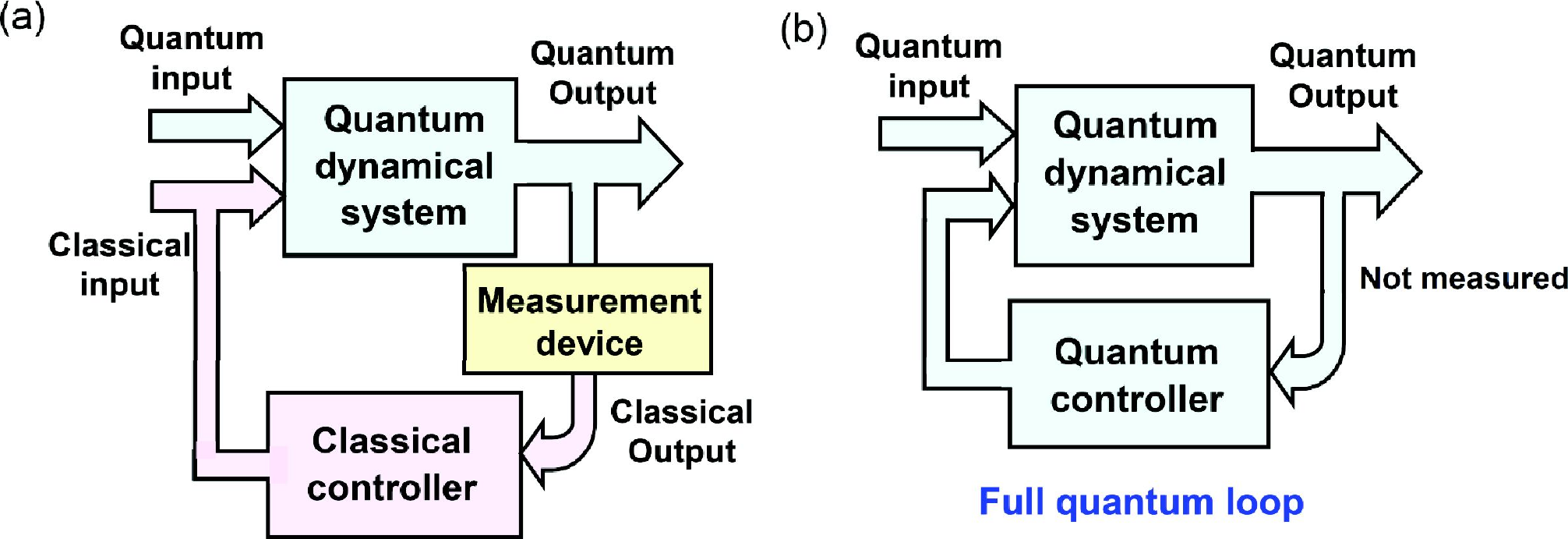}} \caption{(Color online) Comparison of (a)
measurement-based feedback and (b) coherent feedback. In
measurement-based feedback in (a), the system (in blue) is
controlled by a classical feedback loop (in pink); while in
coherent feedback (b) the system is coherently controlled by a
fully quantum feedback loop.}\label{Fig of the measurement-based
feedback and coherent feedback}
\end{figure}

\subsection{Direct coherent feedback}\label{s31}

In general, the action of a controller that is coupled to a system
via a unitary interaction may not break down into clearly defined
processes which involve the extraction of information and use of
this information to apply forces to the system. Nevertheless, it
is interesting to construct an interaction that does perform these
individual processes.  As an example, let us consider the control
of a single qubit by a controller that is also a qubit. The qubit
to be controlled (the primary) is initially in some unknown state
$ | \phi \rangle = \alpha | 0 \rangle + \beta | 1 \rangle $, and
we want to place it in the state $|0\rangle$. If the state of the
primary is completely unknown, then from the point of view of any
observer, and the controller, the state of the qubit is the
density matrix $\rho = (1/2) | 0 \rangle \langle 0 | + (1/2) | 1
\rangle \langle 1 |$. We cannot do this by executing a unitary
operation on the system, because to chose the right unitary we
would need to know the initial state. If we were using
measurement-based feedback, then we could perform a projective
measurement on the primary, at which point we would know what
unitary to apply and execute a unitary operation according to the
measurement output. Note, however, that if we did this we would
have destroyed the initial state, so that no other information can
be extracted from it.

To use a unitary interaction to prepare the primary in the state
$| 1\rangle$, we need the controller to be in a pure state.
Starting the controller in the state $|0\rangle$, we turn on an
interaction that will transform the controller to the state
$|1\rangle$ only if the primary is in state $|1\rangle$. If we
write the states of the primary on the left of the tensor product,
and those of the controller on the right, then the unitary
operator required is
\begin{equation}
   U = |0\rangle \langle 0 | \otimes I + |1\rangle \langle 1 | \otimes ( |0\rangle \langle 1 | + |1\rangle \langle 0 | ) ,
\end{equation}
where "$\otimes$" is the tensor product. This unitary transforms the state of the two systems as
\begin{equation}
   U |\psi \rangle \otimes |0\rangle = U (\alpha | 0 \rangle + \beta | 1 \rangle) \otimes |0\rangle =  \alpha | 0 \rangle \otimes |0\rangle + \beta | 1 \rangle \otimes |1\rangle .
    \label{upsi}
\end{equation}
The two qubits are now correlated, since the controller is in
state $|1\rangle$ if and only if the primary is in state
$|1\rangle$. The controller now ``knows'' the state of the system,
and can act accordingly. To do this, we need an interaction that
performs a different action on the primary for each of the states
of the controller. In particular, we need to transform the state
of the primary from $|0\rangle$ to $|1\rangle$ only if the state
of the controller is $|0\rangle$. The unitary that does this is
\begin{equation}
   U_{\mss{fb}} =  ( |0\rangle \langle 1 | + |1\rangle \langle 0 | ) \otimes |0\rangle \langle 0 |  + I \otimes |1\rangle \langle 1 |   .
\end{equation}
Acting on the joint state in Eq.(\ref{upsi}) with this unitary produces the final state
\begin{equation}
   U_{\mss{fb}} | 0 \rangle \otimes |0\rangle + \beta | 1 \rangle \otimes |1\rangle  =  |1\rangle \otimes ( \alpha |0\rangle +  \beta |1\rangle ) = |1\rangle \otimes  |\psi \rangle .
\end{equation}
This completes the feedback procedure, placing the primary in the
state $|1\rangle$ for every value of $\alpha$ and $\beta$.
Interestingly the initial state of the primary has not been
destroyed. This state, and thus the ``quantum information'' in the
primary has been transferred to the controller.

In the above example, the controller in the measurement-free feedback procedure performs essentially the same action as a measurement that projects the primary onto the basis $\{ |0\rangle, |1\rangle \}$. The controller becomes correlated with the system in the basis $\{ |0\rangle, |1\rangle \}$, and then performs an action depending on whether the system is in state $|0\rangle$ or $|1\rangle$, just as the measurement-based feedback would. But because the projection is not actually performed, and the whole process is ``coherent'', the quantum information in the primary is not destroyed. We note that direct coherent feedback was experimentally demonstrated in a nuclear magnetic resonance (NMR) system shortly after it was proposed~\cite{RJNelsonPRL:2000}.

If we perform the feedback coherently (without measurement), then the above procedure, in which a ``correlation step'' is followed by a ``feedback'' step, is not necessary. The same operation can be performed by a single interaction that  follows a different path in state space. The interaction Hamiltonian
\begin{equation}
   H = \hbar \lambda\; ( |0\rangle \langle 1 | \otimes |1\rangle \langle 0 | + |1\rangle \langle 0 | \otimes |0\rangle \langle 1 |\; )
\end{equation}
generates the unitary
\begin{equation}
   U =  |0\rangle \langle 1 | \otimes |1\rangle \langle 0 | + |1\rangle \langle 0 | \otimes |0\rangle \langle 1 |
\end{equation}
in a time $\tau = \pi/(2\lambda)$. This unitary transforms the
initial joint state $|\psi\rangle\otimes|0\rangle$ to the final
state $ |1\rangle \otimes  |\psi \rangle$, but there is no clear
breakdown into information gathering and feedback.

\subsection{Field-mediated coherent feedback}\label{s32}

To explain how traveling-wave fields can mediate interactions
between quantum systems, it may be simplest to begin with an
example. As mentioned in the introduction, feedback mediated by
fields was first introduced by Wiseman and Milburn in the setting
of quantum optics~\cite{HMWisemanPRA:1994}. In Fig.~\ref{Fig of
the all-optical feedback}, we show the configuration considered
in~\cite{HMWisemanPRA:1994}. In this scheme, there are two optical
cavities, one of these is horizontal in the figure (cavity 1), and
the other is vertical (cavity 2). The two cavities are coupled
directly by a nonlinear crystal, but this is not the
field-mediated part of the coupling. The output beam from the
horizontal cavity is fed through a combination of a polarization
beam-splitter and a Faraday rotator, and then into cavity 2. This
combination breaks time-reversal symmetry, and acts differently
depending on the direction that a beam passes through it. Because
of this, it is able to separate the input beam to cavity 2 from
the beam that comes back out in the reverse direction, so that
this output beam does not go back into cavity 1. The output field
from cavity 1 thus travels to cavity 2 but does not travel in the
reverse direction, and, because of this, it is referred to as a
unidirectional, one-way, or cascade coupling between the cavities.
The combination of the polarization beam splitter and Faraday
rotator is called a unidirectional coupler, or isolator, and the
equivalent exists for electrical (microwave) circuits.

In the feedback scheme in Fig.~\ref{Fig of the all-optical
feedback}, cavity 1 is the primary system, and cavity 2 is the
controller. The controller obtains information about the system
from the one-way field, and applies feedback via the direct
coupling. Wiseman and Milburn~\cite{HMWisemanPRA:1994} considered
three kinds of interaction Hamiltonian $V$. If we define $A$ to be
an arbitrary observable of the mode in cavity 1, $B$ an arbitrary
operator of this mode, and $c_2$ is the annihilation operator for
cavity 2, then the three interactions are (i) $V = c_2^\dagger c_2
A$, (ii) $V = (c_2 + c_2^\dagger) A$ and (iii) $V = c_2^\dagger B
+ c_2 B^\dagger$. The first reproduces feedback via photon
counting and the second feedback via homodyne detection. The third
has no equivalent measurement-based feedback protocol, and can
generate nonclassical states in the primary mode. For example, if
we choose $B = i\lambda(c_1 + c^\dagger_1)$, the resulting
feedback produces squeezed states of the primary mode when $0 <
|\mu | < 1$.

\begin{figure} \centerline{\includegraphics[width =
10 cm]{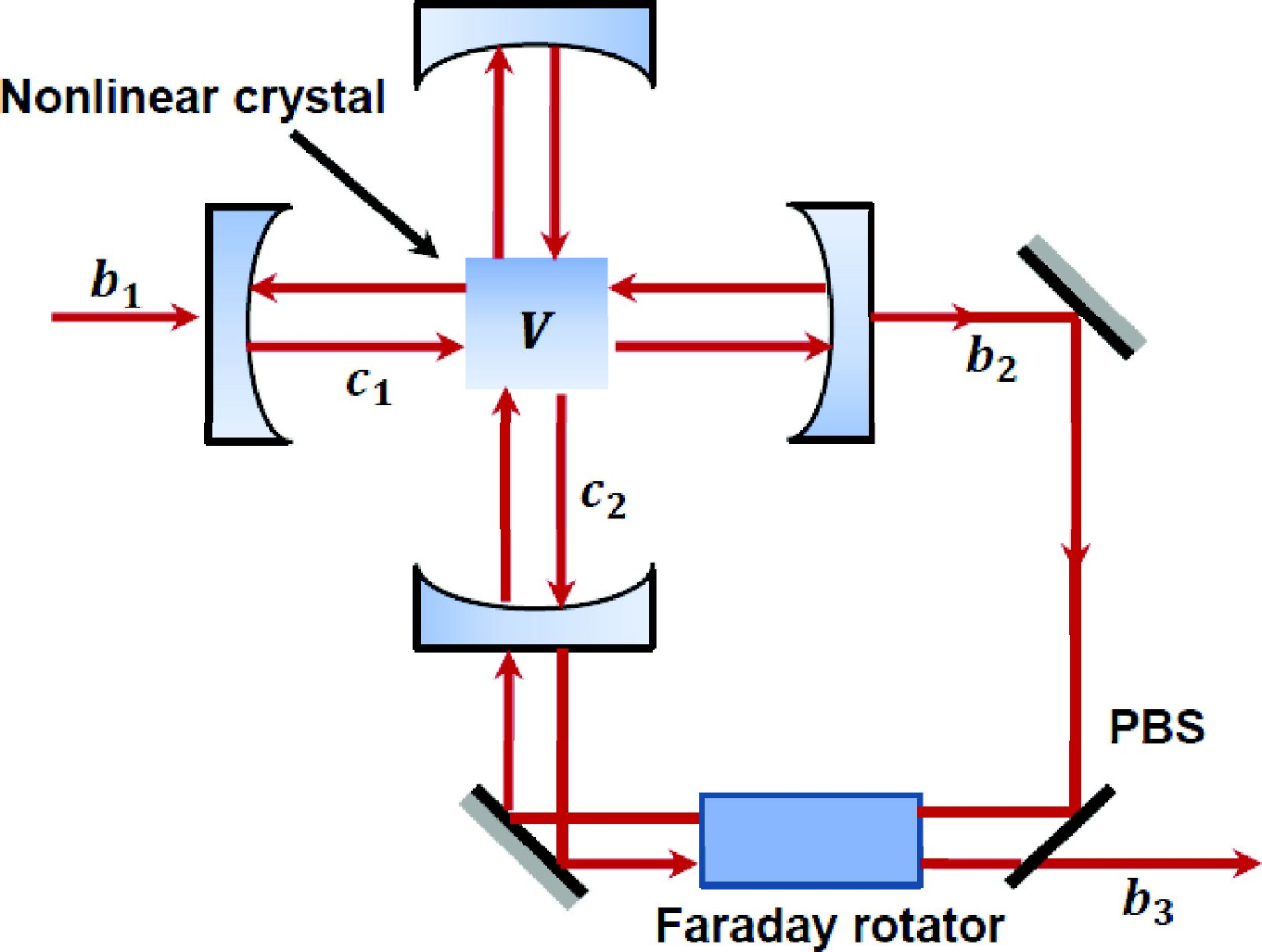}} \caption{(Color online) Diagram
of a theoretical proposal by Wiseman and
Milburn~\cite{HMWisemanPRA:1994} for an all-optical feedback
scheme. An external input field $b_1$ is first fed into the source
cavity $c_1$, and then the output field $b_2$ is directed back to
be fed into another driven cavity $c_2$ via an all-optical
feedback loop. The source (vertical) cavity and the driven
(horizontal) cavity are coupled to each other by a nonlinear
crystal which induces a nonlinear coupling, denoted by the
interaction Hamiltonian $V$. The Faraday rotator and the
Polarization-sensitive Beam Splitter (PBS) are introduced to
generate a unidirectional feedback loop.}\label{Fig of the
all-optical feedback}
\end{figure}

\subsubsection{Networks of quantum systems}\label{s321}

The configuration of the feedback system in Fig.~\ref{Fig of the
all-optical feedback} has a unidirectional connection from the
system to controller, which replaces the measurement in
measurement-based feedback, but does not use a unidirectional
coupling for the feedback part of the loop. We can, however, use a
cascade connection for both, in which case we have a complete
unidirectional loop. What we now need to know is how to describe
these cascade connections mathematically. To do this, we use the
input-output, or ``quantum noise'' formalism of Collet and
Gardiner (CG)~\cite{Collett84,CWGardinerPRA:1985}, also known as
the Hudson-Parthasarathy (HP) model, as the latter independently
derived the same formalism in a more rigorous, measure-theoretic
way~\cite{RIHudsonCMP:1984}. The formalism uses Heisenberg
equations of motion for the operators of the systems, with input
operators that drive these equations in a similar way to that in
which Wiener noise drives classical stochastic equations. The
formalism also contains output operators, and systems are then
easily connected together by setting the input of one system equal
to the output of another.

In the CG/HP formalism, each system is described by a Hamiltonian,
along with the operators through which it is coupled to the
input/output fields. Further, the fields can be coupled to each
other using beam-splitters, which take two inputs and produce two
outputs that are linear combinations of the inputs. By describing
a single ``unit'' as having a Hamiltonian $H$, a vector of input
coupling operators $\mathbf{L}$, and a linear transformation
between inputs and outputs codified by a matrix $\mathbf{S}$,
Gough and James~\cite{GoughTAC:2009} elucidated a set of rules
that covered the ways in which these units, or network elements,
could be combined into networks. We now describe the CG/HP
formalism, and the Gough-James rules~\cite{GoughTAC:2009} for
combining circuit elements.

The dynamics of a system coupled to input fields is given by the
quantum Langevin equations, i.e., Eq.~(2.11), and the output
fields that correspond to each input are given by Eq.~(2.12). As
mentioned above, we describe each unit by a tuple
\begin{equation}\label{SLH}
     G = \left( {\bf S}, {\bf L}, H \right),
\end{equation}
where $H$ is the internal Hamiltonian of the system; $ {\bf S } $
is a $ n \times n $ unitary matrix with operator entries and is
called a scattering matrix; $ {\bf L} = \left( L_1, \cdots, L_n
\right)^{\mss{T}} $ is a vector of operators through which the
system couples to the inputs, with one for each input. We denote
the inputs to the system by ${\bf b}_{\rm in} \left( t \right) =
\left[ b_1 \left( t \right), \cdots, b_n \left( t \right)
\right]^{\mss{T}} $ in which each of the $ b_i \left( t \right),\,
( i=1,\cdots,n ) $ are separate input fields, all initially in the
vacuum state. The notation given in Eq. (3.7) can be used to
describe a wide range of dynamical and static systems. A single
quantum input-output system given by  Eqs.~(\ref{Quantum
stochastic differential equation}) and (\ref{Input-output relation
in the differential form}) can be written as $ G_{\rm LH} = \left(
I, {\bf L}, H \right)$, and a quantum beam splitter is given by $
G_{\rm BS} = \left( {\bf S}, 0, 0 \right) $. Many examples of the
use of this formalism can be found in
Refs.~\cite{MYanagisawaTAC:2003a,MYanagisawaTAC:2003b,GoughTAC:2009,GoughPRA:2008,GoughPRA:2010,GFZhangTAC:2011,GFZhangSJCO:2012}.
We now present the Langevin equations describing input-output
systems in more generality. To begin, we introduce a vector of
quantum Wiener processes $ {\bf B} \left( t\right) $  and a matrix
of quantum Poisson process $ {\bf \Lambda } \left( t \right) $ as
\begin{equation}\label{Quantum Wiener and Poisson processes}
{ \bf B } \left( t \right) = \left(%
\begin{array}{c}
  B_1 \\
  \vdots \\
  B_n \\
\end{array}%
\right), \quad\quad { \bf \Lambda } \left( t \right) = \left(%
\begin{array}{ccc}
  B_{11} & \cdots & B_{1n} \\
  \vdots & \ddots & \vdots \\
  B_{n1} & \cdots & B_{nn} \\
\end{array}%
\right) .
\end{equation}
These noise processes are integrals of the input fields:
\begin{equation}\label{Definition of quantum Wiener and Poisson processes}
B_i \left( t \right) = \int_0^{\mss{t}} b_i \left( \tau \right) d
\tau, \quad\quad B_{ij} \left( t \right) = \int_0^{\mss{t}}
b_i^{\dagger} \left( \tau \right) b_j \left( \tau \right) d \tau.
\end{equation}
The increments of these gauge processes $ {\bf B} \left( t \right),\,{\bf \Lambda} \left( t \right) $ satisfy the quantum stochastic calculus relations given in table~\ref{Table of the quantum Ito rule}.
\begin{table}\large
\centering \caption{Quantum It\^{o} Rule for quantum stochastic
calculus}\label{Table of the quantum Ito rule}
\begin{tabular}{c|cccc}
  \hline
  $ dX/dY $ & $ d {\bf B} $ & $ d {\bf \Lambda} $ & $ d {\bf B}^{\dagger} $ & $ dt $ \\
  \hline
  $ d {\bf B} $ & $ 0 $ & $ d {\bf B} $ & $ dt $ & $ 0 $ \\
  $ d {\bf \Lambda} $ & $ 0 $ & $ d {\bf \Lambda} $ & $ d {\bf B}^{\dagger} $ & $ 0 $ \\
  $ d {\bf B}^{\dagger} $ & $ 0 $ & $ 0 $ & $ 0 $ & $ 0 $ \\
  $ dt $ & $ 0 $ & $ 0 $ & $ 0 $ & $ 0 $ \\
  \hline
\end{tabular}
\end{table}
Let $ V \left( t \right) $ be the unitary evolution operator of the total system composed of the controlled system and the input field, then the evolution equation of the total system can be written as~\cite{GoughTAC:2009}
\begin{equation}\label{Schrodinger equation of SLH}
d V\!\left( t \right)=\left\{ {\rm tr} \left[ \left( {\bf S} - I
\right) d {\bf \Lambda}^{\mss{T}} \right] + d {\bf B}^{\dagger}
{\bf L } - {\bf L}^{\dagger} {\bf S}\,d {\bf B}-\frac{1}{2} {\bf
L}^{\dagger} {\bf L}\,dt - i H dt \right\} V\!\left( t \right)
\end{equation}
with initial condition $ V \left( 0 \right) = I $.

In the Heisenberg picture, the system operator $ X \left( t
\right) = V\!\left( t \right) X\;V^{\dagger}\!\left( t \right) $
satisfies the following quantum stochastic differential equation
\begin{eqnarray}\label{Quantum stochastic differential equation of SLH}
d X \left( t \right) & = & \left\{ \mathcal{L}_{ {\bf L} \left( t
\right) } \left[ X \left( t \right) \right] - i \left[ X \left( t
\right), H \left( t \right) \right] \right\} dt + d {\bf
B}^{\dagger} \left( t \right) {\bf S}^{\dagger} \left( t \right)
\left[ X \left( t \right), {\bf L} \left( t \right) \right] +
\left[ {\bf L}^{\dagger} \left( t \right), X \left( t \right)
\right] {\bf S} \left( t \right) d {\bf B} \left( t
\right) \nonumber \\
& & + {\rm tr} \left\{ \left[ {\bf S}^{\dagger} \left( t \right) X
\left( t \right) {\bf S} \left( t \right) - X \left( t \right)
\right] d {\bf \Lambda}^{\mss{T}} \left( t \right) \right\}, \nonumber \\
\end{eqnarray}
where the Liouville superoperator $ \mathcal{L}_{ {\bf L }
}\left(\cdot \right)$ is defined by
\begin{equation}\label{Dissipation Liouville superoperator}
\mathcal{L}_{ {\bf L } } \left( X \right) = \frac{1}{2} {\bf
L}^{\dagger} \left[ X, {\bf L } \right] + \frac{1}{2} \left[ {\bf
L}^{\dagger}, X \right] {\bf L}=\sum_{j=1}^n \left\{ \frac{1}{2} L_j^{\dagger} \left[ X, L_j
\right] + \frac{1}{2} \left[ L_j^{\dagger}, X \right] L_j
\right\},
\end{equation}
which is of the standard Lindblad form. Similar to
Eq.~(\ref{Input-output relation in the differential form}), the output fields corresponding to the inputs $ {\bf B} \left( t \right) $ and Poisson process $ \Lambda \left( t \right) $ are given by
\begin{eqnarray*}
{\bf B}_{\rm out} \left( t \right) = V^{\dagger}\!\left( t \right)
{\bf B} \left( t \right) V\!\left( t \right), \quad\quad {\bf
\Lambda}_{\rm out} \left( t \right) = V^{\dagger}\!\left( t
\right) {\bf \Lambda} \left( t \right) V\!\left( t \right),
\end{eqnarray*}
from which we obtain the following input-output relation
\begin{eqnarray}\label{Input-output relation of quantum Wiener and Poisson processes}
d {\bf B}_{\rm out} \left( t \right) & = & {\bf S} \left( t
\right) d {\bf B} \left( t \right) + {\bf L} \left( t \right) dt,
\nonumber \\
d {\bf \Lambda}_{\rm out} \left( t \right) & = & {\bf S}^* \left(
t \right) d {\bf \Lambda} \left( t \right) {\bf S}^{\mss{T}} \left( t
\right) + {\bf S}^* \left( t \right) d {\bf B}^* \left( t \right)
{\bf L}^{\mss{T}} \left( t \right) + {\bf L}^* \left( t \right) d {\bf B}^{\mss{T}} \left( t \right) {\bf
S}^{\mss{T}} \left( t \right) + {\bf L}^* \left( t \right) {\bf L}^{\mss{T}}
\left( t \right) dt. \nonumber\\
\end{eqnarray}
It can be verified that the increments $ d {\bf B}_{\rm out} $, $
d {\bf \Lambda}_{\rm out} $ of the output processes also satisfy
the rules of quantum stochastic calculus shown in table~\ref{Table of
the quantum Ito rule}.

For linear quantum systems, the quantum Langevin equations can be
solved directly. In order to perform calculations for nonlinear
quantum systems, one must transform the Heisenberg equations of
the input-output formalism to master equations. The corresponding
master equations are
\begin{equation}\label{Master equation of the input-output system}
\dot{\rho}=-i \left[ H, \rho \right] + \sum_j \left( L_j \rho
L_j^{\dagger} - \frac{1}{2} L_j^{\dagger} L_j \rho - \frac{1}{2}
\rho L_j^{\dagger} L_j \right).
\end{equation}
Although the scattering matrix $ {\bf S } $ does not appear in the
master equation~(\ref{Master equation of the input-output
system}), it affects the input-output relation of the system as
shown in Eq.~(\ref{Input-output relation of quantum Wiener and
Poisson processes}) and thus will affect the dynamics of more
complex quantum input-output systems, such as the quantum cascade
systems which will be specified below.


To connect the outputs of one unit to the inputs of another, so as to form an arbitrary network, we need only two rules. The first is merely a rule that says how to represent a universe that contains more than one separate unit, none of which are connected. If we have the two units $ {\bf G}_1 = \left( {\bf S}_1, {\bf L}_1, H_1 \right) $ and $ {\bf G}_2 = \left( {\bf S}_2, {\bf L}_2, H_2 \right) $, the unit that describes both these units with no connections between them is
\begin{equation}\label{SLH of concatennation product}
{\bf G}_1 \boxplus {\bf G}_2 = \left( \left(%
\begin{array}{cc}
  {\bf S}_1 & 0 \\
  0 & {\bf S}_2 \\
\end{array}%
\right),\left(%
\begin{array}{c}
  {\bf L}_1 \\
  {\bf L}_2 \\
\end{array}%
\right),H_1+H_2 \right).
\end{equation}
Gough and James~\cite{GoughTAC:2009} refer to this rule as the
\textit{concatenation product}.

The second rule for combing circuit elements tells us how to
determine the unit that describes a network in which the outputs
of a unit $ {\bf G}_1$ are connected to the inputs of a unit $
{\bf G}_2$. This rule is
\begin{equation}\label{SLH of series product}
{\bf G}_2 \triangleleft {\bf G}_1 = \left( {\bf S}_2
{\bf S}_1,
{\bf L}_2 + {\bf S}_2 {\bf L}_1, H_1 + H_2 + \frac{1}{2i} \left( {\bf
L}_2^{\dagger} {\bf S}_2 {\bf L}_1 - {\bf L}_1^{\dagger} {\bf
S}_2^{\dagger} {\bf L}_2 \right) \right),
\end{equation}
and is called the \textit{series product}. The concatenation and series products can also be used to decompose a given system into subsystems, and are thus fundamental to feedforward and feedback control.

\subsubsection{Quantum transfer-function model}\label{s322}

The Collet-Gardiner/Hudson-Parthasarathy cascade connections can
be used to model essentially any network. However, for linear
systems, time-delays and quantum amplifiers can be modeled more
easily in frequency space. If we specialize the network formalism
of Gough and James~\cite{GoughTAC:2009} so that all the systems
are linear, and transform the equations of motion to frequency
space, then we have the method of quantum transfer
functions~\cite{MYanagisawaTAC:2003a,MYanagisawaTAC:2003b,GoughPRA:2008,GoughPRA:2010}.

A general linear quantum network described by the tuple
$(\mathbf{S},\mathbf{L}, H)$ satisfies the following
conditions~\cite{GoughPRA:2008}: (i) the entries of the scattering
matrix $\mathbf{S}$ are scalars; (ii) the dissipation operators
$L_j$ are linear combinations of the $a_k$ and $a_k^\dagger$; and
(iii) the system Hamiltonian $H$ is a quadratic function of the
$a_k$ and $a_k^\dagger$. To elucidate the transfer function method
further, we consider a useful special case, in which each system
is a harmonic oscillator, and the field coupling operators are
linear combinations of only the annihilation operators. In this
case, the Langevin equations for the annihilation operators are
not coupled to those for the creation operators. The annihilation
operators for the $n$ oscillators, $ \left\{ a_j: j=1,\cdots,n
\right\}$, satisfy the commutation relations
\begin{eqnarray*}
\left[ a_j, a_k^{\dagger} \right] = \delta_{jk}, \quad\quad \left[
a_j, a_k \right] = \left[ a_j^{\dagger}, a_k^{\dagger} \right] =0.
\end{eqnarray*}
For our special case, the total Hamiltonian is $H = \sum_{ij} \omega_{ij} a^\dagger_i a_j$ and the coupling operators $L_j = \sum_{jk} c_{jk} a_k$, and so we can simplify the SLH formalism, writing the tuple as
\begin{equation}\label{SCOmega}
G = \left( {\bf S}, C, \Omega \right),
\end{equation}
where
\begin{eqnarray*}
C=\left(%
\begin{array}{ccc}
  c_{11} & \cdots & c_{1n} \\
  \vdots & \ddots & \vdots \\
  c_{n1} & \cdots & c_{nn} \\
\end{array}%
\right),\quad \Omega=\left(%
\begin{array}{ccc}
  \omega_{11} & \cdots & \omega_{1n} \\
  \vdots & \ddots & \vdots \\
  \omega_{n1} & \cdots & \omega_{nn} \\
\end{array}%
\right).
\end{eqnarray*}

If we now introduce an operator vector, which we will call the state vector of the system, ${\bf a}= \left( a_1, \cdots, a_n \right)^{\mss{T}}$, then from Eqs.~(\ref{Quantum stochastic differential equation of SLH}) and
(\ref{Input-output relation of quantum Wiener and Poisson
processes}), we can obtain the following Langevin
equation and input-output relation:
\begin{eqnarray}
\dot{{\bf a}}\!\left( t \right) & = & A\,{\bf a}\!\left( t \right)
- C^{\dagger} {\bf S}\;{\bf b}_{\rm in}\! \left( t \right),
\label{Heisenberg-Langevin equation} \\
{\bf b}_{\rm out} & = & {\bf S}\;{\bf b}_{\rm in}\! \left( t
\right) + C\,{\bf a}\!\left( t \right), \label{Input-output
relation}
\end{eqnarray}
where $A=-C^{\dagger}C/2-i\Omega$.

We can now transform these equations to frequency space by taking either the Laplace transform or the Fourier
transform. Using the Fourier transform, defined as
\begin{equation}\label{Laplace transform of the linear equation}
R \left( \nu \right) = \int_0^{\infty}\!\!\exp{ \left( - i\nu t
\right) }\;R \left( t \right) d t ,
\end{equation}
the Langevin equations can be rearranged to obtain
\begin{eqnarray}
& {\bf a}\! \left( \nu \right) = - \left( i\nu I_n - A
\right)^{-1} C^{\dagger} {\bf S}\;{\bf b}_{\rm in}\!\left( \nu
\right), & \label{Heisenberg-Langevin equation in the
frequency domain} \\
& {\bf b}_{\rm out} \left( \nu \right) = {\bf S}\;{\bf b}_{\rm in}
\left( \nu \right) + C {\bf a} \left( \nu \right). &
\label{Input-output relation in the frequency domain}
\end{eqnarray}
From Eqs.~(\ref{Heisenberg-Langevin equation in the frequency
domain}) and (\ref{Input-output relation in the frequency
domain}), we can obtain the input-output relation of the whole
system or network
\begin{equation}\label{Input-output relation by quantum transfer function model}
{\bf b}_{\rm out} \left( \nu \right) = \Xi \left( i\nu
\right)\;{\bf b}_{\rm in} \left( \nu \right),
\end{equation}
where $ \Xi \left( \cdot \right) $ is the transfer function of the
linear quantum system which can be calculated by
\begin{equation}\label{Quantum transfer function}
\Xi \left( i\nu \right) = {\bf S} - C \left( i\nu I_n - A
\right)^{ -1 } C^{\dagger}\;{\bf S}.
\end{equation}
The input-output relation~(\ref{Input-output relation by quantum
transfer function model}) show the linear map between the input
and output of the linear quantum system given by
Eqs.~(\ref{Heisenberg-Langevin equation}) and (\ref{Input-output
relation}).

The quantum transfer function approach is useful for a number of
reasons. While the time-domain network formalism can describe
essentially any network, it cannot be used to incorporate static
models of non-conservative elements, such as quantum amplifiers,
and such components must be treated as dynamical systems. In
frequency space, a static model of a quantum amplifier is simply a
Bogoliubov transformation~\cite{GoughPRA:2010}. Time delays are
also much simpler to include in frequency space, and of course
frequency space has the advantage that the transfer function of
two cascaded systems is merely the product of the transfer
functions of each.

\subsection{Applications}\label{s33}

%

\subsubsection{Noise-reduction in linear systems}\label{s331}

Like measurement-based quantum feedback, the main merit of quantum
coherent feedback is that it can be used to suppress sources of
entropy, such as external noise, uncertainty in the parameters
that define the system, and even to some extent errors in the
modeling of the system. In general, the problem of noise-reduction
can be captured by asking how to minimize the effect of a set of
inputs on a set of outputs. For linear networks, this problem has
been studied by a number of authors. James, Nurdin, and Petersen
developed linear-quadratic-Gaussian
control~\cite{HINurdinAutomatica:2009} and H-infinity
($H_{\infty}$) control~\cite{MRJamesTAC:2008}. Control of linear
systems with squeezers and phase-shifters has been explored by
Zhang \textit{et al.}~\cite{GFZhangSJCO:2012}, and Zhang and
James~\cite{GFZhangTAC:2011} have investigated the relationship
between direct and field-mediated coupling in networks. The
extension of the Collet and Gardiner input-output formalism to
non-Markovian field couplings has been developed by
Diosi~\cite{LDiosiPRA:2012} and Zhang \textit{et
al.}~\cite{JZhangPRA:2013}, and noise suppression via
non-Markovian coherent feedback has been analyzed by Xue
\textit{et al.}~\cite{SBXuePRA:2012}. Coherent noise-reduction for
a single cavity was demonstrated experimentally by Mabuchi
in~\cite{HMabuchiPRA:2008}. It is also shown quite recently the
coherent feedback can be applied to controlling the
quantum-transport properties of a mesoscopic device and optimize
the conductance of a chaotic quantum dot~\cite{CEmary2014}.


\subsubsection{Optical squeezing}\label{s333}

Squeezing as an application of coherent feedback was considered
early on by Wiseman and Milburn~\cite{HMWisemanPRA:1994}. More
recently, a coherent protocol for squeezing was devised by Gough
and Wildfeuer~\cite{JEGoughPRA:2009} which is simpler and allows
more control of the amount squeezing. This
protocol~\cite{JEGoughPRA:2009} has now been experimentally
realized by Furusawa's group in a linear optical
system~\cite{SIidaTAC:2012}. We depict the protocol
in Fig.~\ref{Fig of the optical squeezing by CF}, in which we see
that the coherent feedback loop is composed of a squeezing
component, such as a degenerate parametric amplifier in the
strong-coupling regime, and a beam splitter whose reflectivity can
be adjusted. By tuning this reflectivity, the effective damping
rate of the cavity is modified, and the squeezing effects are
enhanced or suppressed.

\begin{figure} \centerline{\includegraphics[width =
13 cm]{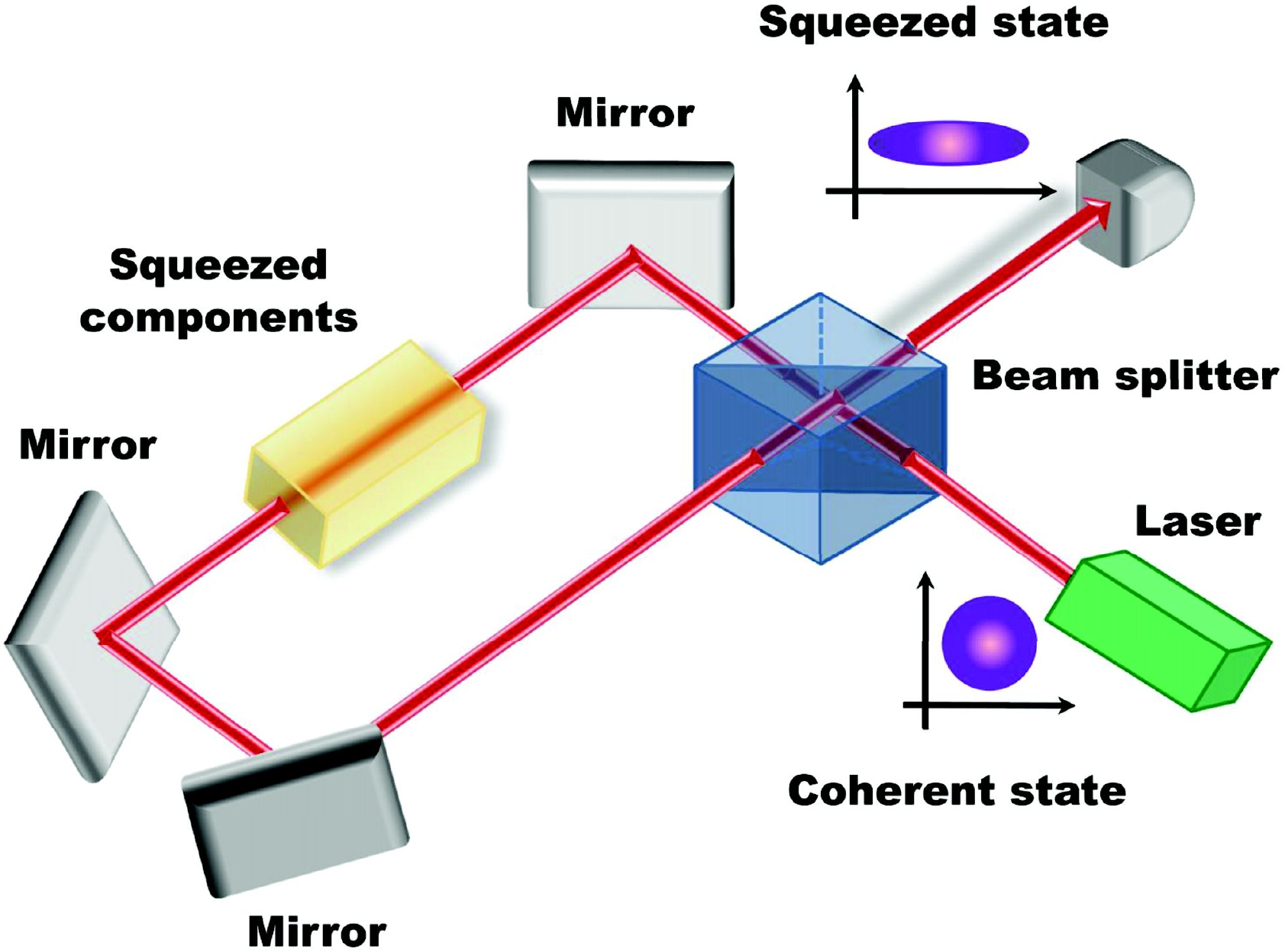}} \caption{(Color online)
Diagram of a theoretical proposal by Gough~\cite{JEGoughPRA:2009}
for tunable optical squeezing by coherent feedback. The optical
squeezing output produced by a squeezing device can be enhanced
and suppressed by tuning the reflectivity of a control beam
splitter within the coherent feedback loop.}\label{Fig of the
optical squeezing by CF}
\end{figure}


\subsubsection{Quantum error correction}\label{s331}

Coherent quantum feedback has been used to implement continuous
quantum error-correction (see the beginning of Section~\ref{s2}
for a brief introduction to quantum error
correction)~\cite{JKerckhoffPRL:2010,JKerckhoffNJP:2011}. In
Ref.~\cite{JKerckhoffPRL:2010}, the authors propose a three-qubit
error correction method to correct single-qubit bit-flip or phase
flip errors using coherent feedback. As shown in Fig.~\ref{Fig of
autonomous quantum error correction network}, the atoms in
cavities $Q_1$, $Q_2$, and $Q_3$ are the three physical qubits
that code for, and thus allow, the single logical qubit to be
corrected. The blue lines are the optical beams for error
detection and the red lines are the laser beams that apply the bit
flips or phase flips. The central components of this autonomous
error correction network are the two relays $R_1$ and $R_2$, which
work as controlled quantum switches~\cite{HMabuchiPRA:2009,
JKerckhoffPRA:2009}. When the ``Reset" (``Set") input port of the
relay receives a coherent input signal, the input from the ``Power
in" port will be transferred to the ``Out" (``$\overline{\rm
Out}$") port. The operating principle of the quantum error
correction network can be summarized as follows. If the qubits
$Q_1$ and $Q_2$ have even (odd) parity, the ``Set" (``Reset")
input port of the relay $R_1$ receives a signal, while the
``Reset" (``Set") input port remains in the vacuum. The same
relationship exists between the relay $R_2$ and the qubits $Q_2$
and $Q_3$. This detected signal controls the power transfer of the
relay from the ``Power in" port to the ``Out" or ``$\overline{\rm
Out}$" port, which is then directed back to the qubits. When a
qubit is simultaneously stimulated by two feedback signals from
the output ports of the relays, the Raman resonance process will
lead to a coherent Rabi oscillation of the qubit and thus correct
the errors. Otherwise, the control signal will only introduce an
ac Stark shift for the qubit. Such a coherent feedback network can
thus automatically correct the bit-flip or phase-flip errors. In
Ref.~\cite{JKerckhoffNJP:2011}, the authors extend this method to
perform corrections for Shor's nine-bit error-correcting code,
which concatenates two three-bit codes so as to correct an
arbitrary error.

\begin{figure} \centerline{\includegraphics[width =
16.8 cm]{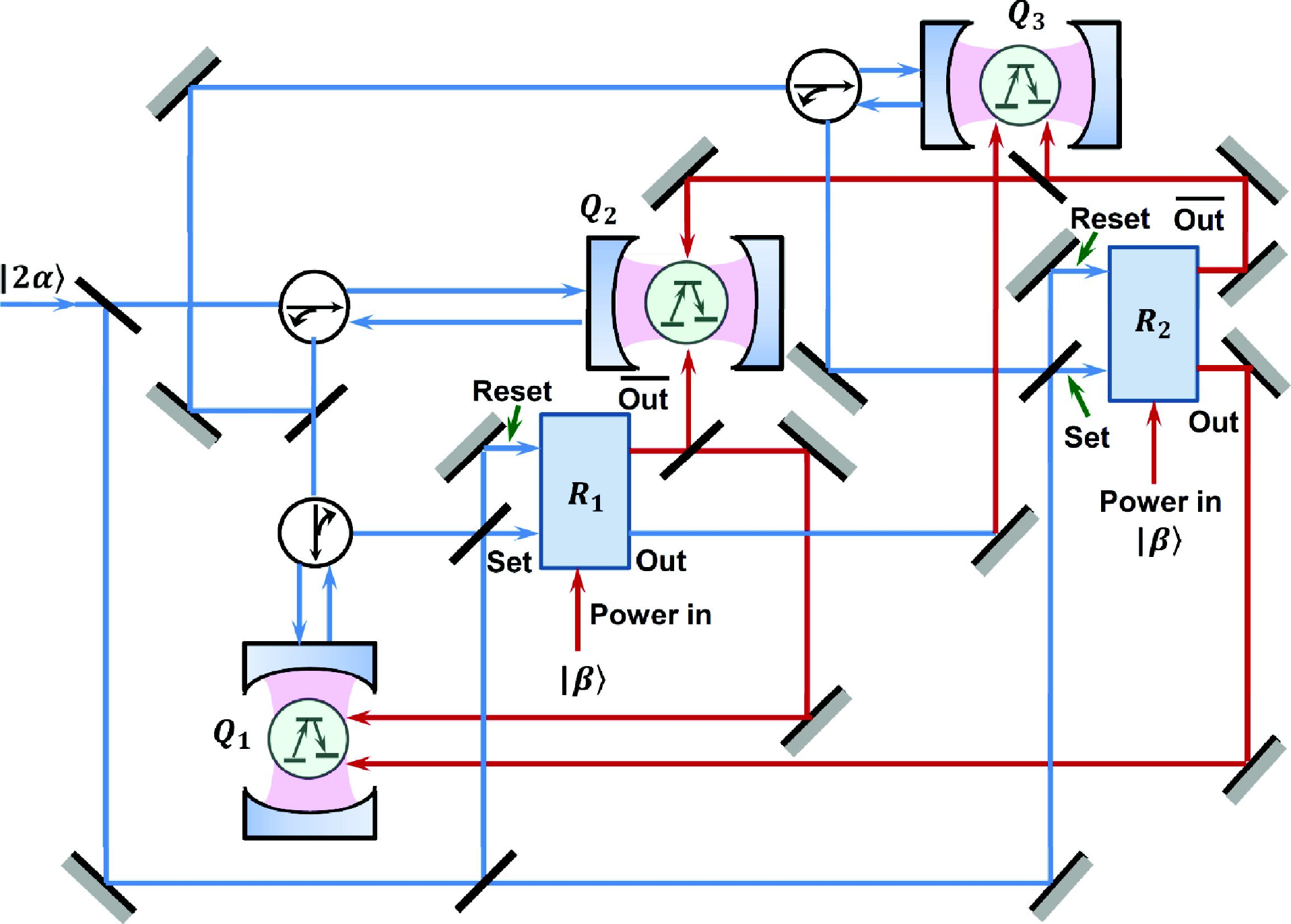}} \caption{(Color online)
Diagram of a theoretical proposal by Kerckhoff \textit{et
al.}~\cite{JKerckhoffPRL:2010} for an autonomous three-qubit
quantum error correction network. This scheme has been proposed to
correct single-qubit bit-flip or phase-flip errors, which includes
register qubits $Q_{1,2,3}$, beam splitters, circulators, and the
relays $R_1$ and $R_2$. The blue lines and red lines represent the
probe beam and the feedback correction beam, respectively. The
relays are quantum switches introduced in
Ref.~\cite{HMabuchiPRA:2009}. $R_1$ and $R_2$ have two outputs
denoted by ${\rm Out}$ and ${\rm \overline{Out}}$.}\label{Fig of
autonomous quantum error correction network}
\end{figure}

\subsubsection{Controlling mechanical resonators}\label{s332}

Mechanical resonators can be built with sufficiently high
frequencies that they will behave quantum mechanically at a
temperature of a few milliKelvin, only an order-of-magnitude from
temperatures that can be reached with dilution refrigerators.
These mechanical resonators can be coupled to optical modes
(``optomechanics'') or cryogenic superconducting circuits
(nano-electro-mechanics) for potential use in more complex
devices. To prepare highly non-classical states, or for the
purpose of using mechanical resonators for quantum technologies,
it is useful to prepare them in the ground state. This is usually
referred to as cooling.

As far as experiments are concerned, the present state-of-the art
in cooling mechanical resonators is a version of
``resolved-sideband'' cooling~\cite{Wineland75, Diedrich89}. This
method is, in fact, an example of coherent feedback. The
mechanical resonator is linearly coupled to another ``auxiliary''
harmonic oscillator, a mode of an optical or superconducting
cavity. These auxiliary resonators have such high frequencies that
they sit in their ground states at cryogenic temperatures. The
coupling is modulated at the difference frequency of the two
resonators, which allows them to exchange excitations as if they
were on resonance. The auxiliary oscillator is arranged to have a
higher damping rate than the mechanical resonator, and because the
former is in its ground state at the ambient temperature, it sucks
the energy out of the mechanical resonator~\cite{Marquardt07,
Wilson-Rae07, Tian09, Jacobs14}. The coupling can be direct or
field-mediated. Sideband cooling is limited by the linear
interaction: to transfer energy without causing heating the
strength of the interaction must be much smaller than the
frequency of the mechanical resonator, so that the rotating-wave
approximation is valid. It has been shown that if the interaction
strength is modulated in a more complex way, then this limitation
can be overcome, and energy (or quantum information) can be
transferred between the resonators within a single period of the
mechanical resonator~\cite{Wang11}.

It has been shown that if one is restricted to a linear
interaction with a resonator, then coherent feedback performs much
better than measurement-based feedback in the quantum regime,
including the regime of ground-state
cooling~\cite{RHamerlyPRL:2012,JKerckhoffPRX:2013}. The
superiority of coherent feedback in this case can be traced to the
projection noise from the position measurement~\cite{Jacobs14b}.
This noise is the change in the quantum state induced by the
measurement, which is the term proportional to the Wiener noise in
the stochastic master equation.


A recent experiment by Kerckhoff \textit{et al.} has shown that coherent feedback can be used as a practical method to tune the damping rate of a superconducting oscillator~\cite{JKerckhoffPRX:2013}.

\subsubsection{Quantum nonlinear optics}\label{s334}

As is well known, it is not easy to deterministically generate
non-classical optical states due to the absence of strong optical
nonlinearities. To solve this problem, an interesting study by
Yanagisawa in Ref.~\cite{MYanagisawaPRL:2009} showed that
non-classical optical states can be produced via linear optical
components by introducing a multi-feedback structure. It is first
shown how a quantum-nondemolition output of $x^2$ can be
constructed by reading out the $x$ quadrature of the optical field
and feeding it back to adjust the system-environment coupling
strength. In this way one can produce eigenstates of $x^2$, which
are superpositions of two eigenstates of position with eigenvalues
of equal magnitude. A further feedback loop is then introduced in
which the Hamiltonian of the controlled system is adjusted by the
quantum nondemolition output of $x$ to increase the probability to
obtain a desired superposition state. However, in this method the
first form of feedback is hard to realize experimentally. To solve
this problem, Zhang \textit{et
al.}~\cite{JZhangTAC:2012,ZPLiuPRA:2013} proposed a method which
they called ``quantum feedback nonlinearization". This enables
strong nonlinear effects in a linear plant by the use of a
weak-nonlinear component and a quantum amplifier. With this method
it is possible to generate strong Kerr effects that are four or
five orders of magnitude stronger than the initial nonlinearity,
and can demonstrate nonclassical optical phenomena such as
sub-Poisson photon counting statistics and photon anti-bunching
effects.

Another potential application of coherent feedback in non-linear
optical systems is classical information processing. An optical
Kerr-nonlinear resonator in a nanophotonic device can exhibit
dispersive bistability effects, and these can be used for
all-optical switching in the attojoule
regime~\cite{MFYanikAPL:2003,XYangAPL:2007,KNozakiNaurePhotonics:2010,RKumarIPTL:2010}.
However, in this regime, the optical logic states are separated by
only a few photons and thus suffer from random switching due to
quantum noise~\cite{MArmenPRA:2006,MArmenPRL:2009}. In
Ref.~\cite{HMabuchiAPL_bistability:2011}, Mabuchi proposed a
coherent feedback method to avoid the quantum noise. In this
scheme, a Kerr-nonlinear ring resonator works as an optical switch
in which the two states have differing numbers of photons. This
switch is connected to a second Kerr-nonlinear ring resonator,
which acts as a controller that suppresses the spontaneous
switching, in a feedback configuration. Since the effective cavity
detuning of a Kerr resonator varies with the driving strength, the
controller induces an amplitude-dependent phase shift $\phi$ on
the optical beam. This leads to a $\phi$-dependent effective
detuning, and a $\phi$-dependent effective cavity decay rate for
the switch (the controlled resonator). One chooses the control
parameter $\phi$ in an optimal way so that the overall feedback
phase is close to $\pi$ when the switch is in the ``low'' state
and close to zero when it is in the ``high'' state. In this way,
the spontaneous switching between the ``low'' and ``high'' states
can be efficiently suppressed.

The proposal~\cite{HMabuchiAPL_bistability:2011} is extended in
Ref.~\cite{HMabuchiAPL:2011} to implement photonic sequential
logic by using optical Kerr resonators, in which interference
effects enable the binary logic gates. Binary logic elements, such
as single-output AND gate and NOT gates with an output fan-out of
two, can be generated in this way~\cite{ZZhouAPL:2012}. This
theoretical proposal has been experimentally realized in
superconducting circuits by Kerckhoff \textit{et
al.}~\cite{JKerckhoffPRL:2012}, in which the emergent bistable and
astable states were used to realize a latch.

\subsubsection{Controlling entanglement}\label{s335}

A proposal in Ref.~\cite{ZHYanPRA:2011} shows that coherent feedback can be used to generate and control continuous-variable multipartite optical entangled states. As shown in Fig.~\ref{Fig of the entanglment control by CF}, in this scheme the multipartite entangled states generated by a non-degenerate optical parametric amplifier
(NOPA), denoted by the output field $a_i^{\rm out}$, are fed into
a coherent feedback control loop. The optical beam $d_i^{\rm in}$ is split into two branches by a controlled beam
splitter (CBS). One branch, denoted by $a_i^{\rm in}=d_i^{\rm out}$, is fed back to the NOPA and the other branch,
$c_i^{\rm out}$, provides the multipartite entangled output. The multipartite entanglement can be controlled by adjusting the transmissivity $t$ of the CBS. The NOPA in Fig.~\ref{Fig of the entanglment control by CF} is
composed of a nonlinear crystal and a bow-tie type ring cavity. The fields $b_i^{\rm in}$ and $e_i^{\rm in}$ are vacuum fields introduced to model the loss in the NOPA and the coherent feedback loop. The piezoelectric transducers (PZTs) are used to lock the cavity length for resonance. The authors evaluated the multipartite entanglement generated by this scheme using the nonseparable criterion developed in Ref.~\cite{RvanLoockPRA:2003}. It was found that the coherent feedback loop can efficiently enhance the multipartite entanglement generated by the NOPA in particular parameter regimes  which can be reached by tuning the transmissivity of the CBS.

\begin{figure} \centerline{\includegraphics[width =
13 cm]{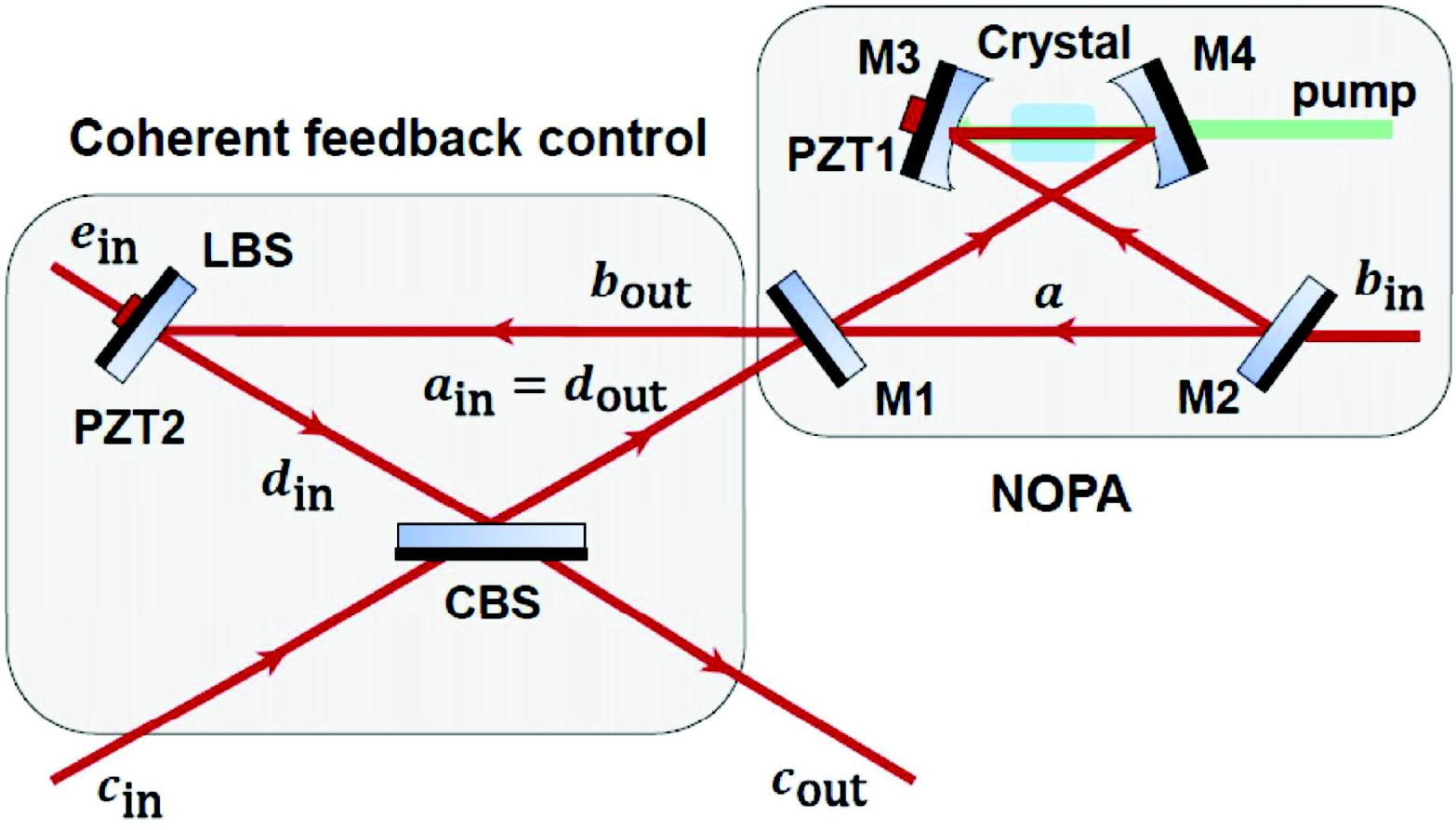}} \caption{(Color online) Schematic
diagram of a theoretical proposal by Yan \textit{et
al.}~\cite{ZHYanPRA:2011} for continuous-variable multipartite
entanglement control. The multipartite entangled states are
generated by Nondegenerate Optical Parametric Amplifier (NOPA) and
enhanced by a coherent feedback loop. The entanglement degree
generated can be tuned by adjusting the transmissivity of the
Control Beam Splitter (CBS).}\label{Fig of the entanglment control
by CF}
\end{figure}

\section{Other Kinds of Quantum Feedback}\label{s4}

\subsection{Adaptive feedback}\label{s41}

``Adaptive feedback'' is a term that was coined by Judson and
Rabitz in 1992 in a now famous paper~\cite{Judson}. This term does
not refer to feedback in the sense use by the classical control
community, in which a measured signal is fed back as it is
received to control a dynamical system. Instead adaptive feedback,
also known as a ``learning control loop'', refers to an iterative
method for searching for open-loop control protocols. The idea is
to start with some arbitrary control protocol, try it out on a
real system to see how well it does, modify the protocol in some
way based on its performance, and repeat this process many times
to obtain increasingly better protocols. One way to do this is to
use a ``genetic algorithm'', in which one tries not one, but $N$
randomly chosen protocols. One then selects from these the $M$
protocols that perform the best, creates a new set of $N$
protocols by making random changes to these $M$ protocols, and
repeats the process. This is the same procedure that is used for
numerical searches to find optimal protocols. The essential point
is that when the system to be controlled is too complex to
simulate on a computer, replacing the computer simulation with the
real system can be a fast and effective way to obtain good control
protocols. We also note that adaptive control has also proved to
be an effective way to obtain protocols that are robust against
imperfections in the control pulse~\cite{CBrif:2010}. This means
that the pulse, or protocol, can sustain a certain level of noise
without significantly affecting its performance.

\begin{figure}
  \centering
  \includegraphics[width=10cm]{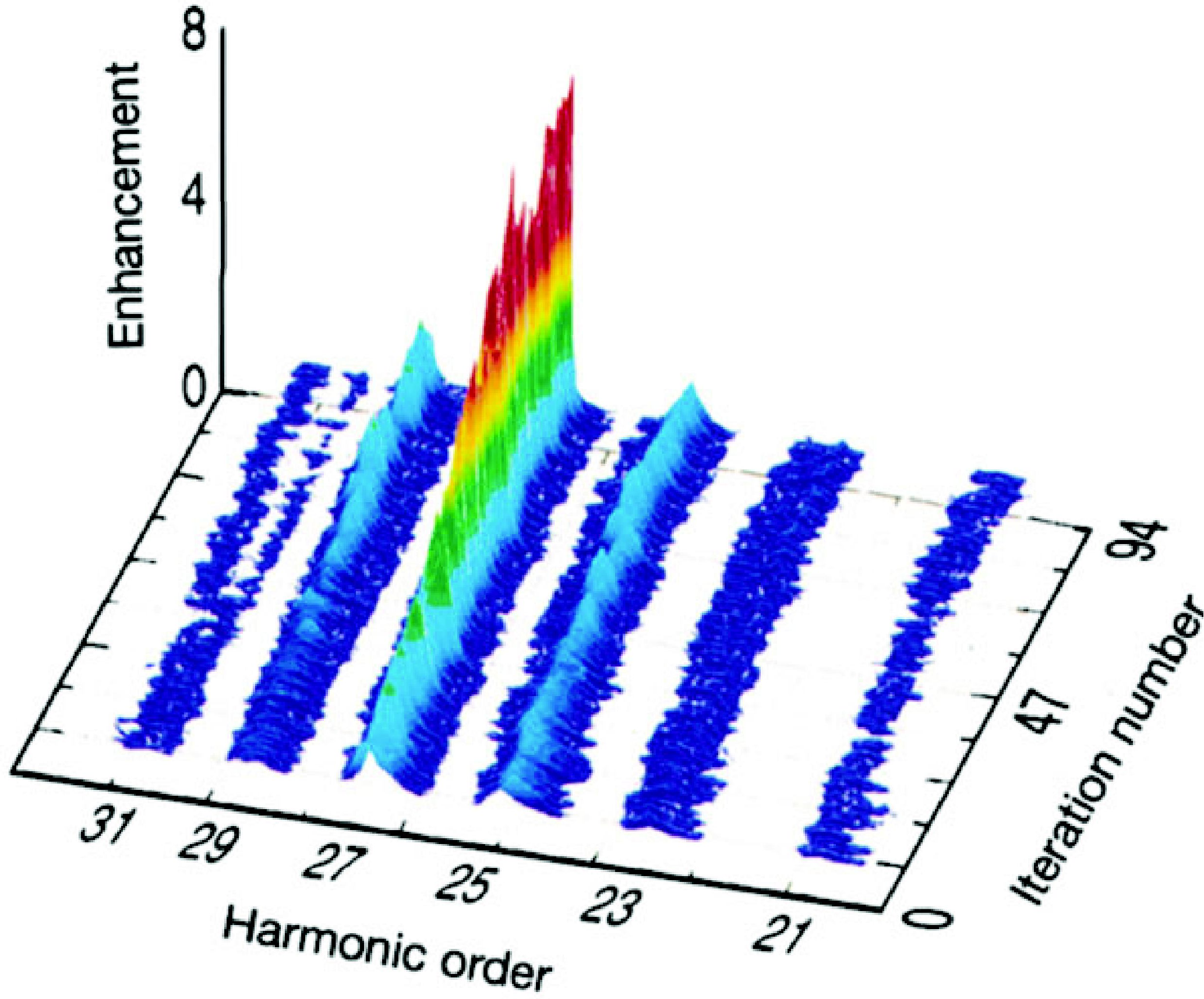}\\
  \caption{(Color online) The improved yield from a High-Harmonic Generation (HHG) experiment by Bartels \textit{et al.}~\cite{Bartels2000}. The Adaptive Feedback Control (AFC) guided search found a control that can selectively enhance the 27th-order harmonic mode. This figure is from Ref.~\cite{Bartels2000}.
}\label{Fig:HHG}
\end{figure}

The adaptive feedback method for designing control protocols has
been successfully applied to a range of tasks in the control of
molecules and chemical reactions, such as the discrimination of
similar molecules, ionization, and molecular isomerization. It has
also been applied to ultrafast optical switching in
semiconductors~\cite{CBrif:2010}, and the production of X-rays
through high-harmonic generation (HHG)~\cite{Bartels2000}. While
second-order harmonics are easily generated, the production of
X-rays requires harmonics to tenth order or higher, which in turn
requires a very high optical nonlinearity and an intense laser. In
the experiment reported in~\cite{Bartels2000}, a shaped ultrafast
and intense laser pulse (with a duration of only $6$-$8$ optical
cycles) is shot into an atomic gas. The authors used an adaptive
feedback loop to shape the laser pulse, and in doing so improved
the efficiency of X-ray generation by an order of magnitude. As
shown in Fig.~\ref{Fig:HHG}, adaptive feedback can be also used to
find pulse shapes that will selectively generate specific
high-order harmonics. To date, over $150$ successful adaptive
feedback control experiments have been reported, and the number
applications is still growing~\cite{Rabitz2004}.


\subsection{Quantum self-feedback}\label{s42}

While the term ``self-cooling'' has been used to refer to
scenarios that are now understood as coherent
feedback~\cite{OArcizetNature:2006}, similar terms have been used
to refer to a situation that is distinct from coherent feedback,
and yet involves a feedback mechanism. This is the method of
cooling a mechanical resonator by using photo-thermal
pressure~\cite{CHMetzgerNature:2004} (as opposed to the radiation
pressure of resolved-sideband cooling~\cite{OArcizetNature:2006}).
In photo-thermal cooling, the motion of the resonator affects a
thermal bath to which the resonator is coupled, and the resulting
effect of the bath back on the resonator is what produces the
cooling. The effect is therefore feedback, but it is not feedback
from a coherent and controllable system that is envisaged in
coherent feedback, but from a many-body environment. For this
reason, we feel that it is reasonable to give this situation
another name, and here we choose to call it ``self-feedback''.

Another setting in which feedback from a many-body environment can
facilitate control is that in which the system to be controlled
interacts with a bath consisting of many nuclear
spins~\cite{AGreilichNature:2007, SGCarterPRL:2009,
XDXuNature:2009, CLattaNatPhys:2009, YSGreenberg, TDLaddPRL:2010,
BSunPRL:2012, DJReillyScience:2008, ITVinkNatphys:2009,
HBluhmPRL:2010, CBarthelPRB:2012, WYaoYLouEPL:2010}. One example
of this is the spin of a single electron trapped in a ``quantum
dot'', and a second is a cantilever (a mechanical oscillator) that
interacts with the nuclear spins via a magnet attached to it. The
ability to use the spin bath for control comes from the fact that
the nuclear spins have a long coherence time. By driving the
system (the electron spin or the cantilever), one can polarize the
nuclear spins so that they act on a single coherent spin, and this
coherence lasts for a time that is long compared to the timescale
on which we wish to control the system. Since the noise from the
nuclear spin-bath comes only from unpolarized spins, we can
greatly reduce this noise in this way.

It is also possible to use the nuclear spin-bath to cool a
cantilever, in a process similar to resolved-sideband cooling. By
polarizing the spins, or merely by using their steady-state
thermal polarization, and driving the electron spin at the
appropriate (blue-detuned) frequency, we can create a net transfer
of excitations from the cantilever to the spin
bath~\cite{YSGreenberg}.

\begin{figure} \centerline{\includegraphics[width =
16.8 cm]{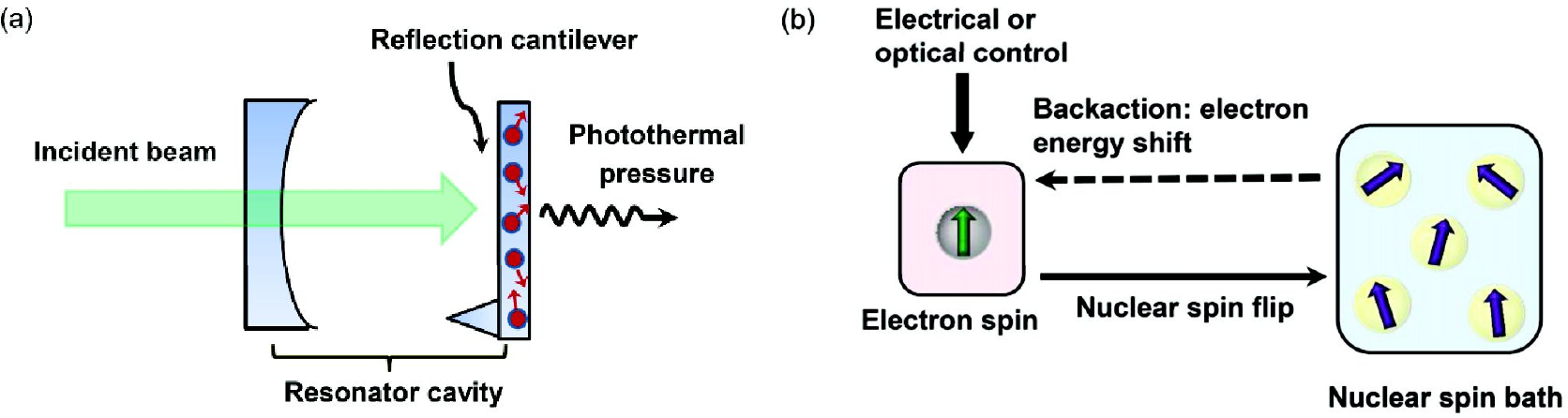}} \caption{(Color online) Schematic
diagram of self-feedback: (a) Passive cooling by photonthermal
pressure in optomechanical systems and (b) Feedback from a nuclear
spin bath to an electron spin in a quantum dot.}\label{Fig of
self-feedback}
\end{figure}

\section{Experiments Realizing Quantum Feedback}\label{s5}

We now give an overview of the experiments that have been
performed to-date realizing feedback in the quantum regime, both
measurement-based and coherent. Feedback has now been realized in
a range of distinct physical settings: atom optics and cavity QED,
opto-mechanics, superconducting circuits, and quantum dots. For
measurement-based feedback, the measurement efficiency is a
crucial factor in determining to what extent control can be
realized in the quantum regime, and the fidelity of this control.
Before we begin, we present a table that shows various key
parameters in current experiments on feedback control, and how
they compare across the various physical realizations. Some of
these parameters, such as the measurement efficiency, represent
the state of the art that we expect will be continually improved.
Other parameters, such as the feedback bandwidth, are merely what
is typically being used in current experiments. The feedback
bandwidth gives the fastest rate at which the control force
applied to the system can change. Thus the timescale of the
feedback control process is necessarily less than the feedback
bandwidth. The \textit{measure rate} also needs some explanations.
First note that the \textit{measurement strength} of a continuous
measurement has units of inverse time, as well as the inverse
square of the observable being measured. It can be thought of as
the rate at which the inverse variance of the observable is
increased (and thus the variance reduced) by the measurement. The
measurement strength becomes a rate only if we choose some
particular units for the variable, and scale the measurement
strength by these units. The measurement rates given in the table
are a result of choosing units that are natural in each case. For
example, if one is measuring the position of a harmonic
oscillator, then the natural unit of position is the uncertainty
in the position for the ground state of the oscillator. For linear
optics, the ``system'' being measured is often a continuous-wave
laser beam, and the measurement rate in this case is merely the
photon flux. Since it is possible to vary this flux over many
orders of magnitude, we have left the entry in the table simply as
``photon flux''. Note that since the continuous measurements are
mediated by coupling to fields, the measurement rate also
corresponds to the rate for coherent feedback that is mediated by
fields. Finally, we note that the time-delay in the feedback loop
and the bandwidth of the control are primarily relevant for
measurement-based feedback. We have not included quantum dots in
the table, since neither measurement-based quantum feedback nor
field-mediate coherent feedback has been realized for those
systems.

\begin{table*}\footnotesize
\centering
\caption{Characteristic parameters for feedback in various physical settings}
\begin{tabular}{|l|c|c|c|c|}
   \hline
                                          & Linear optics     & Optomechanics     & Cavity QED               & Superconding circuits   \\
  \hline
  System energy-scale      & 1 THz                 & 0.1-100 MHz         & 1 THz                         & 1-10 GHz               \\
  Feedback bandwidth      & 1 MHz                & 1 MHz                    & 1 MHz                        & 10 MHz              \\
  Decoherence rate           & 1 MHz                & 10 - 100 Hz           & 1 MHz                       &  10 - 100 kHz           \\
  Measurement efficiency  & 0.9                     & 0.9                         & 0.8                            &  0.4                          \\
  Measurement rate          & photon flux         & 10 Hz - 1 kHz         & 10 Hz - 1 kHz           & 100 kHz                  \\
  Feedback delay              & 0.1-10 $\mu$s    &  0.1-10 $\mu$s      & $1$ $\mu$s-$1$ ms & $250$ ns                 \\
  Ambient Temperature      & 300 K                 & 30 mK                   & 300 K                        & 30 mK                     \\
  \hline
\end{tabular}\label{Table of system parameters in various systems}
\end{table*}

\subsection{Quantum optics}\label{s51}

\subsubsection{Measurement-based feedback}

\textit{Adaptive phase measurement (2002):} The first experimental
demonstration of quantum feedback in optics was the realization of
Wiseman's adaptive phase measurement~\cite{ArmenPRL:2002}. The
limited interactions available to detect light, or in fact any
physical system, makes it impractical to exactly measure optical
phase. If one has a good idea of the phase prior to the
measurement, then homodyne detection can be used to provide good
effect, since the quadrature to measure can be chosen using this
information. But if the phase is completely unknown beforehand,
this is not possible, and it was widely believed prior to 1995
that heterodyne detection gave the best possible phase measurement
in that case~\cite{CMCavesRMP:1994}. In 1995, Wiseman showed that
the use of homodyne detection, when combined with feedback used to
modify the quadrature being measured during the measurement, could
realize a more accurate phase measurement than heterodyne
detection~\cite{HMWisemanPRL_Adaptive:1995}. For a pulse of light
with no more than one photon, this adaptive phase measurement
realizes exactly a measurement of canonical phase, as defined by
the Pegg-Barnett phase operator.

In Fig.~\ref{Fig of the adaptive phase measurement2002}, we show a
diagram of the experimental setup in Ref.~\cite{ArmenPRL:2002}.
The signal consists of noisy weak coherent light from a
single-mode continuous-wave Nd:YAG laser, which first passes
through a high-finesse Fabry-Perot cavity (not shown in the
figure) with a ringdown time $16$ $\mu$s and shot noise limit of
$50$ kHz. This cavity squeezes out the intensity noise in the
signal beam. This beam is then fed into a Mach-Zehnder
interferometor (MZI) to generate interference between the signal
light and a local oscillator to implement homodyne detection. The
local oscillator (LO) has a power of $230$ $\mu$W and is
frequency-shifted by an electro-optic modulator (EOM), which is
driven by an RF-synthesizer (RF$_2$). The two RF synthesizers
RF$_1$ and RF$_2$ are phase-locked to each other to achieve
synchronization between the local oscillator and the signal light.
By changing the amplitude and switching RF$_2$ on and off, the
power of the signal beam can be tuned between $5$ fW and $5$ pW
and a pulse generated with a duration of about $50$ $\mu$s. The
two output ports of the MZI are measured by two photon detectors
and the difference between their respective photo-currents
realizes a homodyne or heterodyne measurement, depending on how
the phase of the local oscillator is modulated. In the experiment,
the shot noise in the difference photocurrent is about $6$ dB
above the noise floor in the range $1$ kHz to $10$ MHz.

To realize an adaptive homodyne measurement, the phase of the
local oscillator, $\Phi$, is modified via RF$_2$ by a feedback
signal as the measurement of the pulse proceeds. It is this phase
that determines the quadrature that is measured.  The feedback
bandwidth, being the maximum rate at which the phase of the local
oscillator can be changed, is about $1.5$ MHz, and is mainly
limited by the maximum slew rate of RF$_2$. The value of $\Phi$ as
specified by Wiseman's adaptive measurement protocol
is~\cite{HMWisemanPRA_Adaptive:1998}
\begin{equation}\label{Feedback signal for adaptive phase estimation}
\Phi \left( t \right) = \frac{\pi}{2} + \int_0^{\mss{t}} \frac{ I \left( s \right) }{\sqrt{s}} ds,
\end{equation}
where $I(s)$ is the photocurrent at time $s$. This integral of the photo-current is calculated by a Field Programmable Gate Array (FPGA). The final estimated phase is given by
\begin{equation}\label{Estimated phase for adaptive phase estimation}
\hat{\phi} = \int_0^{\infty} \frac{ I \left( s \right) }{ \sqrt{s} } ds.
\end{equation}
The experiment demonstrated the superiority of this adaptive measurement over heterodyne detection for pulses with low photon number. When the number of photons in a mode is large enough, heterodyne detection is better than adaptive homodyne detection.

\begin{figure}
\centerline{\includegraphics[width = 15 cm]{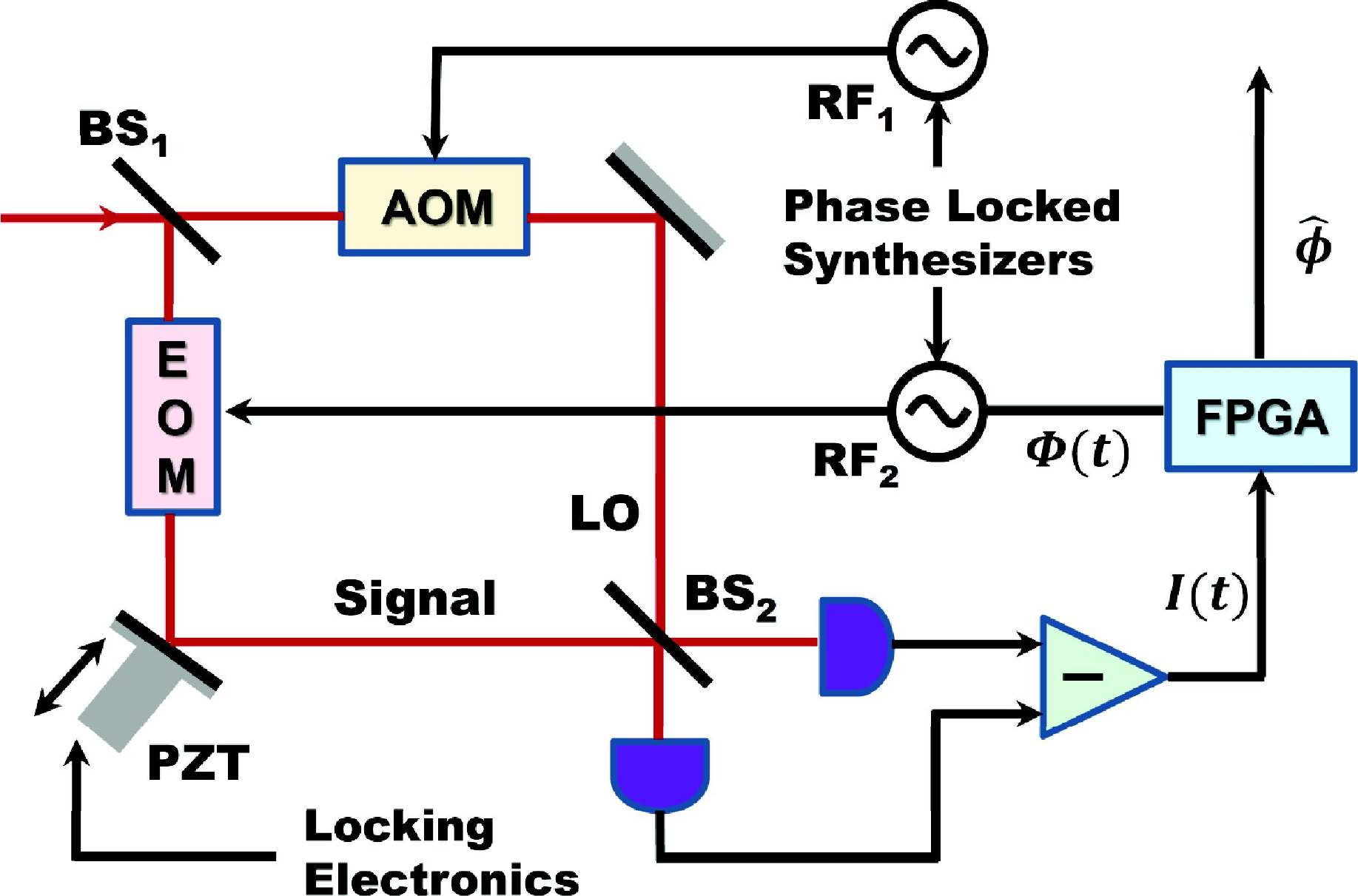}}
\caption{(Color online) Diagram of the adaptive homodyne
measurement of an optical phase in performed by Armen \textit{et
al.}~\cite{ArmenPRL:2002}. Optical beams are indicated by red
lines and electronic beams by black lines. The AOM and EOM
represent the acousto-optic modulator and electro-optic modulator.
BS and RF are the beam splitter and the radio-frequency
synthesizer. PZT is the piezoelectric transducer and FPGA is the
field programmable gate array.}\label{Fig of the adaptive phase
measurement2002}
\end{figure}

\textit{Adaptive phase estimation (2007):} While the previous
adaptive measurement was concerned with measuring phase as
accurately as possible for low-power beams, the 2007 adaptive
measurement of Higgins {\it et al.} was concerned with measuring a
classical parameter, in this case a phase-shift applied to a beam
of light, with the most efficient use of
resources~\cite{HigginsNature:2007}. The resources in question are
the number of photons in the beam. Due to the relationship between
amplitude and phase, the more photons in the beam the more sharply
the phase is defined, and thus the more accurately can an applied
phase shift be determined. If the beam is in a coherent state,
then the error in the measured phase is proportional to
$1/\sqrt{N}$, where $N$ is the average number of photons. But by
using non-classical states, and in particular highly entangled
states~\cite{NOON}, the error can be reduced to $1/N$, which is
known as the Heisenberg limit. The experimental and theoretical
work by Higgins {\it et al.} showed that entanglement was not
necessary to reach the Heisenberg limit, which could instead be
achieved by applying the phase shift to one photon at a time, and
using an adaptive measurement when measuring the sequence of
photons~\cite{HigginsNature:2007}.

The experimental setup used in Ref.~\cite{HigginsNature:2007} is
shown in Fig.~\ref{Fig of the adaptive repreated phase
measurement}. The basic component of this setup is a
``common-spatial-mode'' polarization interferometer. The two arms
of the interferometer correspond to right-circularly and
left-circularly polarized light. A right-polarized photon
experiences a phase shift of $\phi$ which is the unknown phase to
be determined, whereas a left-polarized photon experiences the
phase shift $ p \theta $ which is chosen by the feedback protocol.
To extract the phase information, a measurement of $ \sigma_x $ is
made on the single-photon qubit. This is achieved by using a
Polarizing Beam Displacer (PBD) to discriminate the horizontal and
vertical polarization basis, followed by a  $10$ nm bandwidth
interference filter, and finally by photon detection by the Single
Photon Counting Modules (SPCMs). After each photon is sent through
the interferometer, the factor $p$ is determined for the next
photon from the current estimate of the phase shift $\phi$. The
experiment demonstrated a measurement error well below the
``standard quantum limit'' of $1/\sqrt{N}$ and close to the
Heisenberg limit.

There have been a number of further experiments that have also
demonstrated a phase measurement below the standard quantum limit
without using entangled input
states~\cite{GYXiangNatPhoton:2011,GYXiangNatPhoton:2010,ROkamotoPRL:2012}.
For example, in Ref.~\cite{GYXiangNatPhoton:2011}, Xiang~{\it et
al.} use two $n$-photon states as the two inputs in a Michelson
interferometer, use Bayesian analysis and optimal adaptive
feedback to make full use of these multiphoton states.

\begin{figure}
\centerline{\includegraphics[width= 16.8 cm]{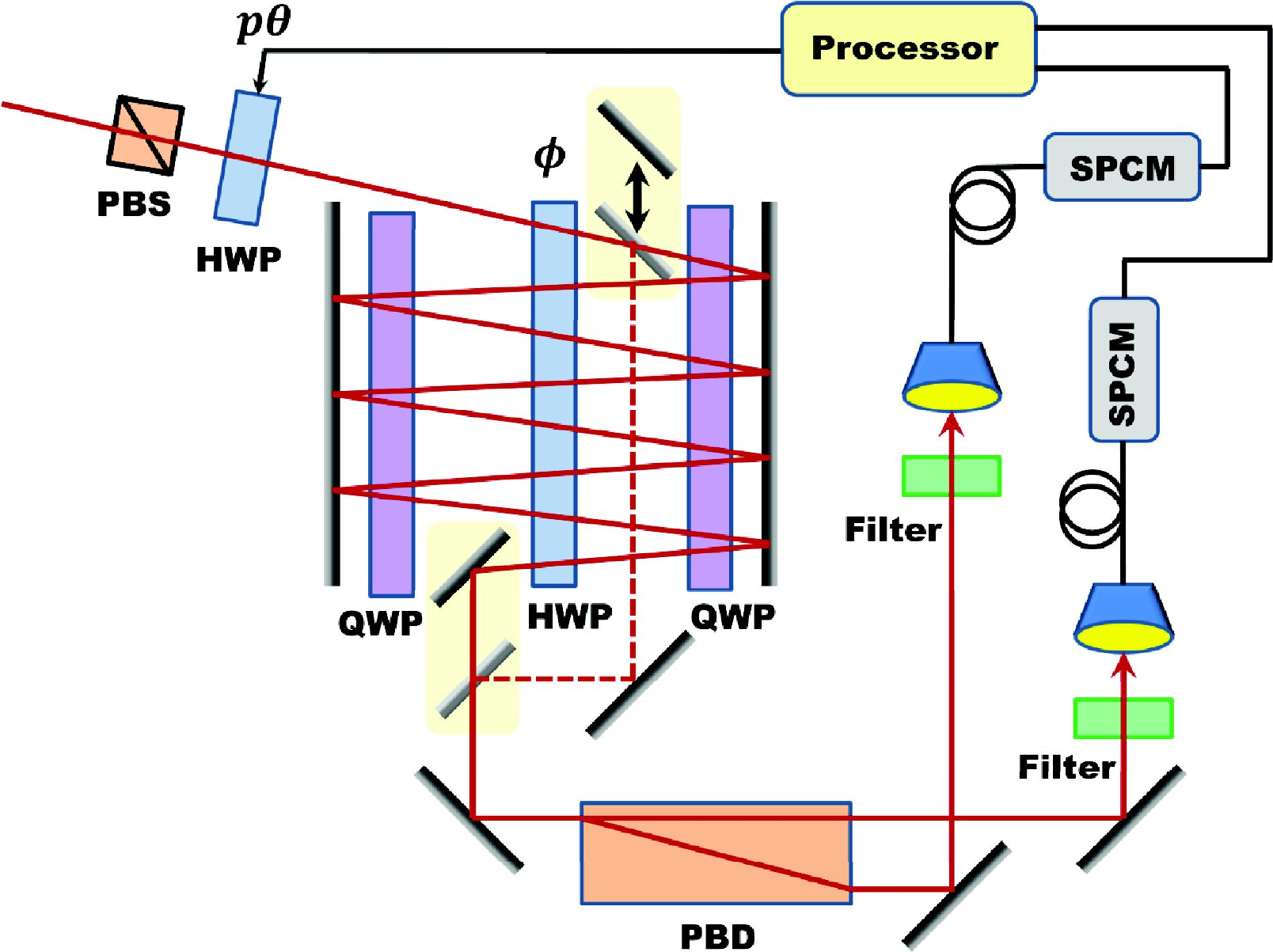}}
\caption{(Color online) Diagram of the Heisenberg-limited phase
estimation by adaptive measurement performed by Higgins \text{et
al.}~\cite{HigginsNature:2007}. The labels PBD and PBS represent
the polarizing beam displacer and the polarizing beam-splitter.
The label SPCM is the single-photon counting module, and QWP and
HWP are the quarter-wave plate and half-wave plate,
respectively.}\label{Fig of the adaptive repreated phase
measurement}
\end{figure}


\textit{Adaptive phase estimation with smoothing (2010):} The
phase estimation experiments we have described so far are
concerned with estimating a single phase shift. This next
experiment involves estimating a time-varying phase shift. To
estimate the phase shift $\phi$ at some time $t$, $\phi(t)$, we
can use the stream of measurement results obtained up until that
time. The procedure of processing the measurement results up until
time $t$ to produce an estimate of a signal at that time is called
\textit{filtering}. If we are prepared to wait until time $t+\tau$
for our estimate, then we can use the measurement results up until
time $t+\tau$ to obtain our estimate of $\phi(t)$. The process of
obtaining an estimate from measurement results obtained both
before and after the time of the estimate is called
\textit{smoothing}. In Ref.~\cite{WheatleyPRL:2010}, Wheatley
\textit{et al.} use an adaptive homodyne measurement procedure,
combined with smoothing, to estimate a time-varying phase shift.
They then compare this with the performance of a ``dual-homodyne''
(or ``eight-port homodyne'') measurement. The smoothed estimate
is, in theory, able to improve on non-adaptive filtering by a
factor of $2\sqrt{2}$.

We show a diagram of the experimental setup in Fig.~\ref{Fig of
the adaptive dual-homodyne phase measurement} (a) and (b). In
fact, it is the dual-homodyne  measurement that is shown in
Fig.~\ref{Fig of the adaptive dual-homodyne phase measurement}. To
obtain the single homodyne measurement, one merely deletes one of
the homodyne set-ups in the diagram. A $860$ nm Ti:sapphire laser
is used as the source to drive the experiment. The two
acousto-optic modulators (AOMs) are used to place sidebands on the
beam so as to generate a much weaker beam at those sidebands, on
which the phase estimation will be performed. This weak beam has a
photon flux of $|\alpha|^2\approx10^6$ photons per second.

The phase signal $\phi(t)$ is imprinted on the beam by the
electro-optic modular (EOM). This signal is chosen to be an
Ornstein-Uhlenbeck (OU) noise process source $\phi\left( t
\right)$ given by
\begin{equation}\label{Classical phase noise process}
d \phi \left( t \right) = - \lambda \phi \left( t \right) dt +
\sqrt{\kappa} d W \left( t \right),
\end{equation}
where $dV$ is a Wiener increment and $1/\kappa$ gives the
coherence time of the phase signal. There are approximately $100$
photons in the beam within a single coherence time, meaning that
the estimate of the phase at any given time can draw information
from roughly $100$ photons. The beam carrying the signal $|\alpha|
e^{i \phi(t)}$ and the output of the local oscillator generated in
Fig.~\ref{Fig of the adaptive dual-homodyne phase measurement}(a)
are both fed into the measurement circuits in Fig.~\ref{Fig of the
adaptive dual-homodyne phase measurement}(b). The homodyne
detection has an effective efficiency of no less than $89\%$.

Given a photo-current $I\left(t\right)$ from a homodyne detection,
the forward (+) and backward (-) estimates of the phase at time
$t$ are given by
\begin{equation}\label{Time symmetric estimated phases}
\Theta_{\pm} \left( t \right)=\pm 2 \sqrt{\kappa} | \alpha |
\int_t^{\pm \infty} \exp{\left[ \omega_0 \left( s - t \right)
\right]} I \left( t \right) / 2 |\alpha| d s,
\end{equation}
where $\omega_0=1.5\times 10^5$ s$^{-1}$ is the cutoff frequency
of the linear Low-Pass Filter (LPF) and is far smaller than the
feedback gain $2\sqrt{\kappa}|\alpha|$. The final smoothed
estimate $\Theta\left( t \right) = \left[ \Theta_+ \left( t
\right) + \Theta_- \left( t \right) \right]/2 $ of the unknown
phase $\phi \left( t \right) $ is merely the symmetric weighted
average of the forward and backward estimates.

\begin{figure}
\centerline{\includegraphics[width=15 cm]{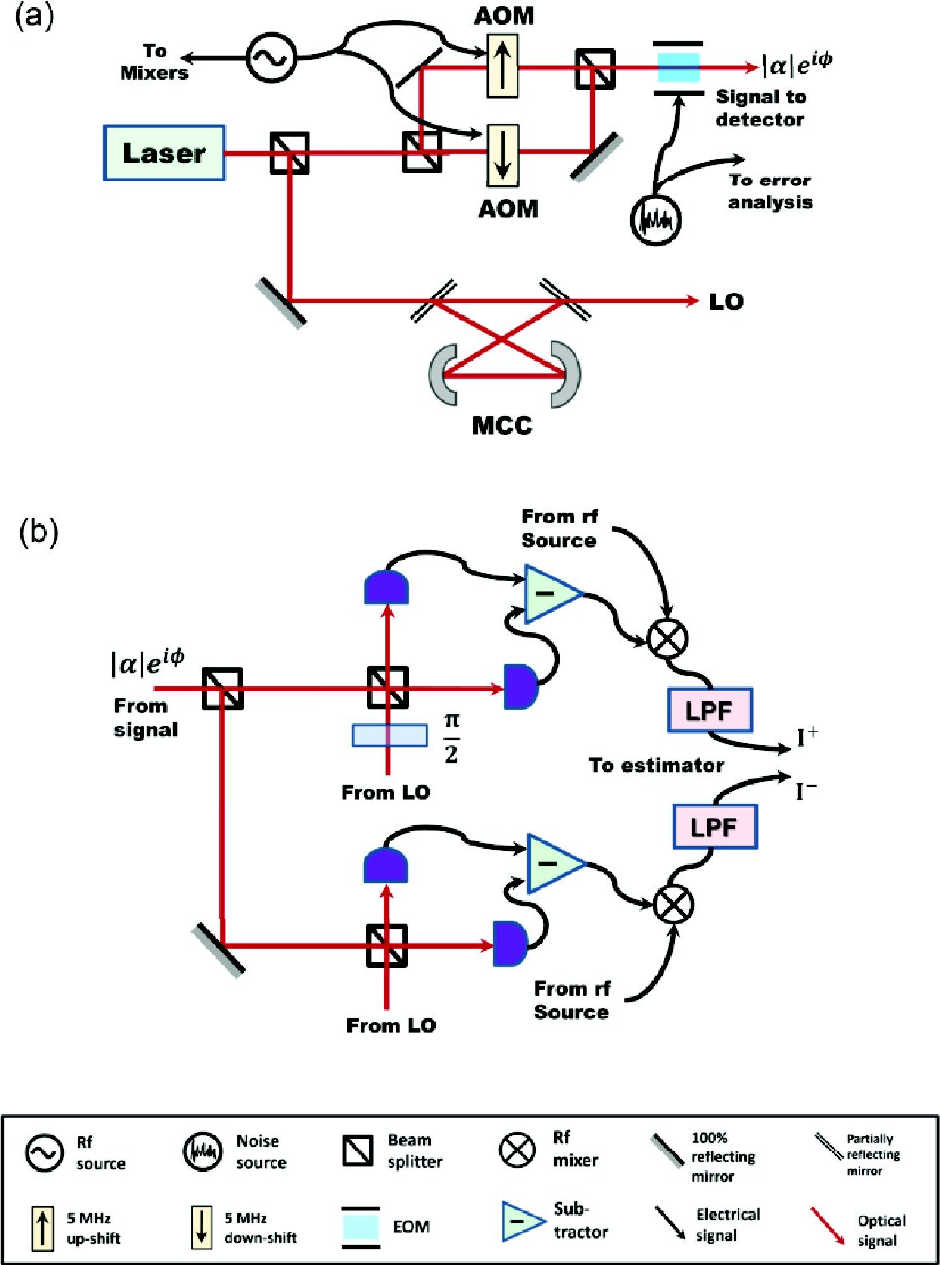}}
\caption{(Color online) Diagrams of the experimental
configurations for (a) signal and local oscillator generation and
(b) adaptive dual-homodyne phase estimation in the experiment of
Wheatley \textit{et al.}~\cite{WheatleyPRL:2010}. The label AOM
denotes the acousto-optic modulator, and LPF and MCC denote the
low-pass filter and mode-cleaning cavity, respectively.}\label{Fig
of the adaptive dual-homodyne phase measurement}
\end{figure}

\textit{Adaptive phase estimation with squeezed light (2012):} One
way to beat the standard quantum limit for phase estimation is to
used squeezed light, since the phase error in this light is less
than $1/\sqrt{N}$, where $N$ is the number of photons. The
experiment of Yonezawa \text{et al.} implemented this method using
a continuous beam of squeezed light~\cite{HYonezawaScience:2012}.
Their experimental setup is shown Fig.~\ref{Fig of the adaptive
phase measurement for squeezing light}. A $860$ nm Ti:sapphire
laser is used to drive the experiment, from which a squeezed beam
is produced using an optical parametric oscillator (OPO). To
achieve this, the beam from the Ti:sapphire laser is first
up-converted to $430$ nm using second-harmonic generation, and
this beam is used to pump the OPO which then produces the squeezed
beam at $860$ nm. The squeezed light has approximately $3$ dB of
squeezing, the anti-squeezed quadrature is about $5$ dB above the
equivalent coherent source, and the beam power is $|\alpha|^2 =
10^6$ photons per second.

The phase signal $\phi(t)$ that is imprinted on the squeezed beam is chosen to be
\begin{equation}\label{Classical phase noise process}
\phi\left(t\right)=\sqrt{\kappa}\int_{-\infty}^{\mss{T}}
\exp{\left[-\lambda\left(t-s\right)\right]}dW\left(s\right),
\end{equation}
where $dW\left(s\right)$ is a classical Wiener process,
$\lambda\approx 6\times 10^4$ rad/s is the inverse correlation time of $\phi\left(t\right)$,
and $\kappa \approx 2 \times 10^4$ rad/s is the magnitude of the phase variation.

An adaptive homodyne method is used to measure the phase of the
beam, having an overall efficiency of $\eta=0.85$. The quadrature
measured by the homodyne detector at time $t$ is chosen to be
$\Phi\left( t \right) = \hat{\phi}\left( t \right) + \pi/2$, where
$\hat{\phi}\left( t \right)$ is the current estimate of the phase
$\phi(t)$. The experiment achieved a phase estimation error that
was $15\pm4\%$ below the ideal limit achievable with a coherent
beam.

\begin{figure} \centerline{\includegraphics[width =
16.8 cm]{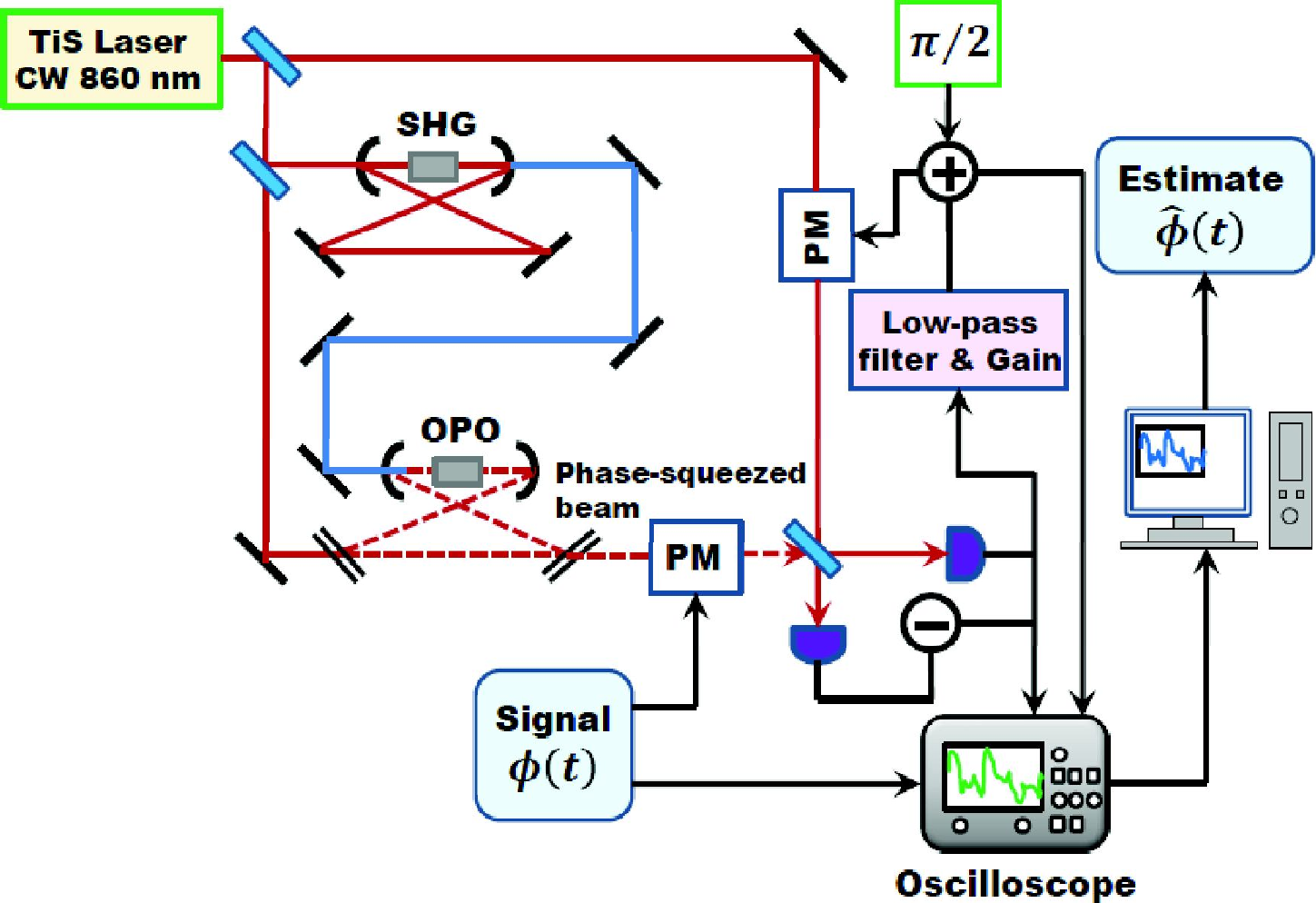}} \caption{(Color online) Schematic
diagram of the experimental setup of Yonezawa \textit{et
al.}~\cite{HYonezawaScience:2012}. The acronyms TiS and CW refer
to titanium sapphire and continuous wave, respectively. The labels
PM and SHG denote the phase modulator and the
second-harmonic-generator. The squeezed signal light generated by
the optical parametric oscillator (OPO) is modulated by a
stochastic phase $\phi\!\left(t\right)$ and then is interfered
with a local oscillator. The output field is measured by homodyne
detection and then fed back to update the phase of the local
oscillator.}\label{Fig of the adaptive phase measurement for
squeezing light}
\end{figure}

\textit{Correcting a single-photon state (2010):} So far all the
experiments we have described in linear optics are adaptive
measurements of optical phase. This next experiment is an
exception. Here Gillett \textit{et al.} use a weak measurement to
optimally correct the state of a single qubit which is initially
prepared in one of two non-orthogonal states. In this case, the
qubit is encoded in the polarization of a single
photon~\cite{GGGillettPRL:2010}. This correction procedure was
suggested and analyzed in~\cite{AMBranczykPRA:2007}. Recall that a
weak measurement is one in which the measurement operators are not
rank-1 projectors, so it does not reduce a mixed state to a pure
state, and does not provide full information about the final
state.

A diagram of the experimental setup is shown in Fig.~\ref{Fig of the qubit stabilization by measurement-based feedback}. The two states of the qubit are labelled as $ | H \rangle \equiv | 1 \rangle $ (horizontal polarization) and $ | V \rangle \equiv | 0 \rangle $ (vertical polarization). To make a weak, single-shot measurement of one of these qubits, Gillett \textit{et al.} perform a gate that partially entangles the qubit with a second ``probe'' qubit, and then perform a projective measurement on the probe.

\begin{figure} \centerline{\includegraphics[width =
16.8 cm]{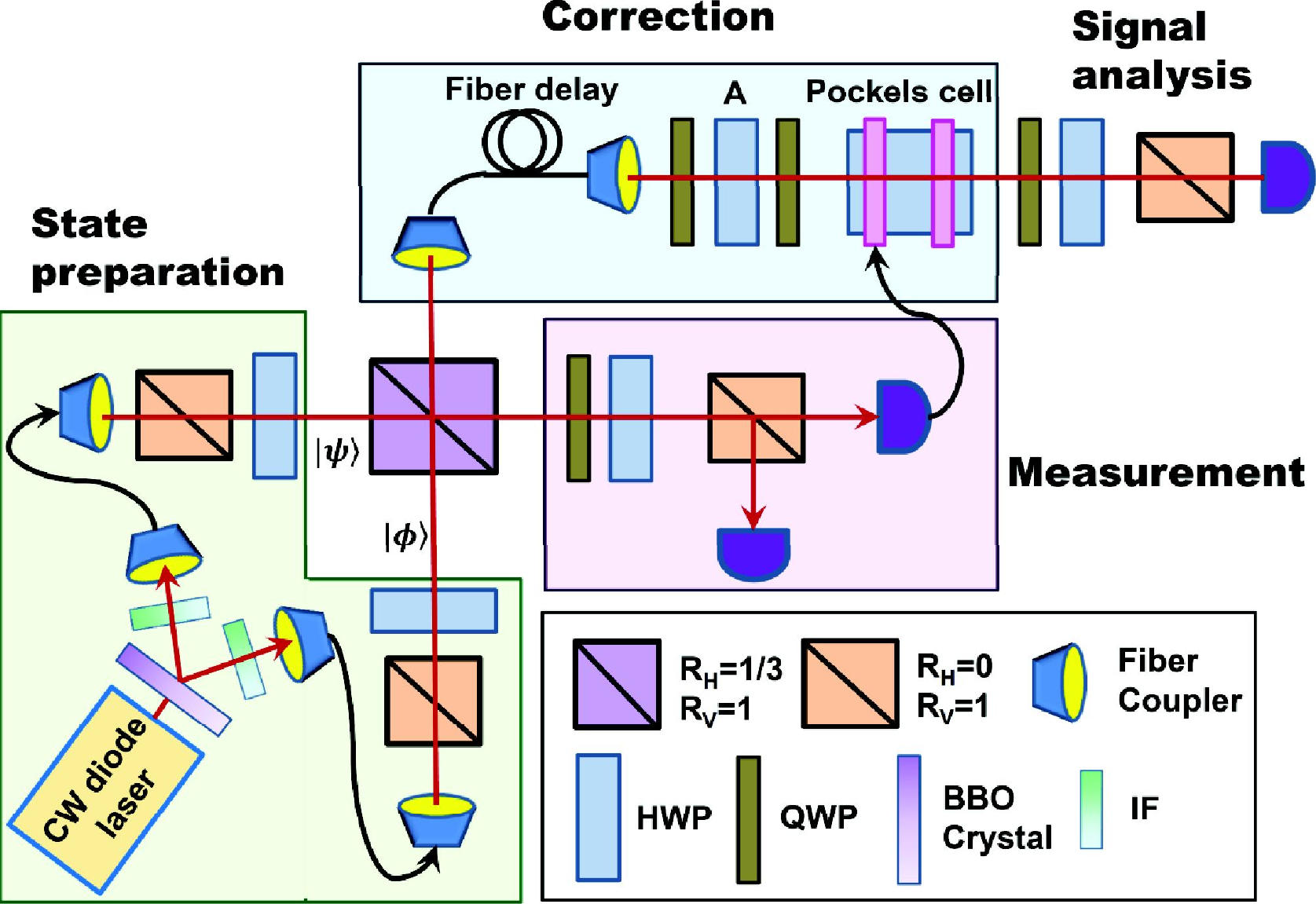}} \caption{(Color online) Schematic
diagram of the experimental setup of Gillett \textit{et
al.}~\cite{GGGillettPRL:2010}. A photon pair composed of a primary
(or ``signal'') photon and an auxiliary photon is prepared by
spontaneous parametric down-conversion in a BiBO crystal. The
primary and auxiliary photons are then prepared in specific
initial states by transmitting through the polarization beam
splitters and half wave plates (HWP). The two photons are combined
at a partially-polarizing beam splitter which applies a controlled
phase gate between the two (conditioned on there being only one
photon in each output). A succeeding projective measurement in the
$+/-$ basis on the auxiliary photon realizes a weak measurement on
the primary. The measurement output is then fed into a Pockels
cell to rotate the signal photon conditional on the measurement
outcome. The acronym QWP labels the quarter wave plate and IF
labels the interference filter.} \label{Fig of the qubit
stabilization by measurement-based feedback}
\end{figure}

The experiment is driven by an $820$ nm Ti:sapphire laser. A 410 nm beam is created from this using second harmonic generation, so that a pair of 820 nm photons can be created from this 410 nm beam using spontaneous parametric down-conversion. The each photon in the pair is then fed into a single-mode fiber. One of these photons carries the qubit to be corrected --- the ``primary'' photon --- and the other will be used as the probe. The primary is prepared in one of the two (non-orthogonal) states
\begin{eqnarray*}
| \psi_\pm\rangle = \cos \frac{\theta}{2} | \pm \rangle \pm \sin
\frac{\theta}{2} | - \rangle
\end{eqnarray*}
by transmitting through a polarizing beam splitter with reflectivity $R_H=0$ and $R_V=1$ followed by a Half Wave Plate (HWP). Here
\begin{eqnarray*}
| \pm \rangle = \frac{1}{\sqrt{2}} \left( | 0 \rangle \pm | 1
\rangle \right) .
\end{eqnarray*}
The probe photon is to be detected, and is prepared in the state
\begin{eqnarray*}
| \phi \rangle= \cos \frac{\chi}{2} | + \rangle + \sin
\frac{\chi}{2} | - \rangle .
\end{eqnarray*}
The primary and probe photons are now interfered through a partially polarizing beam splitter with reflectivity $R_H=1/3$ and $R_V=1$. Conditional on there being only one photon
in each mode, the partially polarizing beam splitter executes a
control-Z gate (given by $| 0 \rangle \langle 0 | \otimes
\mathbf{1} + | 1 \rangle \langle 1 | \otimes Z $), between the qubits with the probe as the control~\cite{ROkamotoPRL:2005,NKieselPRL:2005,NKLangfordPRL:2005}.

The primary qubit is next subjected to a dephasing error with probability $p$.
To correct this error as best as possible, the probe qubit is measured in the basis $\{ |\pm\rangle\}$. This measurement is implemented by rotating the polarizing basis and transmitting through a polarizing beam splitter with reflectivity $R_H=0$ and $R_V=1$. This results in a weak measurement on the primary qubit described by the measurement operators
\begin{eqnarray*}
M_+ = \cos \frac{\chi}{2} | 0 \rangle \langle 0 | + \sin
\frac{\chi}{2} | 1 \rangle \langle 1 |, \quad M_- = \sin
\frac{\chi}{2} | 0 \rangle \langle 0 | + \cos \frac{\chi}{2} | 1
\rangle \langle 1 | .
\end{eqnarray*}
A unitary rotation is then performed on the primary qubit
depending on the result of the measurement. This is implemented
using a Pockels cell, which applies a rotation through an angle of
$4 \eta$ around the $y$ axis if a photon is detected in the
transmission output of the polarization beam-splitter. Combined
with a fixed rotation of $-2\eta$, the result is a rotation of
$\pm 2 \eta$ corresponding to the measurement result $M_\pm$.
Prior to this feedback operation, the primary photon passes
through a $50$ m fiber to allow time for the quantum weak
measurement, as well as a set of plates to compensate the
polarization rotation introduced by this fiber.

The purpose of the weak measurement in this case is that a
projective (or ``complete'') measurement does not achieve the
optimal correction. This is because the two initial states of the
primary are not orthogonal, and so the measurement cannot avoid
disturbing at least one of them.

\subsubsection{Coherent feedback}

\textit{Noise cancellation (2008):} The first all-optical
demonstration of a coherent feedback scenario was performed by
Mabuchi~\cite{HMabuchiPRA:2008}, and was a realization of a
noise-cancellation loop suggested by James~\cite{MRJamesTAC:2008}.
A diagram of the experimental setup is shown in Fig.~\ref{Fig of
the coherent feedback for robust optimal control}. The primary
system (the ``plant'') is a four-mirror ring cavity (top) as is
the auxiliary (the ``controller''). The experiment is driven by an
$852$ nm diode laser, and an external ``noise" signal is injected
into the plant cavity at the input $w$. The output of the primary,
$y$, that is reflected from the plant input coupler, acts as the
error signal, which is processed by the controller to generate a
control signal $u$. This control signal is then fed back into the
plant cavity again to attempt to cancel, as well as possible, the
effect of the noise on the plant output, $z$. The goal of the
feedback loop is to minimize the ratio of the optical power at the
output $z$ to that of the ``noise" input $w$. This quantity can
also be described as the magnitude of the closed-loop transfer
function.

\begin{figure} \centerline{\includegraphics[width =
15 cm]{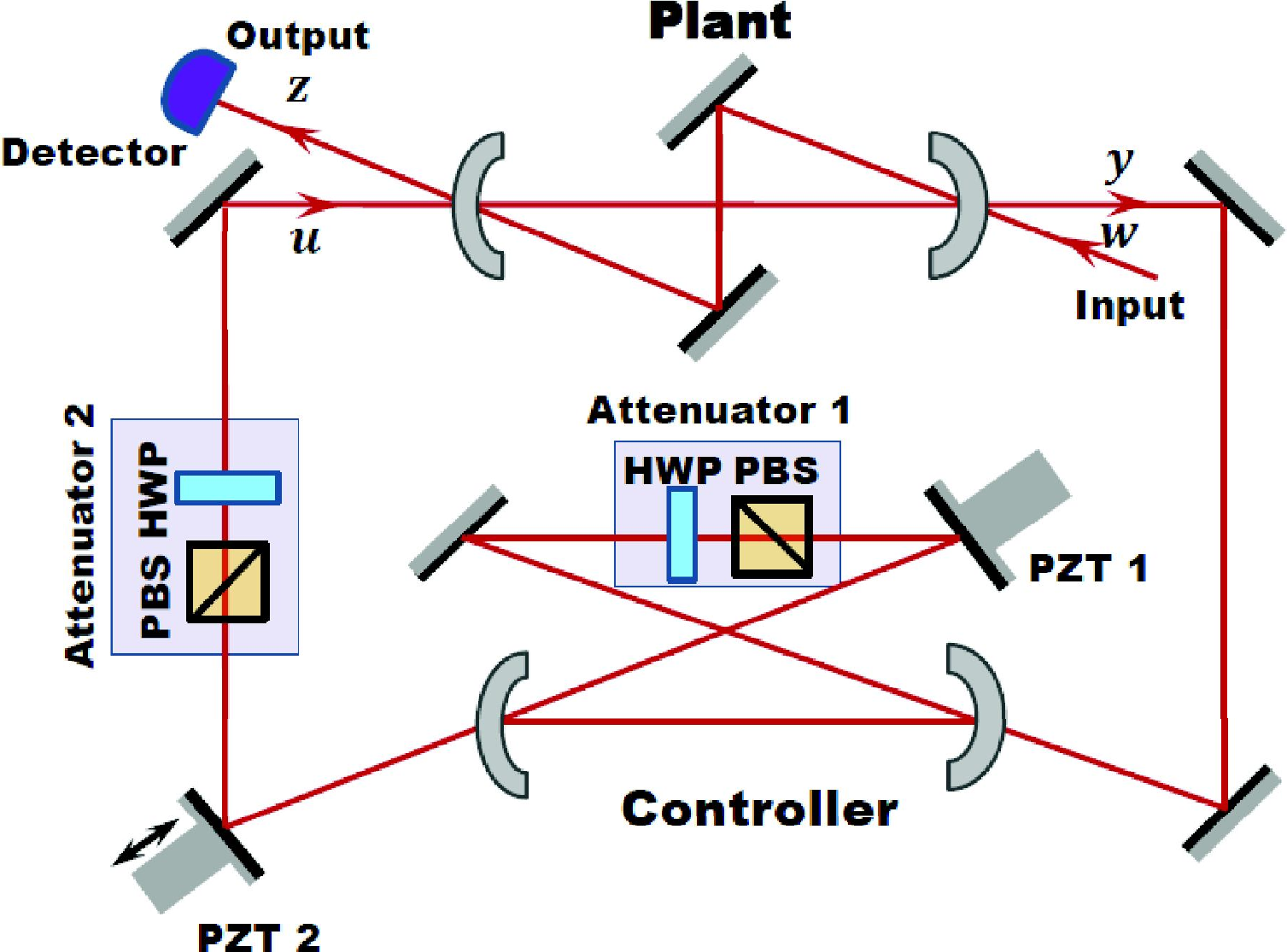}} \caption{(Color online) Diagram of the
experimental setup of Mabuchi~\cite{HMabuchiPRA:2008}. Two
four-mirror ring resonators couple with each other by a
transmitting optical field, which work as the plant cavity and
controller, respectively. The control goal is to tailor the
behaviors of the controller to minimize the power detected at the
system output $z$, when a ``noise" signal $w$ is fed into the
plant cavity. This can be achieved by tuning the parameters via
the two variable attenuators, i.e., attenuator~1 and attenuator~2,
and the two actuator piezoelectric transducers (PZTs), i.e., PZT~1
and PZT~2.}\label{Fig of the coherent feedback for robust optimal
control}
\end{figure}

Four practical ways to tune the parameters of the controller to
achieve the noise cancellation are: (i) adjust its resonance
frequency using the actuator PZT~1; (ii) adjust the phase of its
transfer function using the actuator PZT~2; (iii) adjust its decay
rate using the intracavity variable attenuator (attenuator~1); and
(iv) adjust the magnitude of its transfer function of using
attenuator~2. By optimizing these parameters, Mabuchi was able to
reduce the noise in the output by approximately $7$~dB.

\begin{figure}
\centerline{\includegraphics[width = 16.8
cm]{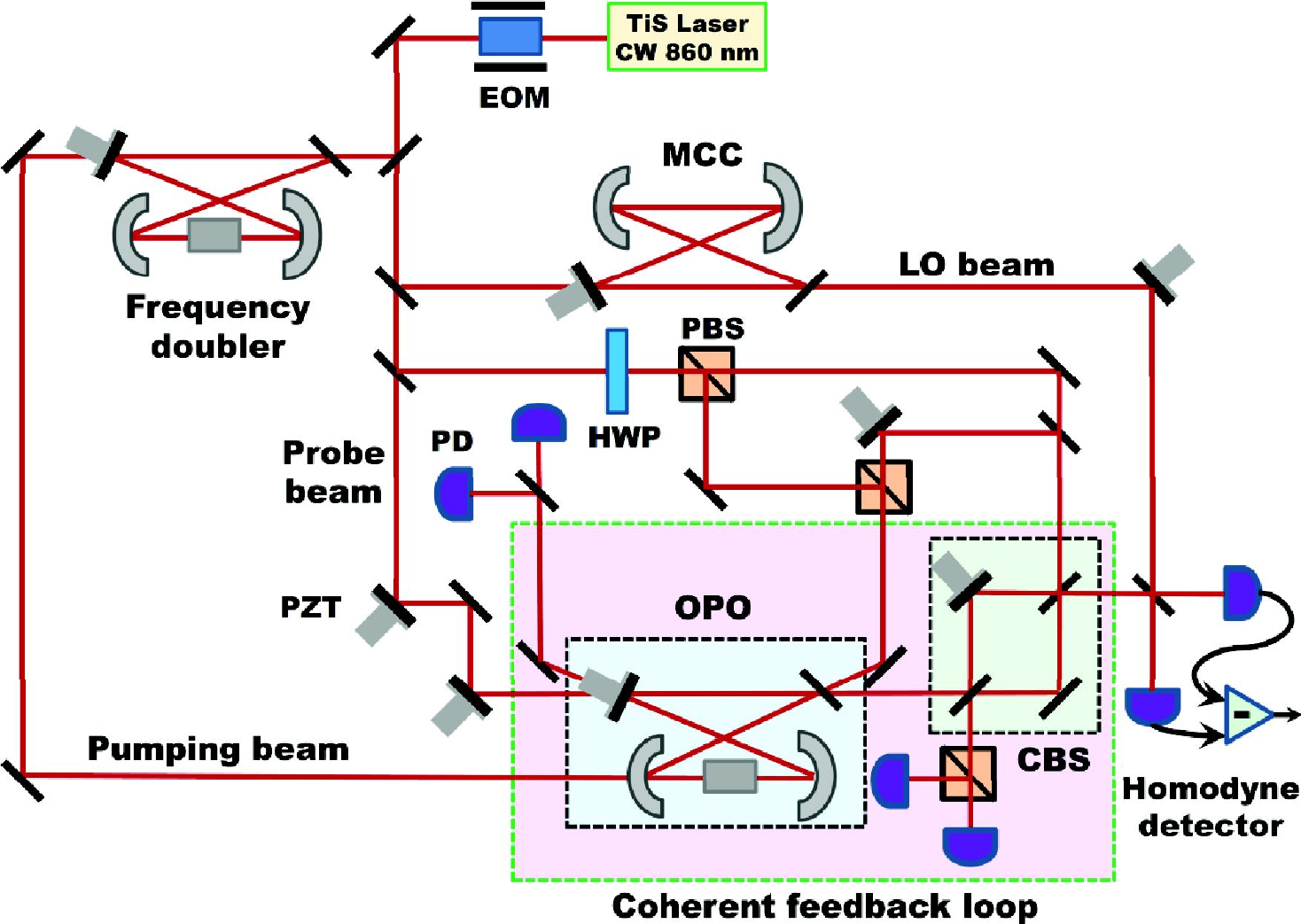}} \caption{(Color online) Schematic
diagram of the experimental setup of Slida \textit{et
al.}~\cite{SIidaTAC:2012}, in which coherent feedback is used to
enhance optical squeezing. The coherent feedback control loop
consists of the optical parametric oscillator (OPO) and a
Mach-Zender interferometer that acts as a beam-splitter with a
tunable reflectivity (the ``controlled-beamsplitter'' (CBS)). The
various labels denote the following optical elements: MCC
(mode-cleaning cavity); EOM (electro-optic modulator); PD
(photo-detector); PZT (piezoelectric transducer); HWP (half-wave
plate); PBS (polarized beam splitter); LO (local oscillator).}
\label{Fig of the coherent feedback for squeezing enhancement}
\end{figure}

\textit{Squeezing light (2012):} The experiment of Slida \textit{et al.} realized a coherent feedback loop to enhance the squeezing of an optical beam. This was an implementation of a protocol devised by Gough~\cite{JEGoughPRA:2009}. The experimental setup is shown in Fig.~\ref{Fig of the coherent feedback for squeezing enhancement}. The optical parametric oscillator (OPO) generates squeezed light, and it is the job of the feedback loop to enhance this squeezing. As such, the primary system is the OPO and auxiliary system is the ``control beamsplitter'' (CBS), which acts as a beamsplitter whose reflectivity can be adjusted. The CBS also acts as the output port through which the squeezed light exits the combined system, and is then evaluated by a homodyne measurement.

This experiment is probably the most sophisticated coherent
feedback loop realized to-date. The feedback loop can be a little
hard to read from the diagram, because there are additional
classical locking loops that share the same mirrors as the
coherent feedback loop, but are separate from it. The feedback
loop is surrounded by the green dashed line, and consists of the
OPO, a Mach-Zender interferometer that acts as the controlled
beam-splitter (CBS), and a loop that connects them.

You will notice that multiple beams (or \textit{branches}) are
split off from the single Ti:sapphire laser that drives the
experiment. The first branch is fed into a frequency-doubler to
generate a second-order harmonic beam of $430$ nm, which is used
as the beam that pumps the OPO to produce squeezed light at $860$
nm. The second branch is used as the local oscillator to implement
the homodyne detection on the squeezed output. The third branch,
the one that goes through the half-wave plate (HWP) is used to
lock the Mach-Zender and the coherent feedback loop. The final
branch works as the ``probe'' beam, which is injected into the OPO
as the ``seed'' that sets the phase of the squeezed light.
Photo-detectors (PDs) and piezoelectric transducers (PZTs) are
used in a classical phase-locking loop to fix the relative phase
between the probe and the pump beams.

The authors were able to demonstrate a squeezing enhancement from
$-1.64\pm0.15$ dB to $-2.20\pm0.15$ dB, in which the corresponding
enhancement of the anti-squeezing was from $1.52 \pm 0.15$ dB to
$2.75 \pm 0.15$ dB.

\textit{Classical logic gates (2012):} We note that an experiment
by Zhou \textit{et al.}~\cite{ZZhouAPL:2012} that used
four-wave-mixing in Rubidium vapor to implement optical OR and NOR
gates used a ``coherent feedback'' loop. However, the loop was not
necessary to implement the logic gates, since a beam generated by
a separate Rubidium vapor cell would also have sufficed in place
of the beam taken from the output and fed back.


\subsection{Trapped particles: Atoms, Ions, and Cavity-QED}\label{s52}

\subsubsection{Measurement-based feedback} \label{secSmith02}

\textit{Stabilization of a quantum state (2002):} The first demonstration of
quantum feedback control of an atom-optical system was that performed by Smith \textit{et al.}\footnote{Earlier experiments in atom-optical systems, while reported as quantum feedback at the time, were not in fact quantum feedback because they could be described by classical analyses~\cite{YamamotoPRA:1986, HAHaus:1982, JHShapiroJOSAB:1987}. The reason for this is that the quantum noise was insignificant compared to the classical noise~\cite{HMWiseman:2009}.} The system to be controlled consists of a stream of atoms interacting with a single mode of an optical cavity, as shown in Fig.~\ref{Fig of the quantum state capture and release by quantum feedback}. If we denote the average number of atoms interacting with the cavity at any given time by $N$, and if the interaction is sufficiently weak that only one of the atoms is excited at any given time, then the system settles down to a steady state given by
\begin{equation}
|\psi_{ss}\rangle\approx|0\rangle |g\rangle+\lambda\left(
|1\rangle |g\rangle-\frac{2g\sqrt{N}}{\gamma}|0\rangle
|e\rangle\right) .
\end{equation}
In this equation, the kets $|0\rangle$ and $|1\rangle$ denote the
states of the cavity with 0 and 1 photons respectively. The ket
$|g\rangle$ denotes the state in which all the atoms are in their
ground states, and $|e\rangle$ denotes the symmetric state in
which one and only one of the atoms is in the excited state. The
parameter $\lambda$ is determined by the intensity of the laser
driving the cavity mode.

The state of the system is changed upon the detection of a photon leaving the cavity. The state at a time $\tau$ after a photon is detected is given by
\begin{equation}
    |\psi_{c}\left( \tau \right) \rangle=|0,g\rangle+\lambda\left[\xi\left( \tau \right) |1,g\rangle - \theta \left( \tau \right)\frac{2g\sqrt{N}}{\gamma} |0,e\rangle \right],
\end{equation}
where $\xi$ and $\theta$ are functions that oscillate at the Rabi
frequency. We now note that when $\tau$ is such that $\xi\left(
\tau \right) = \theta\left( \tau \right)$, then the conditioned
state $|\psi_{c}\left( \tau \right) \rangle$ is precisely the
steady-state $|\psi_{ss}\rangle$, but with a different value of
$\lambda$. Thus if we suddenly switch the intensity of the driving
field by the right amount, the conditioned state \textit{becomes}
the new steady-state, and the evolution of the system is frozen
until we change the driving laser back to its original value.
Switching the driving laser so as to freeze the evolution is a
feedback process, because one must first detect a single photon
emitted by the cavity, and perform the switch based on this
detection. The experiment by Smith \textit{et al.} did just that,
and read out the change in the evolution of the system by
measuring the intensity auto-correlation function of the light
output from the cavity.

Referring again to Fig.~\ref{Fig of the quantum state capture and release by quantum feedback}, the central component of the experiment is an optical cavity composed of two high-reflectivity curved mirrors with separation $l=880$ $\mu$m. A thermal beam of Rb$^{85}$ atoms is produced by an effusive oven
heated to $440$ K. The cavity field is driven by an Ar$^+$-pumped titanium sapphire (Ti:sapphire) laser
which excites the Rb$^{85}$ transition between the states $5S_{1/2}$, $F=3$
and $5P_{3/2}$, $F=4$. The coupling strength between the atom and
the cavity, the decay rate of the cavity, and the decay rate of
the atom are $\left( g, \kappa, \gamma/2\right)/2\pi=\left(5.1,3.7,3.0\right)$ MHz. The output field from
the cavity is split by a beam-splitter and detected by two
(``start" and ``stop") Avalanche Photodiodes (APDs). The output
signal from the start detector is also split into two branches. One branch enters the start channel and is fed into a time-to-digital convertor (TDC) used to measure the second-order
correlation function $g^{\left( 2 \right)} \left( \tau \right) $, and the other is sent to a delay generator and then
fed back to control the strength of the driving laser using an electro-optical modulator.
The delay between the detection of the single photon emitted from the cavity, and the switch of the power of the driving laser is just $45$ ns.

\begin{figure}
\centerline{\includegraphics[width =16.8
cm]{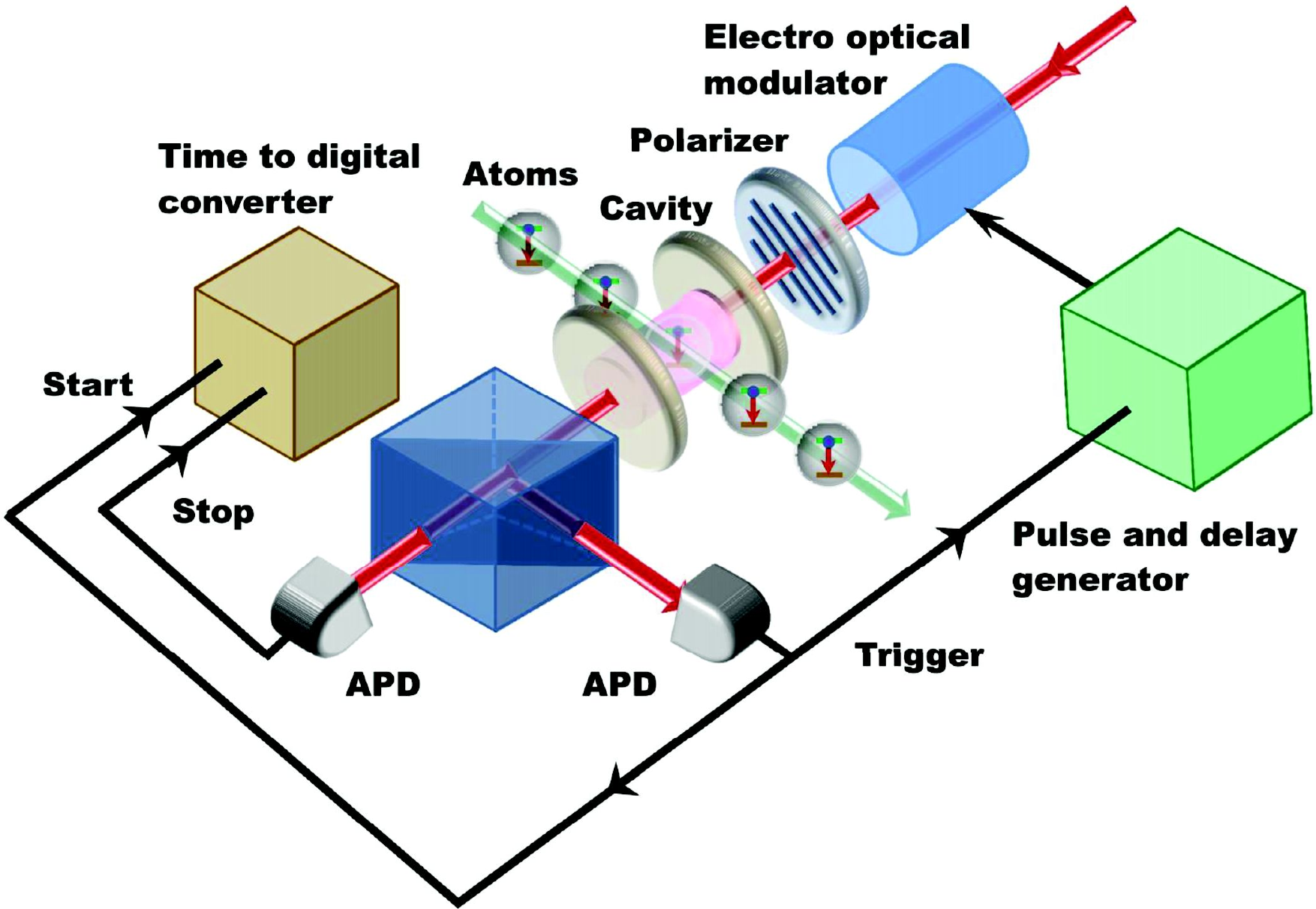}} \caption{(Color online). Diagram of the
experimental setup of Smith \textit{et al.}~\cite{WPSmithPRL:2002}
to capture and release quantum state by feedback. Here, APD
denotes an avalanche photodiode.}\label{Fig of the quantum state
capture and release by quantum feedback}
\end{figure}

\textit{Preparation of Fock-states (2011):} The experiment of
Sayrin \textit{et al.}~\cite{CSayrinNature:2011} uses feedback
based on a sequence of weak measurements to prepare a single
cavity mode in a Fock state (a state with a precise number of
photons and no phase).  Each weak measurement is made by passing a
two-level atom through the cavity and then measuring the internal
state of the atom. This state is measured using a field-ionization
detector which gives a detection efficiency of $35\%$. The
resulting measurement on the cavity mode is given by the
measurement operators
\begin{equation}
M_e=\cos\left[\frac{1}{2}\left(\phi_r+\phi_0\left(a^\dagger a+\frac{1}{2}\right)\right)\right],\quad
M_g=\sin\left[\frac{1}{2}\left(\phi_r+\phi_0\left(a^\dagger a +\frac{1}{2}\right)\right)\right],
\end{equation}
where the phase $\phi_0$ is determined by the initial state of the
atom and the effective basis in which the atomic state is
measured. The operator $a$ is the annihilation operator of the
cavity mode. The initial state of the atom is set by an
interaction with an auxiliary cavity, the effective basis in which
the atomic state is measured is set by a second auxiliary cavity.
Since the measurement operators commute with the number operator
$a^\dagger a$, each measurement provides information about the
number of photons without disturbing it. After each measurement,
the state of the cavity mode is calculated, and the result is used
to modify the amplitude of the coherent light driving the cavity.
This coherent light by itself cannot create a Fock state, and will
in general take the mode further from a Fock state. The role of
the coherent light is merely to help keep the average number of
photons in the cavity constant, while it is the job of the
measurement to narrow the distribution of photons towards a Fock
state.

A diagram of the experiment is shown in Fig.~\ref{Fig of the preparation of Fock state by quantum feedback}, in which $R_1$ and $R_2$ label the auxiliary cavities, each of which performs a chosen unitary operation on the internal state of the atoms. The central component is the microwave cavity $C$, for which the mode in question has a frequency of $51$ GHz and a damping time of $T_{\mss{c}}=65$ ms. The feedback delay in each round is approximately $83$ ns. The experiment was able to stabilize Fock states up to $n=4$. The time taken by the feedback loop to converge to a steady-state with $n=3$ was $50$ ms, which was $5$ times faster than that resulting from an optimized trial-and-error projection method using the same apparatus.

It would clearly be an improvement over the above experiment if
the feedback was able to add or subtract a photon from the cavity,
instead of merely shifting the state in phase-space, because this
kind of feedback would not degrade Fock states. In fact, the atoms
that are passed through the cavity provide a means to do just
this, and a feedback protocol using this method was demonstrated
by Zhou \textit{et al.}~\cite{XZhouPRL:2012}. The experimental
setup is very similar to that in Fig.~\ref{Fig of the preparation
of Fock state by quantum feedback}, with the only difference that
the result of the measurement now controls the initial state of
the atoms prior to entering the central cavity; the atom beam
works both as a quantum non-demolition probe and a means to add or
subtract a single photon to the cavity mode. In this work, the
authors demonstrated the stabilization of Fock states with up to
seven photons.

\begin{figure}
\centerline{\includegraphics[width =16.8
cm]{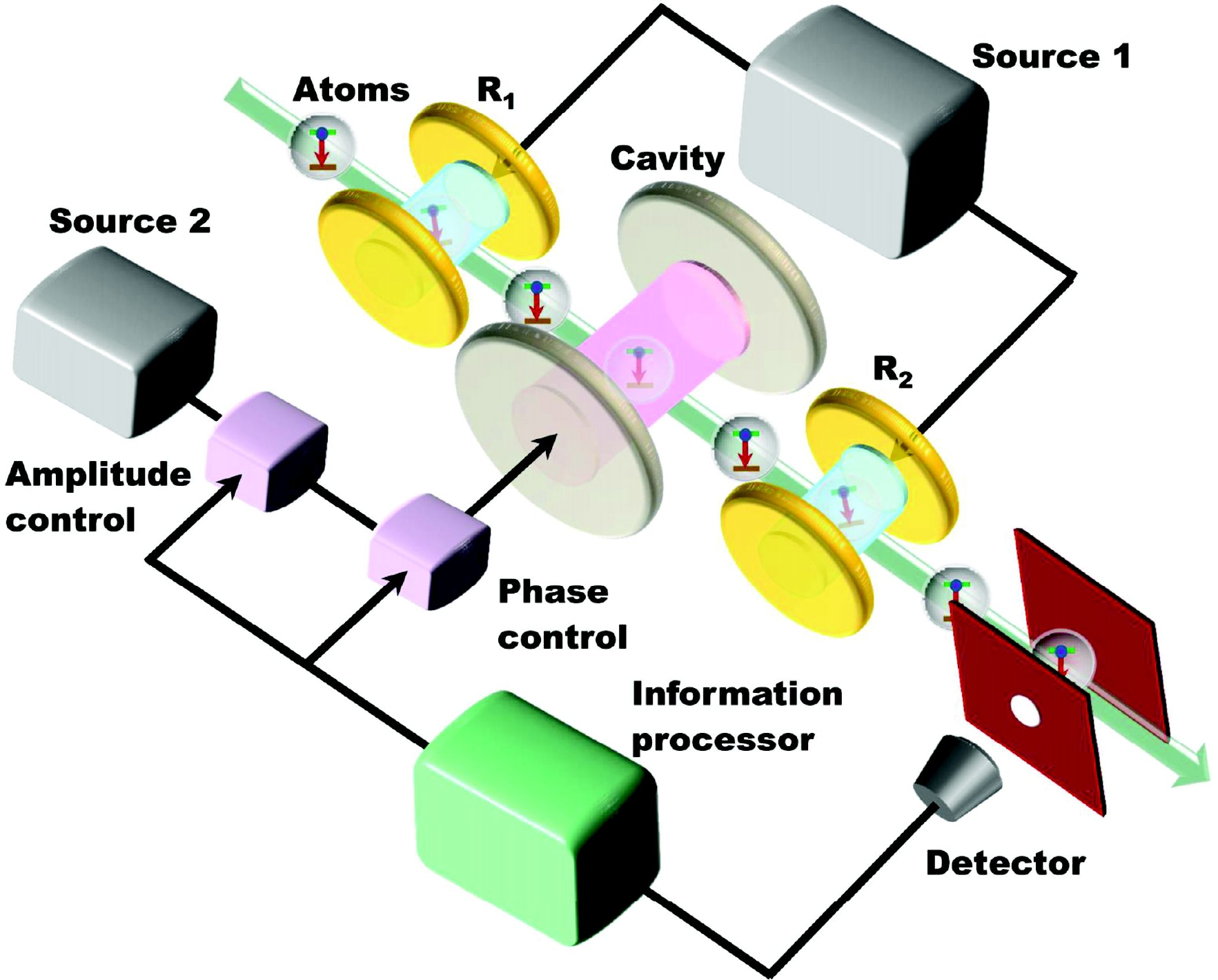}} \caption{(Color online) Diagram of
the experimental set-up of Sayrin \textit{et
al.}~\cite{CSayrinNature:2011}. The state of the cavity mode $C$
is weakly measured by an atomic beam. The atoms in this beam are
resonant with the two cavities $R_1$ and $R_2$, which form an
``atomic Ramsey interferometer'' by setting the initial state in
which the atoms are prepared before entering the cavity, and the
effective basis in which the atomic state is measured by the
field-ionization detector. The output of the measurement is fed
into the information processor to execute the state-estimation and
determine the phase and intensity of the laser that drives the
cavity mode, source 2. }\label{Fig of the preparation of Fock
state by quantum feedback}
\end{figure}


\textit{Classical feedback control of atomic states (2012,2013):} Here we examine some very nice experiments that push the boundaries of feedback control in quantum systems, but are nevertheless not examples of quantum feedback, under our definition. While the present review concerns solely quantum feedback, the experiments here are instructive in helping us to explain more clearly what constitutes quantum feedback.

Recall that out definition of measurement-based quantum feedback is that quantum measurement theory is required to correctly describe it. This means that the change in the state of the system caused by the measurement is (i) important in describing the behavior under the feedback loop, and (ii) that this change is not the same as the change that would be predicted by Bayesian inference. Condition (ii) is often stated by saying that the measurement induces ``quantum back-action''.

We now describe two ways in which a measurement on a quantum
system can be effectively classical, meaning that the measurement
has no dynamical effect on the system, and can therefore be
described purely by Bayesian inference. The first of these is when
the observable being measured commutes at every time with the
density matrix of the system. In this case, the state of the
system is merely a classical probability distribution of the
eigenstates of the observable. Since the measurement does not
disturb these eigenstates, the only change is to the probability
distribution, and since this distribution is classical, this
change obeys Bayesian inference. An experiment that uses this kind
of measurement to control the populations of atomic energy levels
is that by Brakhane \textit{et al.}~\cite{SBrakhanePRL:2012}.
Because the transitions between the levels are incoherent, the
density matrix remains diagonal in the same basis as the
measurement operators, and the feedback is classical.

The second way in which a quantum measurement can avoid
back-action is if the controller has many identical systems that
are all in the same state, $|\psi\rangle$, and undergoing the same
evolution. In this case, the controller can learn $|\psi\rangle$
by extracting only a very tiny amount of information from each
system. As a result, the quantum state $|\psi\rangle$ has been
transformed into a set classical parameters that can be measured
without disturbing them. Feedback that uses this kind of
measurement process to control the internal state of an (ensemble
of) atoms is realized in the experiments by Vanderbruggen
\textit{et al.}~\cite{TVanderbruggenPRL:2013} and Inoue \textit{et
al.}~\cite{RInouePRL:2013}.

\textit{Controlling the motion of trapped particles (2002 --
2009):} Feedback has now been applied to cooling the motion of
single trapped ions~\cite{PBushevPRL:2006}
electrons~\cite{BDUrsoPRL:2003},
nanoparticles~\cite{JGieselerPRL:2012}, and
atoms~\cite{AKubanekNature:2009,TFischerPRL:2002,
MVMorrowPRL:2002}. Only one of these, however, has reached the
quantum domain, the trapped-ion experiment by Bushev \textit{et
al.} The experiment of Morrow~\textit{et al.\ } uses an ensemble
of atoms in identical states, and thus an effectively classical
measurement. Experiments~\cite{BDUrsoPRL:2003, JGieselerPRL:2012,
AKubanekNature:2009,TFischerPRL:2002} make measurements on single
particles, but the effect of the back-action is negligible.
Nevertheless these experiments are interesting as they represent
the state-of-the art in applying feedback control to microscopic
systems. We discuss that of Kubanek~\textit{et al.\ } below.

In Fig.~\ref{Fig of the atomic cooling2006}, we give a diagram of
the experimental configuration used by Bushev \textit{et al.\
}~\cite{PBushevPRL:2006} to cool the external motion of a single
ion in a Paul trap (a 3D quadrupole ion trap). In this experiment,
a single trapped $^{138}$Ba$^{+}$ ion is first cooled to the
doppler limit using laser cooling. The position of the ion along
the $x$-axis is measured by an interferometer, in which the light
reflected from the mirror interfered with the light scattered by
the ion. The signal is filtered and phase shifted, and then
applied directly back to control the voltage on the electrode to
produce a damping force. Thus the experiment implements
essentially Markovian feedback. The authors were able to cool the
ion to $30\%$ below the Doppler limit to yield a steady-state
average phonon number of $12$.

\begin{figure}
\centerline{\includegraphics[width = 16.8
cm]{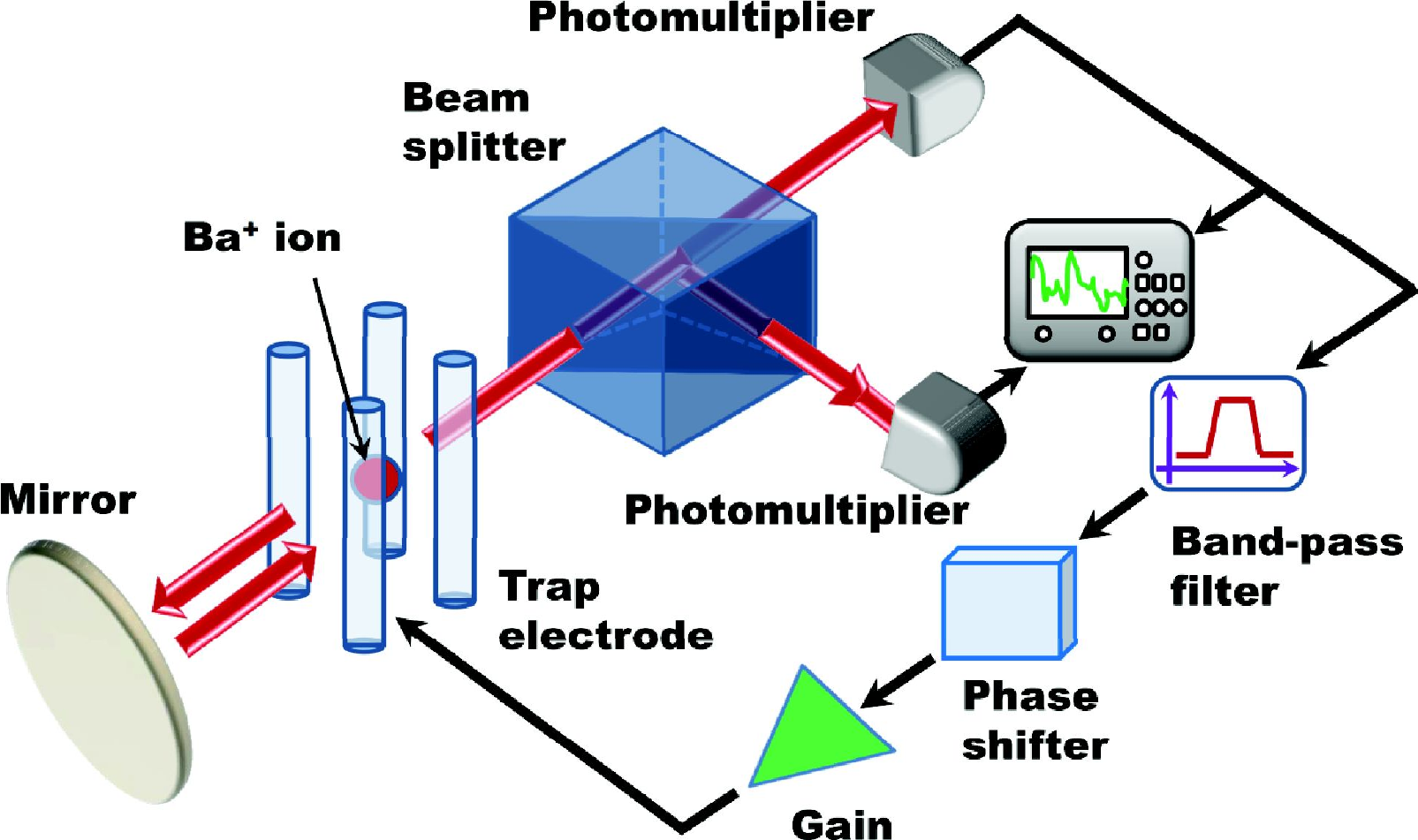}} \caption{(Color online) Diagram of
the experimental setup of Bushev \textit{et
al.}~\cite{PBushevPRL:2006}. A $^{138}$Ba$^{+}$ ion is captured by
a Paul trap and laser cooled to the Doppler limit. The position of
the ion in the trap is measure by a probe beam using an
interferometer: the light that is scattered back by the ion is
interfered with the light that is reflected back by the mirror,
which provides interference fringes with a contrast of $73\%$. The
$1$ MHz oscillation of the ion produces sidebands in the
photomultiplier signal, which is then filtered and phase-shifted
and used to control the voltage on the trap electrodes to create a
damping force. Photomultiplier 2 is not part of the feedback loop,
and is used to monitor the motion of the ion to verify the
effectiveness of the feedback.}\label{Fig of the atomic
cooling2006}
\end{figure}

\begin{figure} \centerline{\includegraphics[width = 16.8
cm]{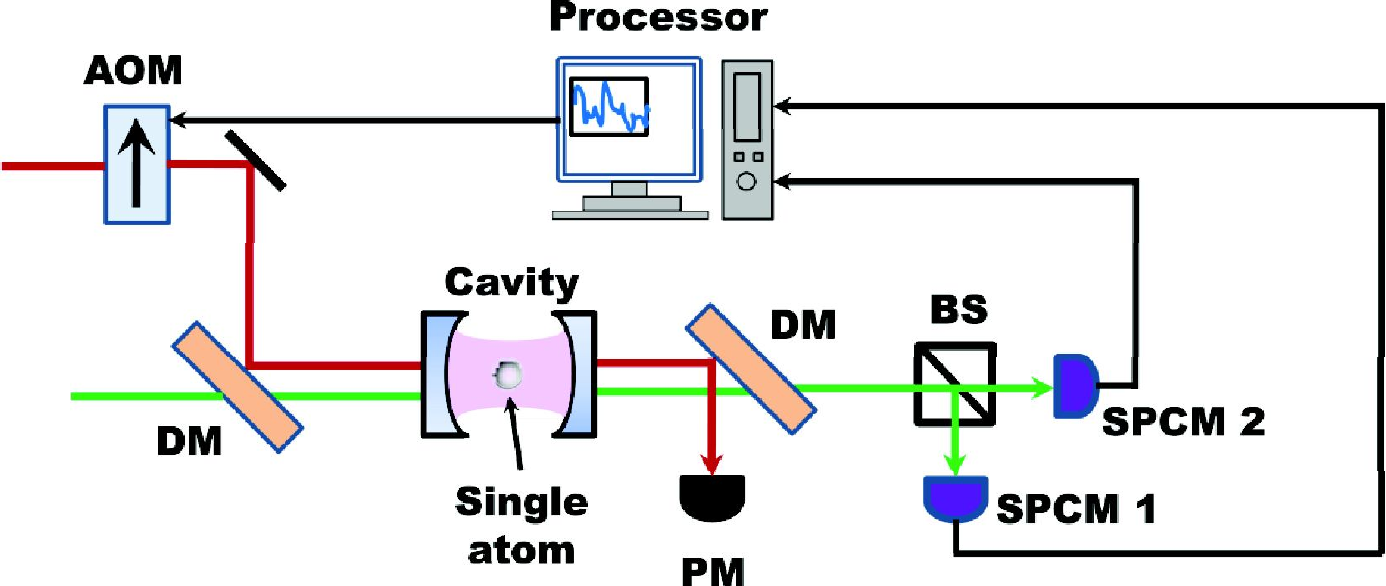}} \caption{(Color online).
Experimental setup of Kubanek \textit{et
al.}~\cite{AKubanekNature:2009}. Both a blue-detuned toroidal
beam, which acts as an actuator to trap a $^{85}$Rb atom, and a
weak probe beam, used to detect the motion of the atom, is fed
into the cavity. The blue-detuned control light is then detected
by a photomultiplier (PM), and the probe light is detected by two
single-photon counting modules (SPCM 1 and SPCM 2). One of the
output signals from the probe light is fed into a processor to
generate a control signal that is used to switch the intensity of
the control beam.}\label{Fig of the atomic cooling2009}
\end{figure}

In Fig.~\ref{Fig of the atomic cooling2009}, we give a diagram of
the experimental configuration used by Kubanek \textit{et
al.}~\cite{AKubanekNature:2009} to cool the external motion of a
single atom trapped by the light in an optical cavity. In this
experiment, an $^{85}$Rb atom is trapped by a $775$ nm toroidal
mode in a high-finesse cavity. The mode is blue-detuned from one
of the atomic transitions, and the Stark shift of this transition,
being dependent on the local intensity of the cavity mode, creates
the trapping potential for the atom. The atom-cavity coupling
strength is $g_0/2\pi=16$ MHz, the atomic decay rate is
$\gamma/2\pi=3$ MHz, and the cavity damping rate of
$\kappa/2\pi=1.25$ MHz. The position of the atom is detected by a
weak probe beam whose transmitted power depends on the Stark shift
of the atomic transition, and thus on the atomic position. The
probe beam is split so that it can be measured independently by
two photon detectors (SPCM 1 and SPCM 2). As in the experiment
above, the signal from one of the detectors is used in the
feedback loop, and the other is used to monitor the position of
the atom and thus the performance of the feedback control.

The authors used one of the control protocols discussed
in~\cite{DASteckPRL:2004, DASteckPRA:2006}, in which the height of
the potential is switched between a high value and a low value (a
``bang-bang'' strategy) to reduce the kinetic energy of the atom.
The potential is switched high when the atom is traveling away
from the center, and low when the atom is traveling towards the
center. The switching is implemented with an acousto-optic
modulator (AOM), which switches the beam power from $50$ nW to
$800$ nW. The feedback delay, which includes a decision-making
time of $1.7$ $\mu$s and a switching-process delay of $3$ $\mu$s,
is much smaller than the oscillation time of the atom in the
potential, which is  $360$ $\mu$s. The authors were able to use
the control protocol to increase the storage time of the atom in
the toroidal beam by a factor of four, and to reduce the amplitude
of oscillation of the atom by a factor of two.

\subsection{Superconducting circuits}\label{s54}

\subsubsection{Measurement-based feedback}

\textit{Stabilizing the state of a single qubit (2012):} It is only very recently that measurement-based feedback has been realized in mesoscopic circuits, because it is only recently that amplifiers have been devised with that have sufficiently low noise~\cite{Beltran07, Yamamoto08, Bergeal10b}. The first two experiments to demonstrate measurement-based feedback control of a single superconducting qubit were those by Vijay \textit{et al.}~\cite{RVijayNature:2012} and Riste \textit{et al.}~\cite{DRistePRL:2012}.

In Fig.~\ref{Fig of quantum continuous-measurement-based feedback
in superconducting systems}, we show the experimental setup used
by Vijay \textit{et al.}. In this experiment, the authors use a
continuous measurement and feedback process to keep the qubit
undergoing Rabi oscillations under its free Hamiltonian. To do
this, the feedback has to continually purify the state of the
qubit and the feedback has to keep it within a given plane of the
Bloch sphere. The qubit is a capacitively shunted Josephson
junction~\cite{JQYouPRB:2007, JKochPRA:2007} with a transition
frequency of $\omega_q/2\pi=5.4853$ GHz. This qubit is
dispersively coupled to a three-dimensional microwave cavity with
cavity resonant frequency of $\omega_{\mss{c}}/2\pi=7.2756$ GHz.
Electrical signals that enable the measurement and feedback
control are fed into the cavity via the weakly-coupled input port,
and the measurement output leaves the cavity via the
strongly-coupled port with decay rate of $\kappa/2\pi=13.4$ MHz.
The qubit is dispersively coupled to the cavity with a strength of
$\chi/2\pi=0.687$ MHz, with the result that the qubit induces a
phase shift of the cavity light that depends on the qubit's
internal state, and this state is therefore measured by making a
homodyne measurement of the signal output from the cavity. This is
a ``cavity-mediated'' (or ``oscillator-mediated'') measurement as
described in~\cite{AClerkNJP:2008, Jacobs14}.

The cavity mode is driven at the frequency of $\omega_r= 2\pi
\times 7.2749$ GHz $\approx \omega_{\mss{c}}-\chi$ to control the
mean cavity photon occupation which sets the measurement strength.
The qubit is driven by the ``Rabi drive'' in Fig.~\ref{Fig of
quantum continuous-measurement-based feedback in superconducting
systems} chosen to give a Rabi frequency of $\Omega_R/2\pi=3$ MHz.
At the output port of the cavity, the output quantum field is sent
through two isolators to protect the qubit from the strong field
driving the parametric amplifier (or ``paramp''). It is then fed
through the paramp (a near-noiseless phase-sensitive quantum
amplifier~\cite{CMCavesPRD:1982, AAClerkRMP:2010}), and again
through the High-Electron-Mobility Transistor (HEMT) amplifier to
produce a macroscopic signal that can easily be manipulated
without significant noise. It is then measured by homodyne
detection. This measurement procedure achieves an efficiency of
$\eta=0.40$. The feedback protocol used in the experiment is
motivated by a classical phase-locked loop. As shown in
Fig.~\ref{Fig of quantum continuous-measurement-based feedback in
superconducting systems}, the output of the homodyne detection is
compared with a $3$-MHz Rabi reference signal and
low-pass-filtered to generate a signal proportional to the sine of
the phase difference, $\theta_{\rm err}$, between the reference
and the homodyne output. This error signal is then fed back to
modulate the amplitude of the Rabi drive, $\Omega_{\rm fb}$, by
$\Omega_{\rm fb}/\Omega_R=-F\sin\left(\theta_{\rm err}\right)$,
where $F$ is the dimensionless feedback gain. This is a Markovian
quantum feedback process. The performance is mainly limited by the
efficiency of the measurement and the time delay in the feedback
loop. There is a tradeoff between the rate at which the feedback
can be performed (the feedback bandwidth) and the noise introduced
by the feedback signal, which results in an optimal measurement
strength. With a finite feedback bandwidth of $10$ MHz and loop
delay of $250$ ns, the optimal measurement strength was found to
be $\Gamma_{\phi}/2\pi=0.134$ MHz.

A more recent experiment from Siddiqi's
group~\cite{SJWeberNature:2014} shows that more complex optimal
control design can be introduced to continuously monitoring a
superconducting qubit to evolve from initial to final states along
the optimal route by reconstructing the individual quantum
trajectories via continuous quantum weak measurement.

\textit{Preparing entanglement between two qubits (2013):} The
experiment by Riste \textit{et al.}~\cite{DRisteNature:2013}
demonstrated the stabilization of entanglement between two
superconducting qubits using measurement-based feedback control.
In Ref.~\cite{DRisteNature:2013}, a Bell state is prepared with a
fidelity of $88\%$ by using a parity measurement, which is a joint
measurement on the both qubits. By introducing a feedback loop
incorporating the parity measurement, the probabilistic
preparation of the Bell state could be replaced by a deterministic
preparation with a fidelity of $66\%$.

\begin{figure} \centerline{\includegraphics[width = 16.8 cm]{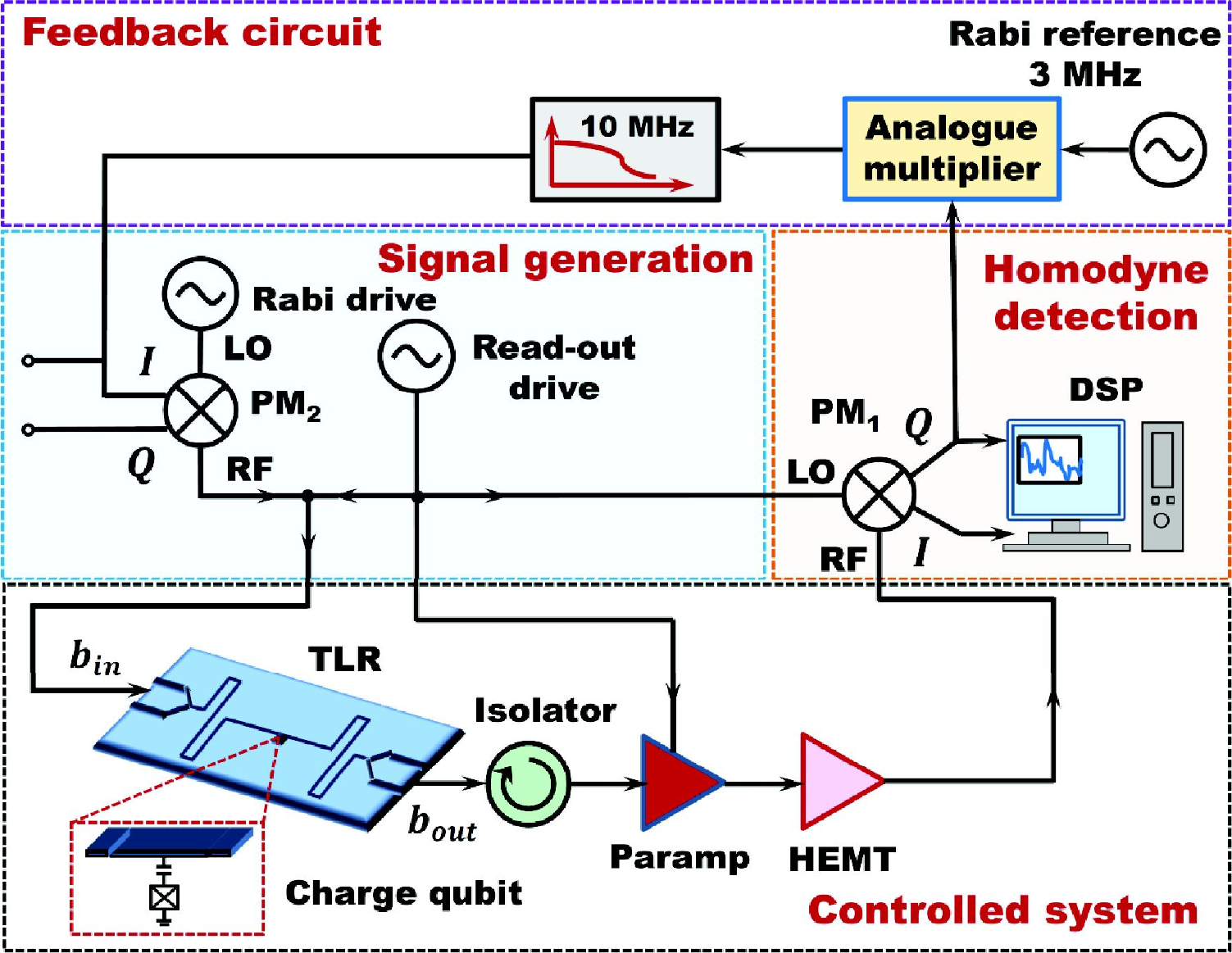}} \caption{(Color online) Diagram of the experimental setup of Vijay \textit{et al.}~\cite{RVijayNature:2012}. The Rabi drive at the ac stark-shifted qubit frequency and the read-out drive at frequency $7.2749$ GHz are both fed into the weakly-coupled input port of a three-dimensional microwave cavity, which is dispersively coupled to a capacitively-shunted Josephson junction working as a superconducting qubit. The output signals leave from the strongly-coupled port of cavity and are then transmitted through two isolators. Afterwards, the output signals
are amplified by a parametric amplifier (paramp) and a
high-electron-mobility transistor (HEMT) amplifier, and then
measured by a homodyne detection setup. The amplified quadrature
$Q$ is then sent to the feedback circuit to be compared to the $3$
MHz Rabi reference signal and filtered by a $10$ MHz
low-pass-filter. The output signal is then fed back to correct the
Rabi frequency imposed on the qubit by the Rabi drive. LO and RF
represent the local oscillator and radio frequency. PM$_1$ and
PM$_2$ are the two photonmultipliers. $I$ and $Q$ are the in-phase
component and quadrature component, respectively.}\label{Fig of
quantum continuous-measurement-based feedback in superconducting
systems}
\end{figure}


\subsubsection{Coherent feedback}

\textit{Engineering dynamics (2012):} Kerckhoff and Lehnert used a coherent feedback network to implement a bistable superconducting circuit, also known as a latch, useful in classical information processing~\cite{JKerckhoffPRL:2012}. This was an experimental realization of a scheme devised by Mabuchi~\cite{HMabuchiAPL:2011}, in which two oscillators with Kerr nonlinearities, when coupled in a loop via a beam-splitter, will generate both bistable and astable dynamics.

\begin{figure} \centerline{\includegraphics[width=16.8 cm]{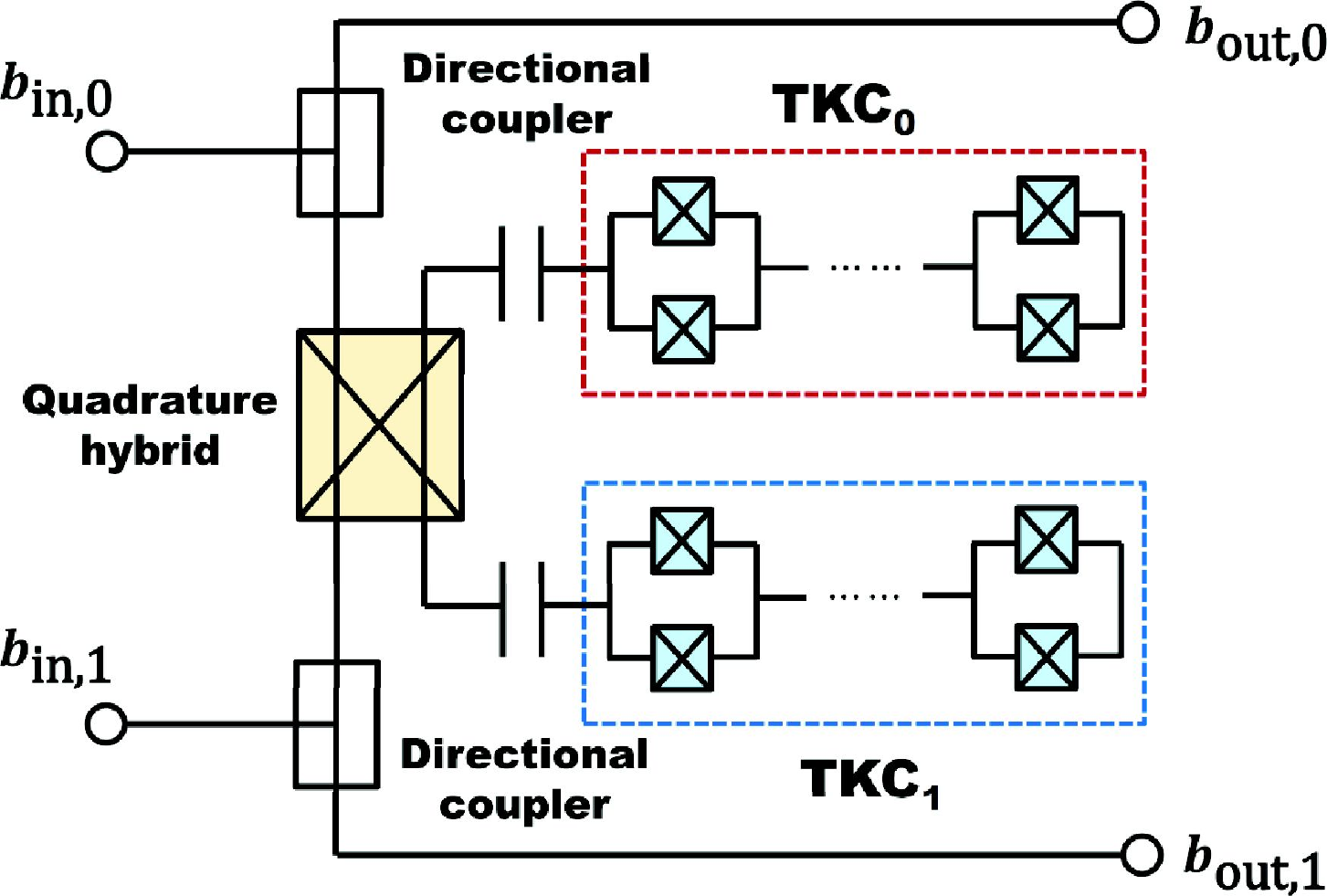}}
\caption{(Color online) Diagram of the two-port coherent feedback circuit used by Kerckhoff and Lehnert~\cite{JKerckhoffPRL:2012}. The input fields are $b_{in,0}$ and
$b_{in,1}$ and the output fields are $b_{out,0}$, $b_{out,1}$. The two
flux-biased tunable Kerr circuits (TKC$_0$, TKC$_1$) are connected
via a beam-splitter which is the ``quadrature hybrid''. Each TKC is composed of a
quarter-wave transmission line resonator interrupted by a series
array of $40$ Josephson junction SQUIDs, and one end of the
transmission line resonator is capacitively coupled to the beam-splitter.
The two directional couplers merely separate the input and output fields. The entire system
is housed in a dilution refrigerator which provides an ambient temperature of $\sim 50$ mK.}
\label{Fig of coherent quantum feedback in superconducting systems}
\end{figure}

A diagram of the experimental configuration of Kerckhoff and Lehnert is shown in
Fig.~\ref{Fig of coherent quantum feedback in superconducting systems}. The system is an input-output circuit with two input fields $b_{in,0}$, $b_{in,1}$ and two
output fields $b_{out,0}$, $b_{out,1}$. The central components of
the coherent feedback loop are two tunable Kerr circuits, TKC$_0$
and TKC$_1$. Each TKC is composed of a quarter-wave transmission
line resonator generated by a coplanar waveguide. In the center of
the coplanar waveguide, a series array of $40$ Josephson Junction
Superconducting Quantum Interference Devices (JJ-SQUIDs) interrupt
the waveguide and generate a Kerr nonlinearity for the
transmission line resonator~\cite{MACastellanos-BeltranNatureP:2008}. One end of each
resonator is capacitively coupled to the $4$-$8$ GHz commercial quadrature hybrid which
acts as a $50\!\!:\!\!50$ microwave beam splitter. The two directional couplers merely
separate the input fields from the output fields.

The above nonlinear coherent feedback network exhibits optical phenomena that neither of the Kerr resonators exhibit by themselves. For example, as shown in
Ref.~\cite{JKerckhoffPRL:2012}, if the two TKCs have
central frequencies equal to $\omega_0/2\pi=6.408$ GHz, and we drive them with
probe fields at the frequency $\omega_p/2\pi=6.39$ GHz, the
output fields of the coherent feedback network exhibit bistable
phenomena. However, if two uncoupled TKCs are individually driven at the
same detuning, nether would be bistable. Since the TKCs and the feedback control circuits typically
contain an average of about $1000$ photons, the experimental results
fit very well with a mean-field model using a semiclassical approximation. Purely quantum effects such as sub-Poisson statistics could potentially also be observed in this feedback circuit.

\textit{Controlling qubits with a cavity controller (2012, 2013):}
The experiment of Shankar \textit{et
al.}~\cite{SShankarNature:2013} demonstrated the use of a
continuous coherent feedback process to maintain two
superconducting circuits in an entangled state. The system
consists of the two qubits and a superconducting cavity that is
used as the feedback controller. The qubits are coupled
dispersively to the cavity, so that the states of the qubits shift
the energy of the states of the cavity, and vice-versa. This means
that many of the joint states of the qubits and cavity are
distinguished by their energy, so that a joint Hamiltonian can be
engineered by driving the system with signals that will drive
selected transitions. The authors apply six driving fields that
implement concurrently the two classic parts of coherent
feedback~\cite{SLloydPRA:2000, Jacobs14c, Jacobs14}. The first
correlates the qubits with the cavity, so that the cavity acts as
a measuring device for the qubits. The second applies a different
unitary operation to the qubits for each of the relevant cavity
states, thus applying an action that is equivalent to an operation
conditional on the result of a measurement. The cavity is
continually reset to its ground state via its own damping. The end
result can be thought of as sideband cooling, in which the
entangled state of the qubits is the ``ground state'' to which the
qubits are ``cooled''. To further examples of coherent feedback in
which a cavity is used to prepare a qubit in a pure state are
given in~\cite{KWMurchPRL:2012, KGeerlingsPRL:2013}.

\subsection{Opto-mechanics and electro-mechanics}\label{s53}

\subsubsection{Measurement-based feedback}

\begin{figure} \centerline{\includegraphics[width =16.8 cm]{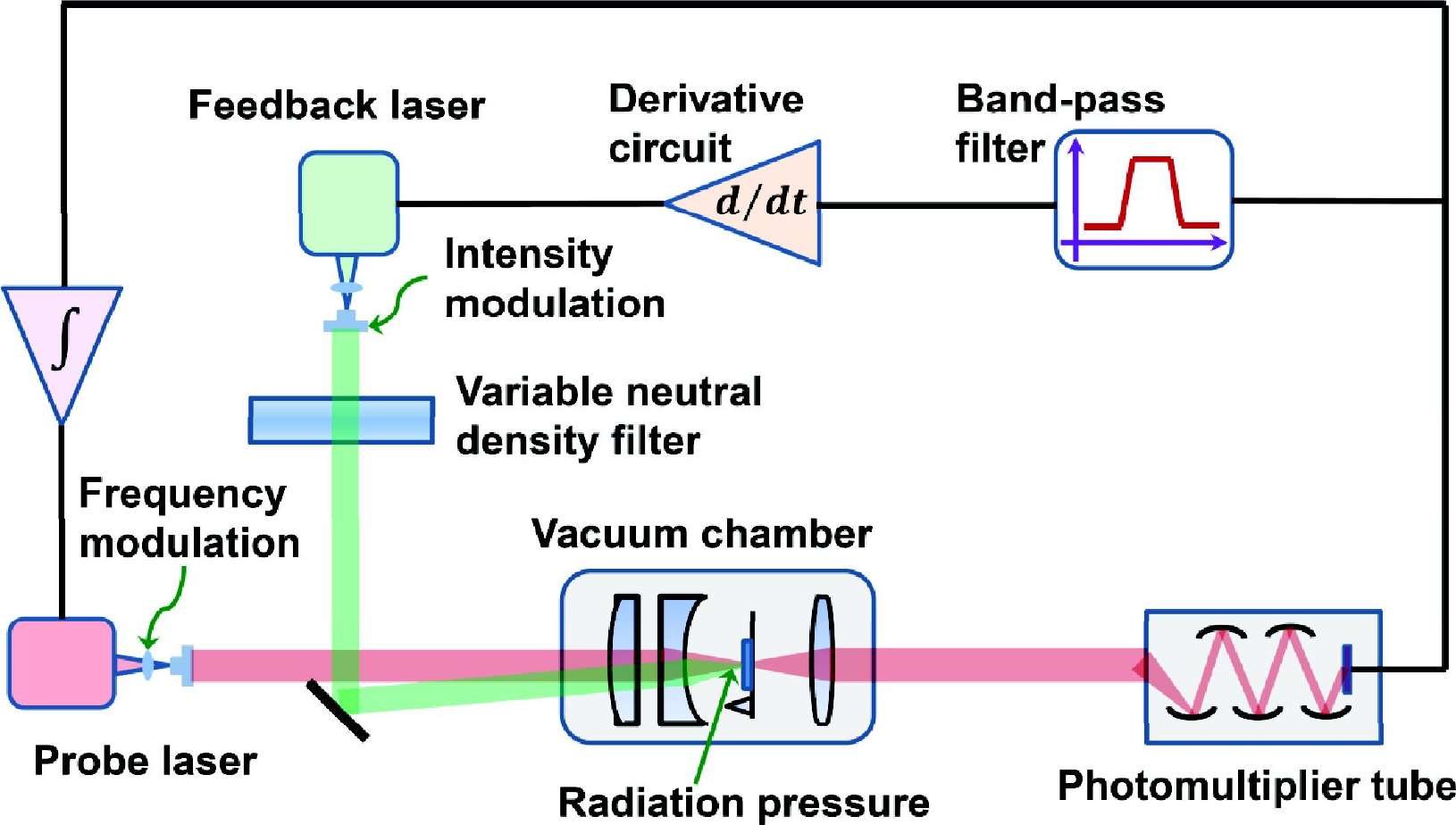}}
\caption{(Color online) Diagram of the experimental setup for measurement-based feedback cooling by Kleckner and Bouwmeester~\cite{DKlecknerNature:2006}. The $780$ nm probe laser is frequency-locked to the optical cavity by the integrated signal emanating from the photo-multiplier tube. The output signal of the photo-multiplier is also sent through a bandpass filter, and then the time derivative of the output signal is used to modulate the amplitude of the $980$ nm feedback laser, which is attenuated with a variable neutral density filter. The feedback laser provides a counteracting radiation pressure, which effectively cools the mechanical mode of the cantilever.}\label{Fig of active feedback cooling of nanomechanical beam}
\end{figure}

\textit{Cooling macroscopic mechanical resonators (1999 - 2012):} A number of experiments have demonstrated cooling of macroscopic mechanical resonators using measurement feedback, but this approach to cooling has not yet reached the quantum regime~\cite{PFCohadonPRL:1999, OArcizetPRL:2006, DKlecknerNature:2006,  TCorbittPRL:2007, Poggio07, EGavartinNaturenanotechnology:2012}. We show the experimental setup used by Kleckner and Bouwmeester~\cite{DKlecknerNature:2006} in Fig.~\ref{Fig of active feedback cooling of nanomechanical beam}, in which the mechanical resonator is a cantilever which is attached to a mirror that forms one end of an optical cavity. The cantilever has dimensions $450\times50\times2\mu$m, a mass of $\left(2.4\pm0.2\right)\times10^{-11}$ kg, a frequency of $12.5$ kHz, and a quality
factor of $Q = 137000$. The spring constant of the cantilever is $0.15\pm0.01$ Nm$^{-1}$, and as a result the rms amplitude of the oscillations of the cantilever due to thermal noise is $\left(1.2\pm0.1\right)\times 10^{-10}$ m.

The cantilever is driven by a $780$ nm probe laser which is locked
to the optical cavity by an external integrating circuit that uses
the signal from a photo-multiplier tube. The output signal of the
photo-multiplier is also fed into a $12.5$ kHz bandpass filter.
The time derivative of the output signal gives an estimate of the
velocity of the cantilever, and this is used to modulate the
amplitude of a diode laser that applies the feedback force to the
cantilever. The authors were able to cool the cantilever from room
temperature to $135$ mK. This indicates that if their cantilever
was at the temperature of a dilution refrigerator, they would be
able to get close to the ground state. Of course, at that point,
the back-action noise and any other sources of classical noise in
the feedback system might provide further obstacles to reaching
the ground state.

\begin{figure} \centerline{\includegraphics[width = 16.8 cm]{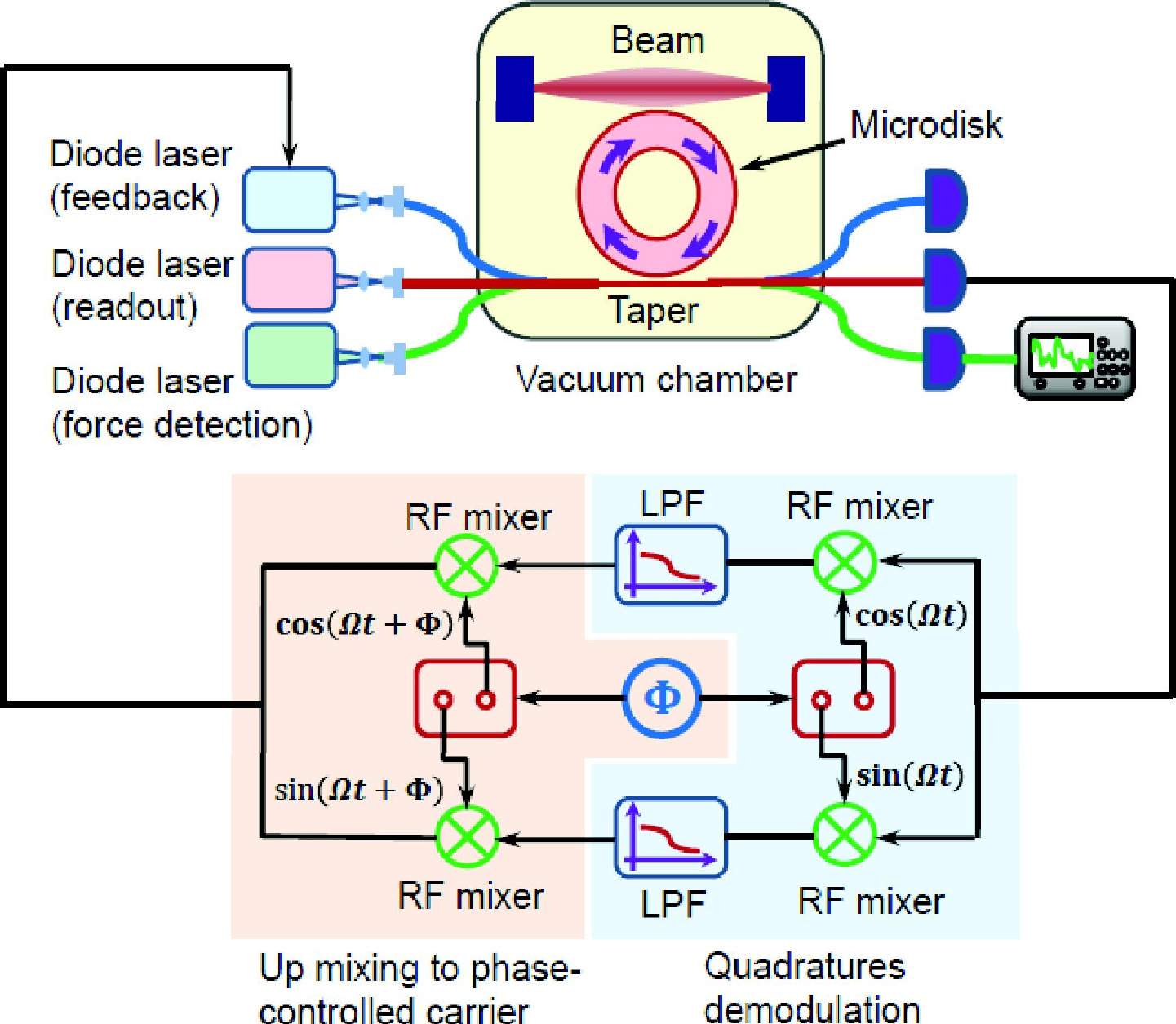}}
\caption{(Color online) Diagram of the experimental setup of the on-chip optomechanical feedback system of Gavartin \textit{et al.}~\cite{EGavartinNaturenanotechnology:2012}. A doubly-clamped nano-beam is coupled to a silica microdisk, which is then coupled to a tapered fiber. A readout laser is fed into the taper to observe the mechanical mode of the beam and fed back to modulate a second laser beam that uses radiation pressure to apply a force to the mechanics.}
\label{Fig of the beam cooling2012}
\end{figure}

As another example of feedback cooling of a macroscopic mechanical
resonator, we consider the experiment by Gavartin \textit{et
al.}~\cite{EGavartinNaturenanotechnology:2012}. We show the setup
for this experiment in Fig.~\ref{Fig of the beam cooling2012}.
Here the resonator is a doubly clamped nano-beam of Si$_3$N$_4$
that has a resonance frequency of $\Omega_M/2\pi=2.88$ MHz. The
effective mass of the beam is $m_{\rm eff}=9\times 10^{-15}$ kg,
and its quality factor is $Q = 4.8\times 10^5$. The beam is placed
in the evanescent field of a toroidal ``microdisk'' optical
cavity. The toroidal cavity has a damping rate of $\gamma/2\pi=3$
GHz and a quality factor $Q=65000$. The movement of the mechanical
resonator changes the frequency of the toroidal cavity mode, an
effect which is described as an ``optomechanical'' coupling. In
this case, the optomechanical coupling strength is
$G/2\pi=d\omega/dx=2.9$ MHz nm$^{-1}$. The phase of the light that
is output from the toroidal cavity provides a readout of the
motion of the mechanical resonator. Light at the telecom
wavelength of $1550$ nm and a power of $400\,\mu$W is fed into the
cavity to measure the mechanical motion, and a second laser to
apply the feedback force. This feedback force was chosen to be a
phase-shifted version of the position, which is one way to obtain
a force that is proportional to the momentum. To obtain the
position spectrum, the authors used a demodulation technique
described by Poot \textit{et al.}~\cite{Poot11}, which is not
limited by the bandwidth of the digital signal processor, and can
therefore be applied to cooling resonators in MHz and even GHz
range. The authors were able to cool the resonator to $0.7$ K. We
note finally that, while the nominal purpose of this experiment,
as described by the authors, was to improve force detection, in
fact it has been shown that force detection cannot be improved by
any linear feedback applied to a
resonator~\cite{AVinanteNaturenanotechnology:2012,
GIHarrisPRL:2013,APontinPRA:2014}.


\subsubsection{Coherent feedback}

\textit{Cooling mechanical resonators (2006-2011):} Cooling of
mechanical resonators using ``resolved-sideband cooling'', a
coherent feedback method, has been realized
experimentally~\cite{OArcizetNature:2006, SGiganNature:2006,
Schliesser08, Schliesser09, Teufel11b, Chan11}. Not only that, but
this method has already achieved cooling to a mean energy below a
single phonon, in experiments performed by Teufel \textit{et al.}
and Chan~\textit{et al.} in 2011. In resolved-sideband cooling,
the mechanical resonator is coupled to an ``auxiliary'' resonator
that may be electrical or optical. The frequency of the auxiliary
is high enough that it sits in its ground state at the ambient
temperature. The coupling is modulated at the frequency difference
between the mechanical and auxiliary resonators, which provides
the energy required to convert the mechanical quanta to electrical
or optical quanta, and vice versa. The interaction therefore
transfers energy and entropy between the two resonators, and since
the auxiliary has neither, energy is transferred out of the
mechanics. So long as the damping rate of the auxiliary is
sufficiently fast, this energy is removed from the auxiliary
quickly so that energy can continue to be sucked out of the
mechanics.

In Fig.~\ref{Fig of self-cooling nanomechanical beam}, we show the
experimental configuration used by Arcizet \textit{et
al.}~\cite{OArcizetNature:2006}). The central component of the
experiment is an optical cavity formed by two mirrors, in which
the back mirror is a coating on the surface of a doubly-clamped
silicon beam with dimensions $1$ mm $\times\,1$ mm $\times\,60$
$\mu$m. The mechanical oscillator to be cooled is chosen to be the
mode of the silicon beam with frequency of $\Omega_m/2\pi=814$
kHz, effective mass $m_{\rm eff}=190~\mu$g, spring constant
$k=5\times10^6$ Nm$^{-1}$, and mechanical quality factor
$Q=10,000$. When the silicon beam vibrates, it changes the length
of the optical cavity, and it is this that couples the optical
modes in the cavity to the mechanical modes. The cavity mode
chosen to be the auxiliary resonator has a frequency of $300$ THz
and a damping rate $\gamma = 2\pi\times 1.05$ MHz. The whole
optomechanical system is put into a vacuum chamber at
room-temperature, with a residual pressure below $10^{-2}$ mbar.
The amplitude of of the mechanical motion is on the order of
$10^{-13}$ m.

The optical cavity is driven by a $1.064 \mu$m Nd:YAG laser stabilized by a voltage-driven optical attenuator with a feedback bandwidth of $1$ kHz. The laser is locked by a triangular
cavity with a bandwidth of $200$ kHz and an optical finesse of $10^4$, introduced to control the shape of the beam and reduce noise. The Pound-Drever-Hall scheme (also called phase-modulation detection~\cite{Jacobs99}) is used to measure the phase of the light that is output from the cavity, and this provides a very sensitive measurement of the position of the mechanical resonator ($\sim 4\times 10^{-19}~\mbox{m}/\sqrt{Hz}$).  The low-frequency component of the phase measurement signal provides information on the drift of the cavity mode from the laser, and so is used to lock the cavity at the desired detuning. The authors were able to cool the mode of the silicon beam from $300$ K to an effective temperature of $10$ K.

\begin{figure} \centerline{\includegraphics[width =
14 cm]{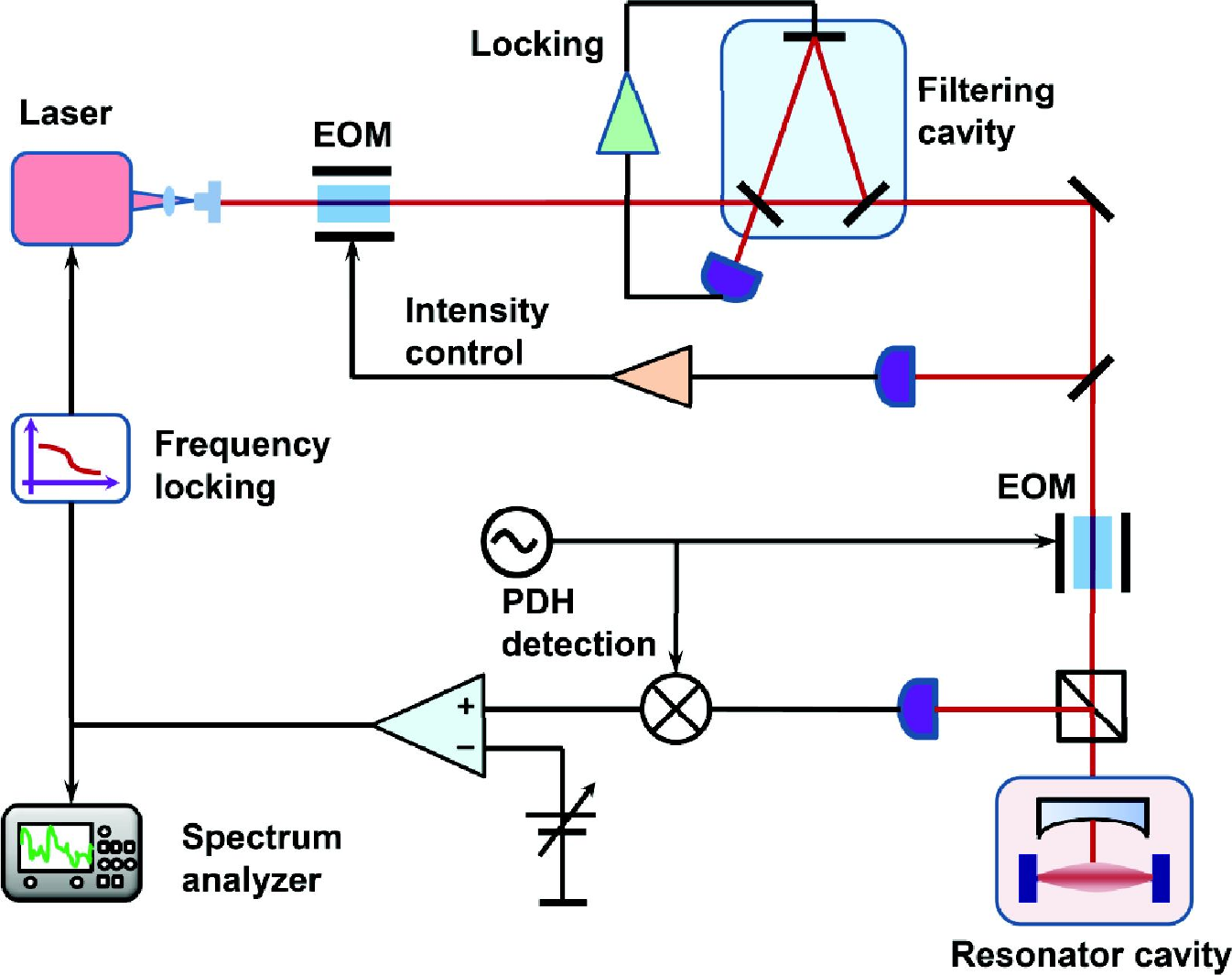}} \caption{(Color
online) Diagram of the experimental realization of
resolved-sideband cooling by Arcizet \textit{et
al.}~\cite{OArcizetNature:2006}.  The optical cavity containing
the mechanical resonator is driven by a $1064$ nm Nd:YAG laser
which is stabilized by an electro-optic modulator and a triangle
filtering cavity. The motion of the mechanics is monitored by
measuring the phase of the light output from the cavity using the
Pound-Drever-Hall (phase-modulation) method.}\label{Fig of
self-cooling nanomechanical beam}
\end{figure}

\textit{Tailoring electro-mechanical systems (2013):} Coherent
quantum feedback has been used by Kerckhoff~\textit{et al.} to
realize a circuit in which a mechanical oscillator is coupled to a
superconducting oscillator, and in which the latter has a tunable
damping rate~\cite{JKerckhoffPRX:2013}. Further, this damping rate
can be switched in less than 100 $\mu$s, much faster than the
one-second damping time of the mechanical oscillator. Since higher
damping rates are better for measuring the mechanical motion, and
lower damping rates better for controlling the mechanics, the
circuit of Kerckhoff~\textit{et al.} is an improvement over
existing technology. In their circuit, the superconducting
oscillator is a lumped-element LC circuit with frequency $4.672$
GHz, output coupling rate of $\kappa_{\mss{c}}/2\pi=2.7$ MHz, and
internal damping rate of $\kappa_l/2\pi=0.1$ MHz. The output of
the LC circuit is fed back to it via a Josephson parametric
amplifier (the ``controller''), which is composed of a $20$-dB
directional coupler and a single-port tunable Kerr circuit with an
output coupling rate of $\gamma/2\pi=50$ MHz and an internal loss
rate of $\gamma_l/2\pi=5$ MHz. This feedback allows the effective
output coupling rate of the LC circuit to be tuned from $300$ kHz
to 3 MHz.


\subsection{Self feedback in quantum dots}
\label{s55}

The self-feedback mechanism discussed in subsection~\ref{s42} has
been used to control the polarized nuclear spin baths of electron
spins in optically pumped quantum dots (QDs), both in ensembles of
dots~\cite{AGreilichNature:2007,SGCarterPRL:2009} and in a single
dot~\cite{XDXuNature:2009,CLattaNatPhys:2009,TDLaddPRL:2010,BSunPRL:2012}.
As an example, we display in Fig.~\ref{Fig of the single quantum
dot self-feedback} the configuration used by Ladd~\textit{et al.}.
In this experiment, a single charged QD is located in a sample
containing $2\times10^9$ cm$^{-2}$ self-assembled InAs QD's grown
by molecular beam epitaxy, situated about 10 nm above
$\delta$-doping layer containing Si donors with a density of $\sim
4\times10^9$ cm$^{-2}$.

The isolated QD is enclosed in a GaAs microcavity in which the
mirrors are composed of several AlAs/GaAs quarter wave-length
layers. The quality factor of the microcavity is $Q=200$. The two
spin states of the quantum-dot electron are split by $4$T external
magnetic field $B_{\rm ext}$ to give a Larmor frequency of $25.3$
GHz. The spin states are optically coupled to a ``trion'' that is
composed of three charged quasiparticles, two electrons and one
hole. The optical pumping flips the electron spin via a stimulated
Raman-transition and the output light can be used to read out the
spin states~\cite{DPressNature:2009,DPressNatPhoton:2010}. After
the pump field flips the electron spin, the recombination of the
trion tends to flip the nuclear spins, thus polarizing them. This
polarization then ``feeds back'' to the electron, in that it
reduces the noise on the electron spin from the spin bath. The
controlled dynamics of the electron spin has a time-scale of $100$
ns, whereas the time it takes the nuclear spins to depolarize is
expected to be milliseconds.

\begin{figure} \centerline{\includegraphics[width = 16.8 cm]{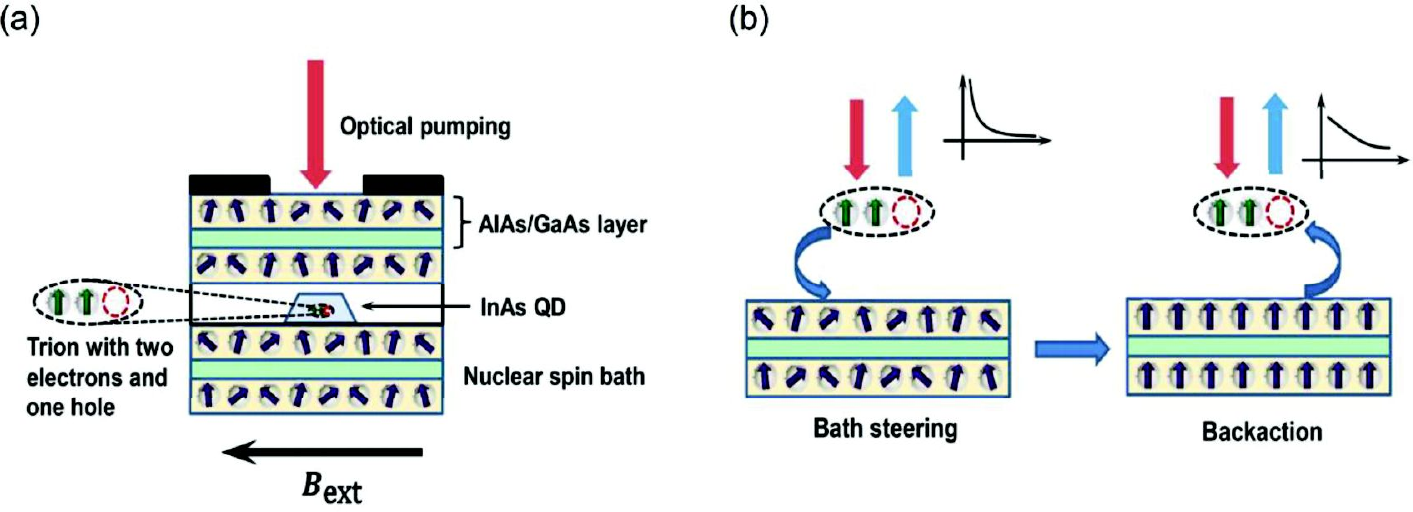}}
\caption{(Color online)
Illustration of the experimental electron-nuclear spin feedback in
optical-pumping single quantum dot systems (see
Ref.~\cite{PBushevPRL:2006}). The self-assembled InAs quantum dots
are embedded in the planar GaAs microcavity with several AlAs/GaAs
$\lambda/4$ layers as the bottom and top mirrors. The pump field
flips the electron spin state, which then suppresses the nuclear
fluctuations. The backaction from the nuclear spin bath protects
the coherence of the electron spin by the hyperfine
interaction.}\label{Fig of the single quantum dot self-feedback}
\end{figure}

Similar quantum self-feedback
protocols~\cite{DJReillyScience:2008,ITVinkNatphys:2009,HBluhmPRL:2010,CBarthelPRB:2012}
have also been realized in microwave-driven double quantum dots
(DQD's). In Fig.~\ref{Fig of the double quantum dot
self-feedback}, we show a diagram of the experiment by
Bluhm~\textit{et al.}~\cite{HBluhmPRL:2010}. An electron is loaded
into each of the quantum dots by locally depleting the
two-dimensional electron gas using electrostatic
gates~\cite{JRPettaScience:2005,SFolettiNatPhys:2009,JRPettaPRL:2009,CBarthelPRL:2009}.
A qubit is formed from the spin states of the two electrons in the
pair of dots, where the two states of the qubit are respectively
the singlet state and the triplet subspace. The states of the
qubit are split by an external magnetic field of $B_{\rm ext}=0.7$
T, applied parallel to the two-dimensional electron gas, and the
inter-dot tunnel coupling is controlled by an external gate with
coupling strength $t_{\mss{c}}=20$ $\mu$eV.

The qubit is measured by the using the fact that the two electrons
cannot occupy the same dot if they have the same spins state, by
virtue of the Pauli exclusion effect. This effect allows the state
of the qubit to be determined by measuring the charge on the dots.
In the experiment by Bluhm~\textit{et al.}, the polarization
induced in the bath of nuclear spins by flipping the electron
spins was able to increase the dephasing time of the electrons
spins from $14$ ns to $94$ ns.

\begin{figure}
\centerline{\includegraphics[width = 12
cm]{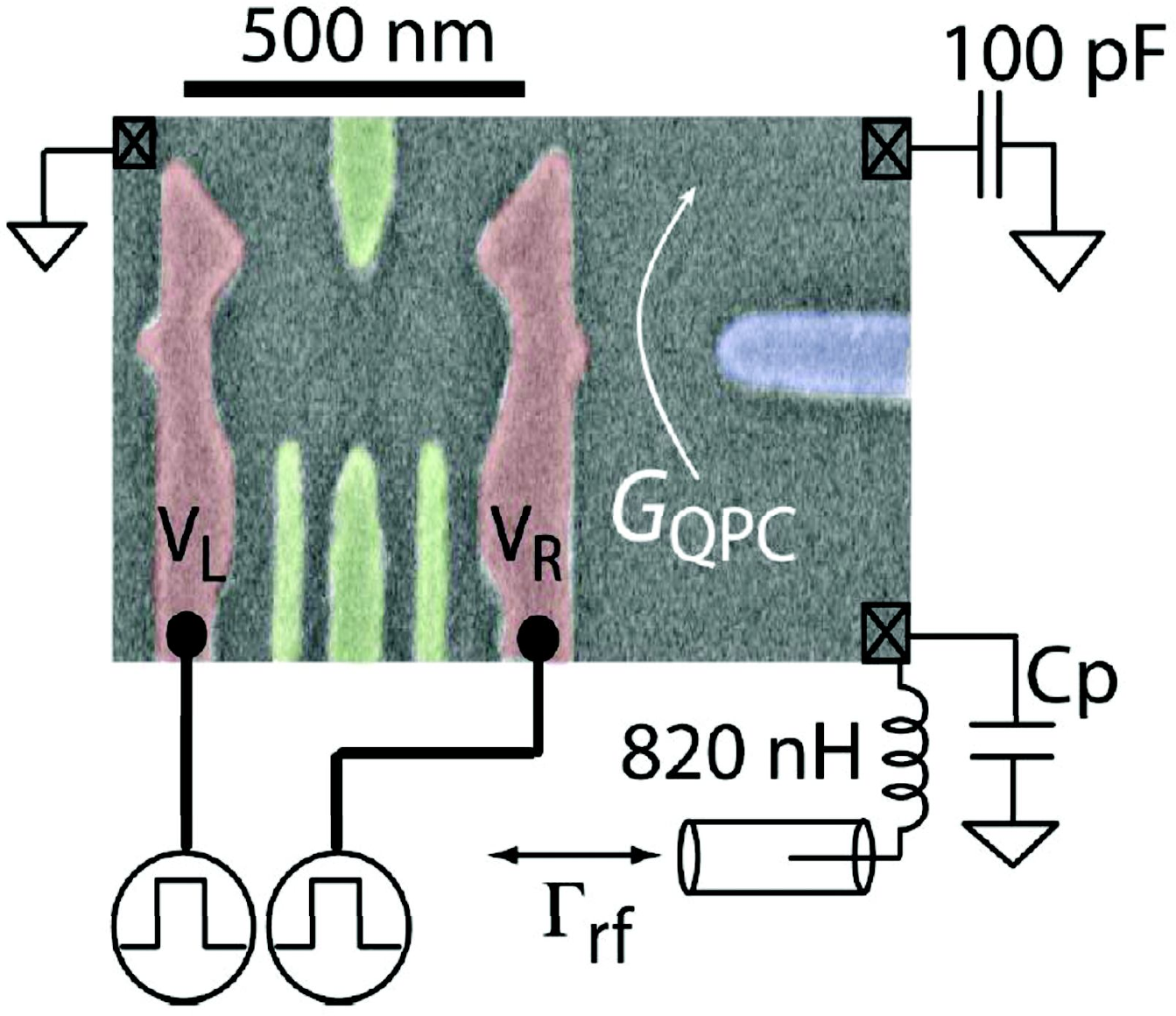}} \caption{(Color online)
Electron-nuclear spin feedback in a microwave-driven double
quantum dot system~\cite{DJReillyScience:2008}. The double quantum
dot is obtained by depleting a two-dimensional electron gas with
electrostatic gates. The electron spins couple to the background
nuclear spins by hyperfine interaction. The hyperfine field
gradient is detected by spin-charge conversion and charge
measurement by a quantum point contact. The control voltage pulses
$V_L$ and $V_R$ are applied to the gates to control the energy
difference of electron spin states. This figure is from
Ref.~\cite{DJReillyScience:2008}.}\label{Fig of the double quantum
dot self-feedback}
\end{figure}

\section{Summary and Outlook}\label{s6}

We have seen in this review that continuous-time feedback can be
implemented with or without measurements, and in the latter case
can be implemented either with direct Hamiltonian coupling between
the system and controller, or with couplings mediated by
irreversible one-way quantum fields. We have also seen that the
range of applications of feedback in quantum systems is rather
broad, and further applications are yet to be discovered.

Measurement-based feedback control of quantum systems was first
demonstrated in quantum optics, and for some time there was no
other field in which this kind of feedback could be realized. It
is only very recently, in 2012, that experimental technology has
allowed measurement-based feedback to be achieved in
superconducting circuits~\cite{RVijayNature:2012, DRistePRL:2012}.
This was possible  because of recent breakthroughs in
quantum-limited amplifiers~\cite{Beltran07, Yamamoto08,
Bergeal10b}. We expect this development to open the door to many
more implementations of measurement-based feedback in both
electrical and electro-mechanical systems.

The situation regarding coherent feedback control is a little
different. Experiments implementing coherent feedback had been
realized for some time --- for example those involving the laser
cooling of atoms and ions --- before the theoretical notion of
coherent feedback was articulated. This notion provides a new way
to think about interacting quantum systems, especially those
coupled via irreversible fields. As pointed out in
Ref.~\cite{JKerckhoffPRL:2012}, coherent feedback provides not
only a tool for controlling quantum systems, but also for changing
the dynamics of a system and thus engineering new dynamics.

While measurement-based feedback control has been studied
theoretically for a little over $20$ years, coherent feedback has
been much less studied, and there are still many basic questions
that remain, particularly to do with the relationship between the
various kinds of feedback. A basic question regarding any kind of
control process is just how well it can perform, given a set of
limitations on the physical control resources, such as the
measurement strength, coupling constant(s), and the speed and
nature of the available control Hamiltonian. Since the dynamics of
most measured quantum systems is nonlinear, and since the question
of the limits to control is essentially one about optimality of
control protocols, it may not be possible to obtain exact answers,
or even numerical answers to these questions for measurement-based
feedback. The connection between measurement-based feedback and
coherent feedback however might provide a new way to analyze such
questions.

There is one question regarding the limits to coherent feedback
control that has been recently answered, at least with
considerable confidence, and that is the limit to the fidelity of
state-preparation given a bound on the rate of the coupling
between the system and auxiliary components~\cite{Wang13}. This
was only solved, however, in the regime of good control, where the
coupling is fast compared to the noise in the system, and in the
regime of weak coupling in which the coupling rate is small
compared to the energy scale of the system (the gaps between the
energy levels of the system). The reason for the weak-coupling
restriction is that without it, the master equation that describes
the noise, even for Markovian weak-damping, is no longer
independent of the control Hamiltonian. The reason that the noise
depends on the control Hamiltonian for strong coupling is that in
this case the energy levels of the system are altered by the
coupling, and the noise that a system experiences usually depends
on the energy levels of the system. As an example, thermal noise
depends on a system's energy levels because the thermal
steady-state depends on these levels. Since the control
Hamiltonian is usually time-dependent, strong control Hamiltonians
imply that the master equation will be time-dependent. Because of
this, up until very recently, theoretical treatments of quantum
control were restricted to the regime of weak coupling/weak
control. Modelling strongly-controlled systems is an important
challenge in quantum control, either by using methods for exact
simulation of open-system dynamics (see e.g.~\cite{Prior10,
Ishizaki05, Bulla03, Thorwart98}) or by obtaining approximate
master equations that correctly model noise for time-dependent
systems~\cite{Albash12}.

A key question regarding the relationship between various kinds of
feedback is whether one kind is superior to the others for certain
applications, or under certain kinds of constraints. Some results
along these lines have already been
obtained~\cite{HINurdinAutomatica:2009,RHamerlyPRL:2012,JZhangTAC:2012,ZPLiuPRA:2013,Jacobs14c}.
It appears that under a constraint on the speed (equivalently the
norm) of the interaction Hamiltonian between the system and
controller, coherent feedback is fundamentally more powerful than
measurement-based feedback, because it allows a larger class of
joint evolutions~\cite{Jacobs14c}. It is not known whether the
same is true under a bound on the measurement strength, equivalent
to a bound on the input-output rate to a field or a damping rate
into a  Markovian bath. What is fairly clear is that a constraint
on an irreversible Markovian damping rate is not equivalent to a
constraint on the norm of an interaction Hamiltonian. This further
suggests that while continuous-time measurement-based feedback can
be compared directly with field-mediated coherent feedback, this
may not be possible with coherent feedback implemented via direct
coupling. Nevertheless the relationship between various forms of
feedback raises questions that are both fundamental and relevant
to potential applications.

Systems driven by white noise obey Markovian master equations,
meaning that the time-derivative of the density matrix depends
only on the density matrix at the current time. One way in which
analyses of feedback control are being extended is to include
systems coupled to baths that induce non-Markovian evolution, or
feedback implemented via fields with a finite or tailored
bandwidth. It turns out that the standard input-output formalism
that we introduced in Section~\ref{s321}, and which is used to
treat field-mediated feedback networks can be extended without
difficulty to couplings with arbitrary
bandwidths~\cite{LDiosiPRA:2012, JZhangPRA:2013}. Interestingly,
with this extension the resulting formalism can handle strong
damping and coupling, something that the standard formalism
cannot. For nonlinear networks, the input-output equations must be
converted to master equations to be solved, and these require
considerable numerical resources. For linear systems, the
input-output formalism provides an efficient means of solution,
and thus appears to be a powerful method for analyzing
non-Markovian feedback for linear systems. It can also be used to
describe feedback in which there is a time-delay in the feedback
control process. With the implementation of feedback in
superconducting circuits, and beyond that to spins in, e.g.,
nitrogen-vacancy centers in diamond or quantum dots, the analysis
of non-Markovian evolution will become increasingly important.

When feedback was introduced into quantum theory in the late 80's
and early 90's, the number of physical systems in which feedback
could be implemented were extremely limited, and such experiments
were very difficult, especially for measurement-based feedback.
With recent breakthroughs in the construction and measurement of
mesoscopic circuits~\cite{ZLXiangRMP:2013}, the number of
experimental applications for feedback control has greatly
increased. We anticipate that the field of quantum feedback
control will expand considerably due to these developments, and
feedback will be realized in an increasing range of mesoscopic
systems, including, e.g., superconducting circuits, quantum dots,
and silicon-based on-chip optical
devices~\cite{KJVahalaNature:2003,TJKippenbergScience:2008,LHeLPR:2013,BPengarxiv:2013,HJingPRL:2014}.

\addcontentsline{toc}{section}{Acknowledgements}
\section*{Acknowledgements}

We would like to thank Prof.~T.~J. Tarn and Dr.~W. Cui for helpful
discussions and Prof.~G.~Y. Xiang for providing related materials.
J. Zhang and R.~B. Wu are supported by the National Natural
Science Foundation of China under Grant Nos. 61174084, 61134008,
60904034, and Y.-X. Liu is supported by this foundation under
Grant Nos. 10975080, 61025022, 60836010. Y.-X. Liu and J. Zhang
are supported by the National Basic Research Program of China (973
Program) under Grant No. 2014CB921401, the Tsinghua University
Initiative Scientific Research Program, and the Tsinghua National
Laboratory for Information Science and Technology (TNList)
Cross-discipline Foundation. K. Jacobs is partially supported by
the US National Science Foundation projects PHY-1005571 and
PHY-1212413, and by the Army Research Office MURI project grant
W911NF-11-1-0268. FN is partially supported by the RIKEN iTHES
Project, MURI Center for Dynamic Magneto-Optics, and a
Grant-in-Aid for Scientific Research (S).

\addcontentsline{toc}{section}{References}
\bibliographystyle{elsarticle-num}

\end{document}